\documentclass[aps,pra,twocolumn,floatfix]{revtex4}

\usepackage{latexsym}
\usepackage{amsmath}
\usepackage{graphicx}
\usepackage{float}
\usepackage[nooneline]{subfigure}

\begin{document}

\title{One-dimensional extended Hubbard model in the atomic limit}

\author{F. Mancini}
\affiliation{ Dipartimento di Fisica "{\it E. R. Caianiello}" -
Unit\`a CNISM di Salerno \\ Universit\`a degli Studi di Salerno,
Via S. Allende, I-84081 Baronissi (SA), Italy}

\author{F. P. Mancini}
\affiliation{ Dipartimento di Fisica and Sezione I.N.F.N.
\\
Universit\`a degli Studi di Perugia, Via A. Pascoli, I-06123
Perugia, Italy}

\begin{abstract}

We present the exact solution of the one-dimensional extended
Hubbard model in the atomic limit within the Green's function and
equation of motion formalism. We provide a comprehensive and
systematic analysis of the model by considering all the relevant
response and correlation functions as well as thermodynamic
quantities in the whole parameter space. At zero temperature we
identify four phases in the plane ($U,n$) ($U$ is the on-site
potential and $n$ is the filling) and relative phase transitions
as well as different types of charge ordering. These features are
endorsed by investigating at $T=0$ the chemical potential and
pertinent local correlators, the particle and double occupancy
correlation functions, the entropy, and by studying the behavior
in the limit $T \to 0$ of the charge and spin susceptibilities. A
detailed study of the thermodynamic quantities is also presented
at finite temperature. This study evidences that a finite-range
order persists for a wide range of the temperature, as shown by
the behavior of the correlation functions and by the two-peak
structure exhibited by the charge susceptibility and by the
entropy. Moreover, the equation of motion formalism, together with
the use of composite operators, allows us to exactly determine the
set of elementary excitations. As a result, the density of states
can be determined exactly and a detailed analysis of the specific
heat allows for identifying the excitations and for ascribing its
two-peak structure to a redistribution of the charge density.

\end{abstract}

\date{\today}
\maketitle

\section{introduction}

It goes without saying that the Hubbard model \cite{Hubbard63} is
of seminal importance in the study of modern condensed matter
theory. Originally introduced to describe the correlations of
electrons in the narrow $d$ band of transition metals, the Hubbard
model has been subsequently used to investigate a large variety of
situations experimentally observed. Upon varying the model
parameters, it is believed that the Hubbard model can describe
many properties of strongly correlated electronic systems.
However, its applicability to real materials is rather limited and
some extensions are needed. Among various generalizations, one of
the most studied is the so-called extended Hubbard model (EHM),
where a nearest-neighbor interaction $V$ is added to the original
Hubbard Hamiltonian, which includes only an on-site interaction
$U$. If one assumes that the interaction parameters arise solely
from Coulomb interactions, then both $U$ and $V$ are repulsive.
However, these parameters may represent effective interactions.
Thus their ranges can be broader, allowing for both positive and
negative values. The electronic properties of the model are
substantially modified by including nonlocal interactions allowing
one to study general features of doped systems like copper-oxide
materials \cite{co_mat}, high-temperature superconductors
\cite{Gabovich02}, manganese compounds \cite{Dagotto02}, and to
describe charge ordering (CO), experimentally observed, in a
variety of systems. Indeed, several studies of the extended
Hubbard model have pointed to the presence of states with phase
separation \cite{pha_sep} and of charge-ordered phases
\cite{pietig_99,co_phas}. Furthermore, the introduction of an
intersite interaction can mime longer-ranged Coulomb interactions
needed to describe effects observed in conducting polymers
\cite{baeriswyl_92}.

Although the Hubbard model is oversimplified and in spite of very
intensive study, both analytical and numerical, exact solutions
exist only in one or infinite dimensions. In 1D the thermodynamics
of the Hubbard model has been widely investigated by means of
different exact approaches for both nearest-neighbor hopping
\cite{nn_hop,juttner_98} and long-range hopping \cite{gebhard_94}.
For higher dimensions ($D>1$), the study is far from being
completed and only few rigorous results are known. In particular,
relevant results were obtained by exact diagonalization on small
clusters \cite{cluster} and by numerical investigations
\cite{note}. For the case of infinite dimension, an exact solution
has been obtained by dynamical field theory \cite{georges_96}.

The situation is more severe for the extended version of the
model: even in 1D there is not a complete exact solution. A
particular extended version has been examined in Ref.
\cite{dolcini_02}, where a kinetic term ($Y$) for paired carriers
 is added to the usual electron hopping
($t$). The model has been solved only in the particular case of
$t=-Y$. Actually, it is known that the EHM is non-integrable for
general values of the model parameters \cite{poilblanc_93}. These
difficulties have led to the investigation of the so-called narrow
band limit, where the hopping $t$ can be neglected with respect to
the interaction couplings $U$ and $V$. This simplified version,
also known as the atomic limit of the extended Hubbard model
(AL-EHM), has been largely studied for several years.

There is a large and diffused interest in the study of this
relatively simple model, motivated by several reasons: (i) the
phase diagram has a very rich structure with a multicritical
behavior; (ii) one observes several kinds of long-range orders
pertinent to commensurate sublattice charge-density alternation
(long-range order has been observed in many narrow-band materials,
such as $BaBi_x Pb_{1-x} O_3 $, $Cs_2SbCl_6 $, $Fe_3 O_4 $,
$CsAuCl_3 $, and several tetracyanoquinodimethanide salts); (iii)
the results for the atomic model may give some information beyond
the zero-bandwidth limit [several approximate studies, as well as
some exact results for the 1D case \cite{Hubbard79}, hint at the
possibility that charge ordering persists in the corresponding
narrow band model]; (iv) the knowledge of the atomic limit can be
used as a starting point for a perturbation expansion in powers of
the hopping $t$.

Without pretending to be exhaustive, a review of the different
approaches available in the literature could be presented as
follows. The phase diagram of the AL-EHM at half filling ($n=1$)
has been investigated in several papers by means of a modified
Hartree-Fock theory \cite{Bari71}, beyond Hartree-Fock
approximation employing a decoupling procedure \cite{Ihle73},
variational approach \cite{Robaszkiewicz79} and Bethe-Peierls
approximation \cite{Rice81}. The molecular field approximation
\cite{micnas_84} has been used to obtain the phase diagram in the
 ($n,U/zV$) plane and in the  ($\mu$, $U/zV$) plane ($z$ being the
coordination number and $\mu$ the chemical potential). Another
approach makes use of the transfer-matrix method that, although
practically employable only for one-dimensional systems, can
provide exact results. In Refs. \cite{Tu74, Beni74} the specific
heat and the static magnetic susceptibility have been studied at
half filling. The case of quarter filling, in the limit $U\to
\infty$, has been considered in Ref. \cite{Beni74}, where exact
expressions for the charge-charge correlation function have been
obtained for $n=1$ and $n=0.5$ ($U\to \infty )$. For $n=1$ and
$U>2V$, the ground state is a Mott insulator with one electron per
atom; when $U<2V$ the ground state is a CO state with one
sublattice doubly occupied and the other empty. The phase
transition (PT) from the CO state to the Mott state can be either
first or second order, and the phase diagram exhibits a
tricritical point.

A method of constructing the ground state phase diagram has been
formulated in Refs. \cite{Frohlich78_80} by exploiting the
reflection positivity property with respect to reflections in
lattice planes exhibited by the AL-EHM. Estimates of the
probability of some events at nonzero temperatures were also
provided. In Refs. \cite{Jedrzejewski85_94} the AL-EHM has been
studied in 2D addressing the problem of determining the zero
temperature phase diagram in the ($\mu /V$, $U/V$) plane. The
ground state exhibits several kinds of charge long-range orders,
which persist also at sufficiently low temperatures, provided one
stays sufficiently far away from the boundary between the
different regions. This study reports the absence of PTs among the
different regimes at finite temperatures, in contrast to the
results of the mean-field approximation \cite{micnas_84}. It is
also suggested that the staggered charge order persists in the
corresponding narrow band model, although there is no rigorous
proof since the reflection positivity property fails to be true
for non-zero $t$.

Another category of studies is based on the use of the Pirogov and
Sinai methods \cite{Pirogov_Sinai}. This approach, originally
conceived to investigate classical spin systems at low
temperatures, has been extended to quantum systems: spin,
fermionic and bosonic models. When applied to the AL-EHM, these
methods allow for a detailed study of the low temperature phase
diagram. Moreover, they allow for an extension to non-zero $t$,
treating the narrow band Hubbard model as a quantum perturbation
of the $t=0$ model. In Ref. \cite{Borgs96} the phase diagram of
the AL-EHM has been derived. At zero temperature, the phase
diagram in the $(U,\mu)$ plane shows six phases: three homogeneous
(characterized by particle densities $n=0,1,2$ and a staggered
charge density $\Delta =0)$ and three staggered (characterized by
particle densities $n=1/2,1,3/2$ and staggered charge density
$\Delta =1/2,1,1/2)$. The staggered order parameter being defined
as $\Delta =\lim_{\Lambda \to \infty }\sum_i \varepsilon
(i)n(i)/\Lambda$, where $\varepsilon (i)$ assumes two values, 1 or
-1, depending on which sublattice of the bipartite lattice
$\Lambda$ the site $i$ belongs to. For any sufficiently low
non-zero temperature, the three staggered phases merge into one,
without any PT. These results have been confirmed in Ref.
\cite{Frohlich01}, where also the case of attractive intersite
potential ($V<0$) has been considered. The narrow-band extended
Hubbard model in $D \ge 2$ has been studied in Ref.
\cite{Borgs00}, where it has been shown that the staggered charge
order persists for sufficiently small values of the hopping
parameter $t$. A comprehensive study of the phase diagram of the
1D EHM was carried out in Ref. \cite{lin_00} using different
methods: perturbation analysis in the strong and weak coupling
limits, quantum Monte Carlo in the intermediate coupling region
and exact diagonalization calculations. The authors individuated
various phases, including charge and spin density waves,
superconductivity, and phase separation. G-ology investigations
\cite{gology,tsuchiizu02}, renormalization group \cite{rg}, and
bosonization \cite{bosons} studies have provided analytic
insights, particularly in the weak coupling regime.

Finally, a large activity by numerical simulation must be
mentioned. The EHM has also been studied by means of quantum Monte
Carlo (QMC) simulations \cite{Hirsch84,sengupta_02}, Lanczos
technique \cite{cannon_Hellberg}, exact diagonalization
calculations \cite{80s,calandra_02}, density-matrix
renormalization group \cite{dmrg}. The phase diagram and some
thermodynamical properties at half filling of the EHM have been
studied by means of the density-matrix renormalization group
applied to transfer matrices \cite{Glocke07}. Recently, some
interesting results in the study of the AL-EHM have been obtained
by means of Monte Carlo (MC) simulation, for dimensions greater
than 1. In Ref. \cite{Misawa06}, by combining MC and mean-field
approximation, the phase diagram at fixed $\mu$ has been derived
both at $T=0$ and finite temperatures, and the critical behavior
in the vicinity of the tricritical points has been analyzed. In
Ref. \cite{pawlowski_06} MC simulation has been used to study the
AL-EHM on a square lattice ($L=10$-40);  a detailed study of the
model at finite temperature has been presented for different
values of the filling and/or chemical potential, evidencing the
presence of different charge orderings and tricritical point
lines, and the possibility of phase separation and formation of
stripes. The variational cluster approach has also been
successfully applied for the EHM in one and two dimensions,
showing that QMC and density-matrix renormalization-group results
can be reproduced with very good accuracy \cite{aichhorn04}. Most
of these analyses have focused on the half filling case, paying
attention to the boundary line separating spin-density wave from
charge-density wave, as well as to the existence and location of a
tricritical point. In the last few years, there have been
discussions galore on the existence of an intermediate bond-order
wave phase \cite{nakamura}, whose presence seems widely confirmed
by now \cite{tam_06}. The case of quarter filling has been
examined in several works, providing precise estimates of the
charge gap \cite{sano_07}, hinting at the possibility of
superconductivity in an unexpected region of the ($U$, $V$) plane
\cite{mila_93_4} and the existence of a CO transition
\cite{calandra_02,pietig_99}.

As it emerges from the above review, an exact, rigorous and
detailed study of the EHM and, in particular, of the
one-dimensional case, for different values of the parameters $n$,
$T$, $U$, and $V$, is still missing in the literature. Most of the
existing studies have been restricted to the case of half filling.
The analysis given in Refs. \cite{Jedrzejewski85_94, Borgs96,
Borgs00, Frohlich01} allows for a complete description of the
phase diagram at zero temperature, but in the ($\mu$, $U$) plane.
Moreover, in several works, many properties of the system are
presented as a function of the chemical potential $\mu$. Although
from a theoretical point of view the chemical potential and the
particle density $n$ are conjugated variables, experimentally it
is possible to tune the density in a controlled way, by varying
the doping. The chemical potential is an internal parameter,
dynamically determined by the system itself according the values
of the external parameters. It is very difficult to induce
transformations at fixed chemical potential. Moreover, by fixing
the particle density or the chemical potential one can obtain
different results, if the canonical and grand canonical ensembles
are inequivalent due to a divergence  of the compressibility. For
instance, if the chemical potential is considered as an external
parameter, one finds that a solution exists only for discrete
values of the particle density ($n=1/2$, 1, 3/2,2). Thus, all the
studies where $\mu$ is considered as an independent variable are
doubtless interesting from a mathematical point of view, but are
not directly related to experimental situations. It is just more
physical to consider the particle density number $n=N_e/N$ - where
$N_e$ is the total number of electrons and $N$ is the number of
sites - as an external parameter, determined by the particular
material under study. In real materials $n$ can vary in a large
range: in cuprate materials and in conducting polymers, $n=1$
(half filling) in the insulating state, but largely varies under
doping \cite{baeriswyl_92}; in charge transfer salts $n$ varies by
varying the pressure in the neighborhood of $n=2/3$ \cite{su_81};
Bechgaard salts have $n=1/2$ (quarter filling) \cite{mazumdar_90}.

It is worthwhile to recall that there is a correspondence between
the atomic limit of the extended Hubbard model and the spin-1
Ising model defined on the same lattice. A transformation from the
fermionic to the spin Hamiltonian can be performed by defining the
pseudospin variable $S(i)=n(i)-1$. Under this transformation, one
ends up with a Hamiltonian pertinent to a spin-1 Ising model with
nearest-neighbor exchange $J=-V$, in the presence of an effective
crystal field $\Delta=U/2+k_B T \log 2$ and of an external
magnetic field $h=\mu -2V-U/2$. The temperature dependence of the
crystal field is due to the spin degeneracy induced by the mapping
between $S$ and $n$. The literature on the spin-1 Ising model is
vast and we refer the interested reader to the bibliography given
in Ref. \cite{mancini05b}. However, here we would like to comment
on the equivalence between the two models. There is a dramatic
difference from a physical point of view because, for a spin
system, it is possible to tune the magnitude of the applied
magnetic field $h$ and vary the temperature $T$. The system
responds to such variations by rearranging the spins in a
pertinent configuration, described, in average, by the
magnetization $m= \langle S \rangle$. For the fermionic model, the
external thermodynamic parameters are $n$ and $T$: the system
responds to variations of these parameters by adjusting the
chemical potential $\mu$.

In this work we  present a study of the 1D AL-EHM by using a new
method based on the equations of motion and Green's function
formalism. Recently it has been shown \cite{Mancini05_06} that a
system built up of $q$ species of localized Fermi particles
subject to finite-range interactions, is exactly solvable in any
dimension. By exactly solvable we mean that it is always possible
to find a complete set of eigenenergies and eigenoperators of the
Hamiltonian closing the hierarchy of the equations of motion.
Thus, one can obtain exact expressions for the relevant Green's
functions and correlation functions depending on a finite set of
parameters, whose knowledge is required in order to have a
complete solution of the system. In the case of a one-dimensional
lattice, for $q =1,2,3$ \cite{mancini05b,Mancini05a,Avella06}, and
in the case of a Bethe lattice for $q =1$ \cite{Mancini06a}, it
has been shown how it is possible to exactly fix such parameters
by means of algebra constraints. The analytical exact solution of
the 1D AL-EHM was reported in Ref. \cite{mancini05b}, where
preliminary results were given for vanishing on-site potential.
Here we present a systematic and detailed study of the model. We
study the properties of the system as a function of the external
parameters $n$, $T/V$, $U/V$ (throughout the paper we set $V>0$ as
the unit of energy) allowing for the on-site interaction $U$ to be
both repulsive and attractive. Owing to the particle-hole
symmetry, it is sufficient to explore the interval $[0,1]$ for the
parameter $n$. The chemical potential is then self-consistently
determined as a function of the external parameters. We also
address the problem of determining the zero temperature phase
diagram in the ($U/V$, $n$) plane. Relevant local quantities, such
as the double occupancy, the next-nearest correlation functions,
the density of states, and the correlation functions for the
charge and double occupancy are systematically computed both at
$T=0$ and as functions of the temperature.

For specific values of the particle density, we find regular
spatial distributions of the electrons determined by the two
competing terms in the Hamiltonian, i.e., the on-site and the
intersite interactions. For particle densities less than quarter
filling we do not observe CO states by varying the on-site
potential $U$. At $U=0$ there is a CO transition at quarter
filling from a spatial disordered configuration where the
electrons are coupled in pairs to a checkerboard distribution of
singly occupied sites for $U>0$. For particle densities between
quarter and half filling, we find three possible spatial
arrangements of the electrons by varying $U$. For attractive
on-site potential the electrons are still disorderly coupled in
pairs; at $U=0$ one observes a transition to a peculiar ordered
state with alternating empty and occupied sites. The peculiarity
consists in the fact that, upon fixing the particle density, some
of the singly occupied sites become doubly occupied - with the
constraint that the double occupancy $D$ is equal to $(n-1/2)$ -
and the possible distributions along the chain are energetically
equivalent. Further increasing the on-site potential, a second
transition to a disordered distribution of only singly occupied
sites is observed at $U=2V$. At half filling, we find the
well-known transition at $U=2V$ from a state where every second
site is doubly occupied to a state with one electron per site
\cite{Bari71}. We also find the interesting result that the
correlation functions are controlled by two correlation lengths
$\xi_n$ and $\xi_D$, though the latter is relevant only at $T \neq
0$.

On the other hand, if one fixes the on-site potential and allows
the particle density to vary, one observes PTs only at quarter and
half filling, signalled by the divergence of the correlation
length $\xi_n$. For fixed attractive on-site potential, the
correlation length $\xi_n$ diverges only at half filling and thus,
a long-range order is established. For values of on-site potential
$0<U<2V$ the correlation length $\xi_n$ diverges for all values of
the filling in between quarter and half filling, but only at these
extremal values does one observe a CO state. For large repulsive
on-site potential, $\xi_n$ diverges only at quarter filling, where
one finds a regular alternating distribution of singly occupied
sites. The further increase of the particle density leads to an
increasing single occupation of the empty sites until the
homogeneous distribution is reached at half filling.

At finite but low temperatures, one expects that the scenario
previously depicted still holds. This wisdom is strengthened by
the behavior of thermodynamic quantities such as the
compressibility, the charge and spin susceptibilities, the
internal energy, the specific heat, and the entropy. The
computation of the entropy requires some attention since the
ground state is macroscopically degenerate (unless for $U<2V$ and
$n=1$) and we find, thus, a finite zero-temperature entropy.

Before ending this section, we would like to stress the
motivations of our work. Firstly, we use a new method. The
transfer matrix method is the most common approach used in the
literature to obtain analytical and exact solutions of the AL-EHM
and, thus, allowing for performing explicit calculations of the
relevant correlation functions and thermodynamic properties.
However, this method is basically restricted to 1D systems. There
are great difficulties in extending the method to higher
dimensions, as the famous Onsager solution of the 2D spin-1/2
Ising model shows \cite{onsager_44}. Our approach has the
advantage of offering a general formulation for any dimension
\cite{Mancini05_06}. The problem within  our approach is that the
exact (in principle) solution is expressed in terms of a set of
parameters not easily computable by means of the dynamics. The
solution of this problem has been found for 1D systems
\cite{mancini05b,Mancini05_06,Mancini05a,Avella06} and Bethe
lattices \cite{Mancini06a,mancini07}; more complicated lattice
structures are currently under investigation. Secondly, our
approach provides us with an insight into the thermodynamic
properties of the AL-EHM. Besides the exact determination of the
zero-temperature phase diagram in the full range of the
parameters, one is also capable to analyze in detail the behavior
of several thermodynamic quantities such as the entropy, the
specific heat and the charge and spin susceptibilities.
Furthermore, the study of these quantities allows one also to
ascertain the different PTs occurring at zero temperature. Last
but not least, the Green's function formalism and the equation of
motion method allows one to exactly determine the energy spectra
and, as a result, to identify the excitation energies contributing
to the specific heat. As a consequence, we are able to enlighten
the issue of the nature of the peaks in the specific heat: we show
that the double-peak structure is only due to charge excitations
described by the Hubbard projection operators.

The paper is organized as follows. In Sec. \ref{sec_II}, upon
introducing the Hubbard composite operators, we briefly review the
analysis leading to the algebra closure and to analytical
expressions of the retarded Green's functions (GF) and correlation
functions (CF). Since the composite operators do not satisfy a
canonical algebra, the GF and CF depend on a set of internal
parameters, leading only to exact relations among the CF.
According to the scheme of the composite operator method
\cite{manciniavella}, it is possible to determine these parameters
by means of algebra constraints fixing the representation of the
GF. By following this scheme, one can obtain extra equations
closing the set of relations and allowing for an exact and
complete solution of the 1D AL-EHM. In Sec. \ref{sec_III} we
analyze the properties of the system at zero temperature. We
characterize the different phases emerging in the phase diagram
drawn in the ($U,n$) plane, by studying the behavior, as functions
of $n$ and $U$, of the chemical potential and of various local
properties (double occupancy, next-nearest correlation functions);
the density of states and the charge and double occupancy
correlation functions are also presented for various values of $n$
and $U$. In the following, we shall denote with charge and double
occupancy, the particle-particle and the double occupancy-double
occupancy correlation functions, respectively. Section
\ref{sec_IV} is devoted to the study of the finite temperature
properties. Further to the study of all the quantities analyzed in
Sec. \ref{sec_III}, we also present results for the
compressibility, the charge and spin susceptibilities, the
specific heat, and the entropy. Finally, Sec. \ref{sec_V} is
devoted to our conclusions and final remarks, while the appendices
report some relevant  but rather cumbersome computational details.

\section{Exact solution of the 1D atomic extended Hubbard model
} \label{sec_II}

The exact solution of the 1D AL-EHM was presented in Ref.
\cite{mancini05b}, where the central issue was the case of
vanishing on-site potential ($U=0$). In this section we shall
briefly review the results previously obtained, providing a
simpler framework of calculation and extending the solution to
nonlocal correlation functions. This new framework allows one to
study the properties of the system in the entire region of the
external parameters $n$, $T/V$, $U/V$.

By considering the atomic limit, the Hamiltonian of the extended
Hubbard model in one dimension is given by
\begin{equation}
  \label{EHM_1}
 H=\sum_i \left[-\mu n(i)+UD(i)+V n(i)n^\alpha
(i)\right],
\end{equation}
where $n(i)=c^\dag (i)c(i)$ is the charge density operator, $c(i)$
$[c^\dag (i)]$ is the electron annihilation (creation) operator -
in the spinor notation - satisfying canonical anticommutation
relations. We use the Heisenberg picture: $i=({\bf i},t)$, where
{\bf i} stands for the lattice vector ${\bf R}_i$. The Bravais
lattice is a linear chain of $N$ sites with lattice constant $a$.
$\mu$ is the chemical potential; $U$ and $V$ are the strengths of
the local and intersite interactions, respectively.
$D(i)=n_\uparrow (i)n_\downarrow (i)=n(i)[n(i)-1]/2$ is the double
occupancy operator. Hereafter, for a generic operator $\Phi(i)$,
we shall use the following notation
\begin{equation*}
\Phi ^\alpha (i,t)=\sum_j \alpha _{ij} \Phi (j,t),
\end{equation*}
where $\alpha_{ij}$ is the projection operator over nearest
neighboring sites
\begin{equation*}
\alpha_{ij} =\frac{1}{N}\sum_k e^{ik  (R_i -R_j )}\alpha (k)\quad
\quad \quad \alpha (k)=\cos (ka),
\end{equation*}
and the sum is over the first Brillouin zone.

\subsection{Eigenoperators and eigenvalues}
\label{II_A}

Upon introducing the Hubbard operators $\xi (i)=[n(i)-1]c(i)$ and
$\eta (i)=n(i)c(i)$, one may define the composite field operator
\begin{equation}
\label{EHM_2} \psi (i)=\left( \begin{array}{*{20}c}
 {\psi ^{(\xi )}(i)}   \\
 {\psi ^{(\eta )}(i)}   \\
\end{array}  \right),
\end{equation}
where
\begin{equation}
\label{EHM_3}
\begin{split}
\psi ^{(\xi )}(i)&=\left( {{\begin{array}{*{20}c}
 {\psi _1^{(\xi )} (i)}   \\
 {\psi _2^{(\xi )} (i)}   \\
 \vdots   \\
 {\psi _{5}^{(\xi )} (i)}   \\
\end{array} }} \right)=\left( {{\begin{array}{*{20}c}
 {\xi (i)}   \\
 {\xi (i)[n^\alpha (i)]}   \\
 \vdots   \\
 {\xi (i)[n^\alpha (i)]^{4}}   \\
\end{array} }} \right), \\
\psi ^{(\eta )}(i)&=\left( {{\begin{array}{*{20}c}
 {\psi _1^{(\eta )} (i)}   \\
 {\psi _2^{(\eta )} (i)}   \\
 \vdots   \\
 {\psi _{5}^{(\eta )} (i)}   \\
\end{array} }} \right)
=\left( {{\begin{array}{*{20}c}
 {\eta (i)}   \\
 {\eta (i)[n^\alpha (i)]}   \\
 \vdots   \\
 {\eta (i)[n^\alpha (i)]^{4}}   \\
\end{array} }}
\right).
\end{split}
\end{equation}
By exploiting the algebraic properties of the operators $n(i)$ and
$D(i)$, it is easy to show that the fields $\psi ^{(\xi )}(i)$ and
$\psi ^{(\eta )}(i)$ are eigenoperators of the Hamiltonian
\eqref{EHM_1} \cite{mancini05b}:
\begin{equation}
\label{EHM_4}
\begin{split}
 i\frac{\partial }{\partial t}\psi ^{(\xi )}(i) &=[\psi ^{(\xi
)}(i),H]=\varepsilon ^{(\xi )}\psi ^{(\xi )}(i),
\\
 i\frac{\partial }{\partial t}\psi ^{(\eta )}(i)&=[\psi ^{(\eta
)}(i),H]=\varepsilon ^{(\eta )}\psi ^{(\eta )}(i).
 \end{split}
\end{equation}
In Eq. \eqref{EHM_4}, $\varepsilon ^{(\xi )}$ and $\varepsilon
^{(\eta )}$ are the $5\times 5$ energy matrices \cite{mancini05b}
\begin{equation}
\label{EHM_5} \varepsilon ^{(\xi )}=
\begin{pmatrix}
 {-\mu }   & {2V}   & 0   & 0   & 0   \\
 0   & {-\mu }   & {2V}   & 0   & 0   \\
 0   & 0   & {-\mu }   & {2V}   & 0   \\
 0   & 0   & 0   & {-\mu }   & {2V}   \\
 0   & {-3V}   & {\frac{25}{2}V}   & {-\frac{35}{2}V}
& {-\mu +10V}   \\
\end{pmatrix},
\end{equation}
\begin{equation}
\label{EHM_6} \varepsilon ^{(\eta )}= \left(
\begin{array}{ccccc}
 {U-\mu }  & {2V}  & 0  & 0  & 0  \\
 0 & {U-\mu }  & {2V}  & 0  & 0  \\
 0  & 0  & {U-\mu }  & {2V}  & 0  \\
 0  & 0  & 0  & {U-\mu }  & {2V}  \\
 0 & {-3V}  & {\frac{25}{2}V}  & {-\frac{35}{2}V}
& {U-\mu +10V}  \\
\end{array} \right),
\end{equation}
whose eigenvalues, $E_m^{(\xi )} $ and $E_m^{(\eta )}$, are given
by
\begin{equation}
\label{EHM_7}
\begin{split}
 E_m^{(\xi )} &=-\mu +(m-1)V  ,   \\
 E_m^{(\eta )} &=-\mu +U+(m-1)V
\end{split}
\quad \quad \quad \quad \{m=1,\cdots ,5\}.
\end{equation}

\subsection{Retarded Green's functions and correlation functions}
\label{II_B}

The knowledge of a complete set of eigenoperators and
eigenenergies of the Hamiltonian allows for an exact expression of
the retarded Green's function
\begin{equation}
\begin{split}
\label{EHM_8} G^{(s)}(t-t')&=\theta (t-t')\langle\{\psi
^{(s)}(0,t), {\psi^{s}}^\dag
 (0,t')\}\rangle
\\
  &=\frac{i}{(2\pi
)}\int\limits_{-\infty }^{+\infty } d\omega {\kern 1pt}e^{-i\omega
(t-t')}G^{(s)}(\omega ),
\end{split}
\end{equation}
and, consequently, of the correlation function
\begin{equation}
\label{EHM_9}
\begin{split}C^{(s)}(t-t')&=\langle \psi^{(s)}(0,t) {\psi^{(s)}}
^\dag (0,t')\rangle
\\
&=\frac{1}{(2\pi )}\int\limits_{-\infty }^{+\infty } d\omega
{\kern 1pt}e^{-i\omega (t-t')}C^{(s)}(\omega ).
\end{split}
\end{equation}
In the above equations, $s=\xi ,\eta$ and $\langle\cdots \rangle$
denotes the quantum-statistical average over the grand canonical
ensemble. One finds \cite{mancini05b}
\begin{equation}
\begin{split}
\label{EHM_10_11} G^{(s)}(\omega )&=\sum_{m=1}^5 \frac{\sigma
^{(s,m)}}{\omega -E_m^{(s)} +i\delta },
\\
 C^{(s)}(\omega )&=\pi \sum_{m=1}^5 \sigma ^{(s,m)} \, T_m^{(s)}
\, \delta (\omega -E_m^{(s)} ),
\end{split}
\end{equation}
where $T_m^{(s)} =1+\tanh \big( \beta E_m^{(s)}/2 \big)$, $\beta
=1/k_B T$, and the spectral density matrices $\sigma ^{(s,m)}$ are
given by
\begin{equation}
\label{EHM_12} \sigma ^{(s,m)}=\Sigma_m^{(s)} \Gamma ^{(m)},
\end{equation}
where $m=1,\cdots ,5$; $\Gamma^{(m)}$ are matrices of rank $5
\times 5$:
\begin{equation*}
\begin{split}
\Gamma_{l,k}^{(1)} &=\delta_{l,1} \delta _{k,1} ,  \\
\Gamma_{l,k}^{(m)} &=\left({\frac{m-1}{2}} \right)^{l+k-2},
\end{split}
\end{equation*}
where $m=2,\cdots,5$ and the $\Sigma _m^{(s)}$ are given by
\begin{equation}
\begin{split}
\Sigma _1^{(s)} &=\frac{1}{6}\left(6I_{1,1}^{(s)} -25I_{1,2}^{(s)}
+35 I_{1,3}^{(s)} -20I_{1,4}^{(s)} +4I_{1,5}^{(s)} \right),
\\
\Sigma _2^{(s)} &=\frac{4}{3}\left(6I_{1,2}^{(s)} -13I_{1,3}^{(s)}
+9I_{1,4}^{(s)} -2I_{1,5}^{(s)} \right),
 \\
\Sigma _3^{(s)} &=-6I_{1,2}^{(s)} +19I_{1,3}^{(s)}
-16I_{1,4}^{(s)}+4I_{1,5}^{(s)},
\\
 \Sigma _4^{(s)} &=\frac{4}{3}\left(2I_{1,2}^{(s)} -7I_{1,3}^{(s)}
+7I_{1,4}^{(s)} -2I_{1,5}^{(s)} \right),
 \\
 \Sigma _5^{(s)} &=\frac{1}{6}\left(-3I_{1,2}^{(s)} +11I_{1,3}^{(s)}
-12I_{1,4}^{(s)} +4I_{1,5}^{(s)} \right).
\end{split}
\label{EHM_13}
\end{equation}
$I_{a,b}^{(s)} $ are the elements of the normalization matrix
$I^{(s)}=\langle\{\psi ^{(s)}(i), {\psi ^{(s)}}^\dag (i)\}\rangle$
which can be expressed as
\begin{equation}
\label{EHM_14}
\begin{split}
 I_{1,k}^{(\xi )} &=\kappa ^{(k-1)}-\lambda ^{(k-1)},
 \quad \quad  \kappa ^{(p)}=\langle[n^\alpha (i)]^p\rangle ,
  \\
 I_{1,k}^{(\eta )} &=\lambda ^{(k-1)},
 \quad \quad \quad \quad \quad \quad
 \lambda ^{(p)}=\frac{1}{2}\langle n(i)[n^\alpha (i)]^p\rangle .
\end{split}
\end{equation}
Thus, the CFs depend on the  external parameters $n=\langle n(i)
\rangle$, $T$, $U$, $V$ and on the internal parameters: $\mu$,
$\kappa ^{(p)}$, and  $\lambda ^{(p)}$. It is easy to show that
the CFs obey the following self-consistent equations:
\begin{equation}
\label{EHM_15} C^{(\xi)}_{1,k}+ C^{(\eta)}_{1,k}=
 \kappa ^{(k-1)}-\lambda ^{(k-1)}
 \quad (k=1,\cdots 5).
\end{equation}
The number of these equations is not sufficient to determine all
the internal parameters, and one needs other equations. This
problem will be considered in the next subsection, where a
self-consistent scheme, capable to compute the internal
parameters, will be presented.

\subsection{Self-consistent equations}
\label{II_C}

The previous analysis shows that the complete solution of the
model requires the knowledge of the parameters $\mu ,\kappa
^{(k)}$ and $\lambda ^{(k)}$. These quantities may be computed by
using algebra constraints and symmetry requirements
\cite{mancini05b}. By recalling the projection nature of the
Hubbard operators $\xi(i)$ and $\eta(i)$, it is easy to see that
the following algebraic properties hold:
\begin{equation}
\label{EHM_16}
\begin{split}
\xi^\dag(i) n(i)&=0 \quad  \quad \eta^\dag(i) n(i)=\eta^\dag(i) ,
\\
\xi^\dag(i) D(i)&=0 \quad  \quad \eta^\dag(i) D(i)=0 .
\end{split}
\end{equation}
These are fundamental relations and constitute the basis to
construct a self-consistent procedure to compute the various
unknown parameters of the solution. One first fixes an arbitrary
site, say $i$, of the chain, then splits the Hamiltonian
\eqref{EHM_1} in the sum of two terms:
\begin{equation}
\label{EHM_17}
\begin{split}
 H &=H_0(i) +H_I(i) ,\\
 H_I(i) &=2Vn(i)n^\alpha (i),
 \end{split}
\end{equation}
and introduce the $H_0(i)$ representation: the statistical average
of any operator $O$ can be expressed as
\begin{equation}
\label{EHM_18}
 \langle O\rangle =\frac{\langle Oe^{-\beta H_I(i)
}\rangle _{0,i}} {\langle e^{-\beta H_I(i) }\rangle_{0,i} },
\end{equation}
where $\langle \cdots \rangle _{0,i}$ stands for the trace with
respect to the reduced Hamiltonian $H_0(i)$: i.e., $\langle \cdots
\rangle_{0,i} ={\rm Tr}\{\cdots e^{-\beta H_0(i)}\}/{\rm
Tr}\{e^{-\beta H_0(i)}\}$. In the following, we shall drop the
index $i$ since, by requiring translational invariance, all the
sites are equivalent. As it is shown in Appendix \ref{App.A}, the
parameters $\kappa ^{(k)}$ and $\lambda ^{(k)}$ can be written as
a function of two parameters $X_1 =\langle n^\alpha (i)\rangle _0$
and $X_2 =\langle D^\alpha (i)\rangle _0$, in terms of which one
may find a solution of the model. By exploiting the translational
invariance along the chain, one can impose
\begin{equation}
\label{EHM_19}
\begin{split}
 \langle n(i)\rangle &=\langle n^\alpha (i)\rangle , \\
 \langle D(i)\rangle &=\langle D^\alpha (i)\rangle,
 \end{split}
\end{equation}
finding thus two equations allowing one to determine $X_1$ and
$X_2$ as functions of $\mu$.
\begin{equation}
\label{EHM_20}
\begin{split}
 X_1 &=2e^{\beta \mu }(1-X_1 -dX_2 )(1+aX_1 +a^2X_2 )
 \\
&+e^{\beta (2\mu -U)}[2+(d-1)X_1 -2dX_2 ](1+dX_1 +d^2X_2 ),
\\
 X_2 &=e^{\beta (2\mu -U)}[1+dX_1 -(2d+1)X_2 ](1+dX_1 +d^2X_2 )
\\
 &-2e^{\beta \mu }(1+d)X_2 (1+aX_1 +a^2X_2 ),
\end{split}
\end{equation}
where $a=e^{-\beta V}-1$ and $d=e^{-2\beta V}-1$. The chemical
potential $\mu$ can be determined by means of the equation
\begin{equation}
\label{EHM_21}
\begin{split}
n&=\frac{1}{\Xi} \big[(X_1 -2X_2 )(1+aX_1 +a^2X_2 ) \\
&+2X_2 (1+dX_1 +d^2X_2 )\big] ,
\end{split}
\end{equation}
where $\Xi=1-X_1 +X_2+(X_1 -2X_2 )(1+aX_1 +a^2X_2 )+X_2 (1+dX_1
+d^2X_2 )$. Thus  Eqs. \eqref{EHM_20} and \eqref{EHM_21}
constitute a set of coupled equations allowing one to ascertain
the three parameters $\mu$, $X_1$, and $X_2$ in terms of the
external parameters of the model ($n$, $U$, $V$, and $T$). Once
these quantities are known, all the properties of the model can be
computed. In particular, the local CFs can be expressed in terms
of the parameters $D=\langle D(i)\rangle$, $\kappa^{(p)}$, and
$\lambda^{(p)}$, defined in Eq. \eqref{EHM_14}, and
$\pi^{(p)}=\langle [D^\alpha (i)]^p \rangle$,
$\delta^{(p)}=\langle n(i)[D^\alpha (i)]^p \rangle/2$,
$\theta^{(p)}=\langle D(i)[D^\alpha (i)]^p \rangle/2$. Indeed, it
is easy to derive the following relations for the two-point
correlation functions
\begin{equation}
\label{EHM_22}
\begin{split}
\langle n(i) n(i+1) \rangle &=2\lambda^{(1)}
\\
\langle n(i) n(i+2) \rangle &=2\kappa^{(2)}-n-2D
\\
\langle n(i) D(i+1) \rangle &=2\delta^{(1)}
\\
 \langle n(i) D(i+2) \rangle
&=\frac{2}{3}\kappa^{(3)}-\kappa^{(2)}+\frac{1}{3} n \\
\langle D(i) D(i+1) \rangle &=2\theta^{(1)} \\
\langle D(i) D(i+2) \rangle &=2\pi^{(2)}-D ,
\end{split}
\end{equation}
and for the three-point correlation functions
\begin{equation}
\label{EHM_23}
\begin{split}
\langle n(i) n(i+1) n(i+2) \rangle &=4\lambda^{(2)}- 2
\lambda^{(1)}-4\delta^{(1)}
\\
\langle n(i) n(i+1) D(i+2)\rangle
&=\frac{4}{3}\lambda^{(3)}-2\lambda^{(2)}+\frac{2}{3}\lambda^{(1)}
\\
\langle D(i) n(i+1)D(i+2) \rangle &=4\delta^{(2)}-2\delta^{(1)}
\\
\langle D(i) D(i+1)D(i+2) \rangle &=4\theta^{(2)}-2\theta^{(1)} .
\end{split}
\end{equation}
The double occupancy, in terms of $X_1$ and $X_2$, is given by
\begin{equation}
\label{EHM_24}
 D=\frac{1}{\Xi} \big[X_2(1+d X_1+d^2
X_2) \big] .
\end{equation}
In  Appendix \ref{App.A} we show how the parameters
$\kappa^{(p)}$, $\lambda^{(p)}$, $\pi^{(p)}$, $\delta^{(p)}$,
$\theta^{(p)}$ can be expressed in terms of the two basic
parameters $X_1$ and $X_2$. At half filling, the system of
equations \eqref{EHM_20} and \eqref{EHM_21} can be solved
analytically and one has
\begin{equation}
\label{EHM_25}
 \mu= \frac{1}{2}\, U+2V,
\end{equation}
\begin{equation}
\label{EHM_26}
\begin{split}
X_1 &=1-\frac{1}{4 G K (K-1)} \big\{(K+1)[1+2G K+K^2
\\
&-\sqrt{(1+2G K+K^2)^2-8 G K(K-1)^2}]\big\},
\\
X_2 &= \frac{1}{4 G K (K-1)^2}\big\{1+2G K+K^2
\\
&-\sqrt{(1+2G K+K^2)^2-8 G K(K-1)^2}\big\},
\end{split}
\end{equation}
where $K=e^{- \beta V}$ and $G=e^{\beta U/2}$. As a consequence,
one finds
\begin{equation}
\label{EHM_27}
\begin{split}
D &=\frac{1}{2+2G(1-a^2 X_2)^2}
\\
\lambda^{(1)}&= \frac{1-2a^2 X_2+(1-4K+2K^2+K^4)X_2^2}{2[1-2a^2
X_2+a^2(1+K^2)X_2^2]}.
\end{split}
\end{equation}

\subsection{Nonlocal correlation functions}
\label{II_D}

We shall now consider the problem of computing nonlocal
correlation functions $\langle O_1(i) O_2(j)\rangle$, where
$O_1(i)$ and $O_2(j)$ are generic operators and $j$ is a site $m$
steps ($m \ge 1$) away from the site $i$. To compute these
quantities one needs only to know the local correlators
\eqref{EHM_22} thanks to some recurrence relations. We refer the
interested reader to Appendix B for analytical details. In
particular, for the charge and double occupancy correlation
functions, one finds
\begin{equation}
\label{EHM_28}
\begin{split}
\langle n(i) n(j)\rangle &=n^2+ A p^{\vert i - j \vert} +B
q^{\vert i - j \vert},
\\
\langle D(i) D(j)\rangle &= D^2+ C p^{\vert i - j \vert} +E
q^{\vert i - j \vert},
\end{split}
\end{equation}
where the coefficients $A$, $B$, $C$, $E$, and the exponents $p$
and $q$ - all defined in Appendix B - are functions only of the
short-range correlation functions $\lambda^{(1)}$, $\kappa^{(2)}$,
$\theta^{(1)}$, and $\pi^{(2)}$.

\subsection{Density of states}
\label{II_E} By noting that the cross GFs $\langle
\psi^{(\xi)}(i,t) {\psi^{(\eta)}}^{\dag}(i,t) \rangle$ vanish, the
electronic density of states (DOS) is given by
\begin{equation}
\label{EHM_29}
 \begin{split}
 N(\omega )&=\left( {-\frac{1}{\pi }} \right) Im [G_{11}^{(\xi)} (\omega
)+G_{11}^{(\eta)} (\omega )] \\
&=\sum_{m=1}^{5} [\Sigma^{(\xi)}_m \delta (\omega -E_m^{(\xi )}
)+\Sigma^{(\eta)}_m \delta (\omega -E_m^{(\eta )} )] .
\end{split}
\end{equation}
As one can immediately see, the DOS is a sum of delta's functions
weighted by the corresponding $\Sigma^{(s)}_m$'s. It is worthwhile
to observe that the expressions of the $\Sigma^{(s)}_m$'s
\eqref{EHM_13} lead to the following sum rule:
\begin{equation}
\label{EHM_30}
 \sum_{n=1}^{5} [\Sigma _n^{(\xi)} +\Sigma _n^{(\eta)}]=\int_{-\infty
}^{+\infty } d\omega  N(\omega ) =1
 .
\end{equation}
In the following, we shall use a Lorentzian broadening for the
delta functions
$\delta(\omega-\omega_0)=\varepsilon/[(\omega-\omega_0)^2+\varepsilon^2]\pi$
with $\varepsilon=0.1$.

\subsection{Summary}
\label{II_F}

Summarizing, in this section we have shown that the model can be
exactly solved by using the Green's function formalism and the
equation of motion method. The central point in this approach lies
in the fact that there exists a closed set of eigenoperators of
the Hamiltonian \eqref{EHM_1} [see Eqs. \eqref{EHM_4}] allowing
for an exact determination of the correlation functions, as it is
shown in Secs. \ref{II_B}, \ref{II_D} and in the Appendices. Exact
expressions for the GFs and CFs can be written in terms of few
local correlators. By using algebraic properties of the relevant
operators and symmetry properties, these correlators can be
expressed in terms of the chemical potential $\mu$ and of the
parameters $X_1$ and $X_2$. These three quantities are determined
as functions  of the external parameters $n$, $U$, $V$ and $T$ by
solving the system of equations \eqref{EHM_20} and \eqref{EHM_21}.
Once these quantities are known, all properties of the model can
be computed. In the next section we shall present a comprehensive
study of the properties of the system at zero-temperature.

\section{Zero temperature results}
\label{sec_III}
\begin{figure}[t]
\begin{center}
\vspace{12mm}
\includegraphics[scale=0.45]{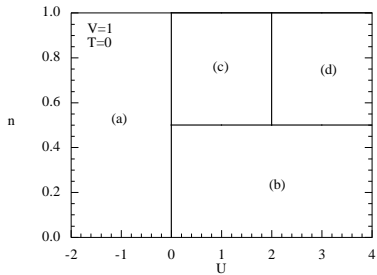}
\caption{\label{fig1} The phase diagram in the plane $(U,n)$ at
$T=0$ and $V=1$. One can identify four phases.}
\end{center}
\end{figure}

In this section we derive the phase diagram of the AL-EHM in the
$(U,n)$ plane. More specifically, we shall numerically solve the
set of equations \eqref{EHM_20} and \eqref{EHM_21} at $T=0$ and
study the behavior of relevant correlation functions and of the
density of states in order to envisage the distribution of the
particles for different densities. The results obtained are
displayed in Fig. \ref{fig1}: one may distinguish four different
phases, according to the values of the internal parameters.

{\it Phase (a)}. The phase ($a$) is observed in the region $(0\le
n\le 1$, $U<0)$ and is characterized by the following values of
the parameters:
\begin{equation*}
\begin{split}
  X_1 &=\frac{2n}{2-n}, \\
 X_2 &=\frac{n}{2-n} ,\\
 D &=n/2 ,\\
 \kappa ^{(2)} &=\frac{2n}{2-n},
\end{split}
\quad
\begin{split}
\kappa ^{(3)}&=\frac{2n(1+n)}{2-n}, \\
\kappa ^{(4)}&=\frac{2n(1+3n)}{2-n}, \\
\pi ^{(2)}&=\frac{n}{2(2-n)} ,\\
\lambda ^{(k)}&=\theta ^{(1)}=0,
\end{split}
\quad
\begin{split}
 A&=(2-n)n ,\\
 B&=E=0, \\
 C&=\frac{(2-n)n}{4}, \\
 p&=-\frac{n}{2-n}, \\
 q&=0 .
\end{split}
\end{equation*}
The chemical potential takes the value $U/2$ for $n<1$, whereas at
$n=1$ one naturally finds $\mu = 2V+ U/2$.
\begin{figure}[th]
 \centering
 \subfigure[]
   {\includegraphics[width=1.48in,height=1.10in]{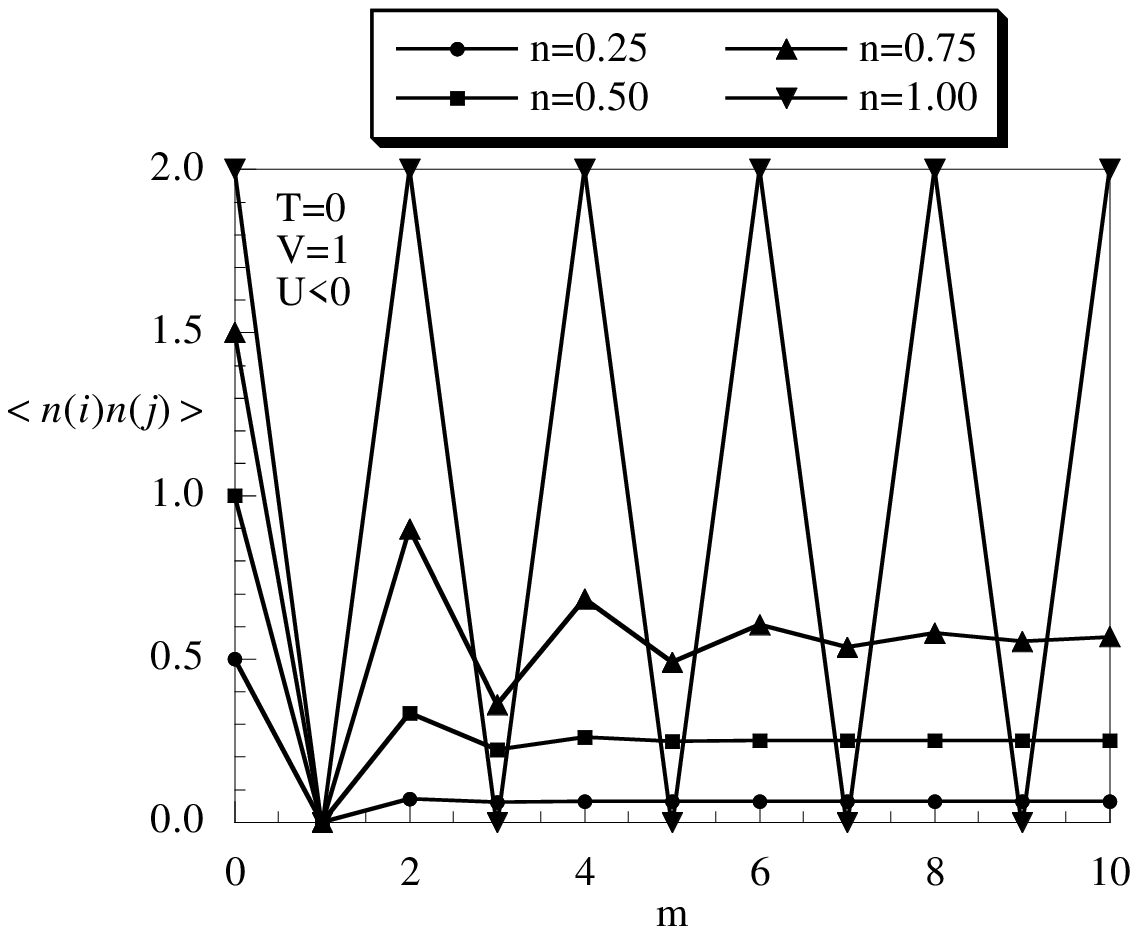}}
 \hspace{1mm}
 \subfigure[]
   {\includegraphics[width=1.48in,height=1.10in]{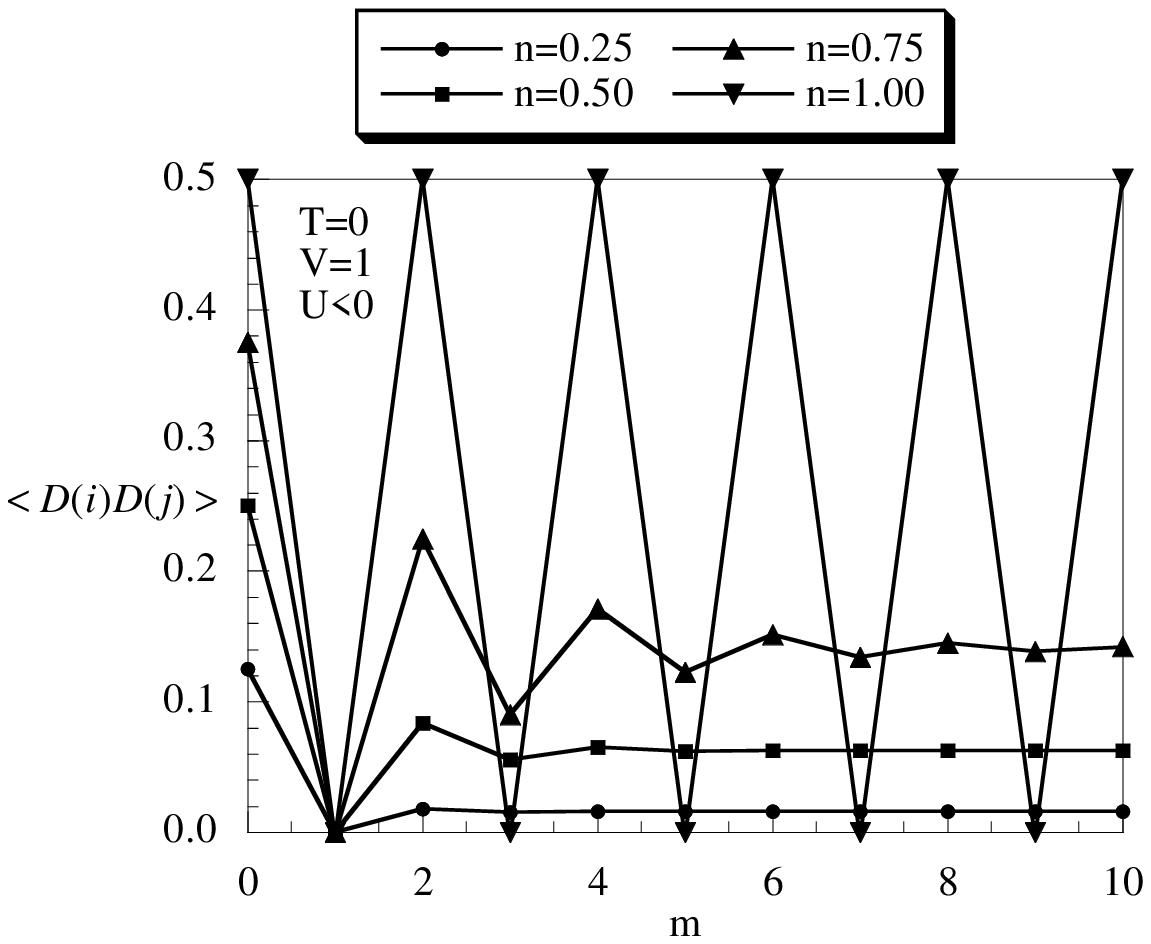}}
 \caption{\label{fig2}The charge (a) and the double occupancy
 (b) correlation functions as functions of the distance
$m=\left| {i-j} \right|$ for $U<0$ at $T=0$, $V=1$ and $n$=0.25,
0.5, 0.75, 1.}
 \end{figure}
The attractive on-site potential favors the formation of doubly
occupied sites. At the same time, the repulsive intersite
potential disfavors the occupation of neighboring sites. The
expectation values $D =n/2$ and $\langle n(i)[n^\alpha
(i)]^k\rangle =0$ suggest that, in the thermodynamic limit, there
are neither singly occupied sites nor neighboring sites occupied.
For $n<1$ one does not find any charge ordered state and the
energy necessary to add two electrons is $2\mu =U$. At half
filling $(n=1)$, the chemical potential jumps from the value $\mu
=U/2$ to the value $\mu =2V+U/2$, as required by particle-hole
symmetry. Indeed, one needs to provide an energy amount of $2V$
for an extra electron to occupy a nearest neighbor site. At $n=1$
half of the sites are doubly occupied: by varying the particle
density, the system undergoes a PT to an ordered state
characterized by a checkerboard distribution of doubly occupied
sites. This can be clearly seen in Figs. \ref{fig2}a-b, where we
plot the correlation functions
\begin{equation}
\begin{split}
 \langle n(i)n(j)\rangle &=n^2+n(2-n)(-1)^m \, e^{-m/\xi_{n}},
\\
 \langle D(i)D(j)\rangle &=D^2+\frac{n(2-n)}{4}(-1)^m \,
e^{-m/\xi_{n}}
 \end{split}
 \label{EHM_31}
\end{equation}
as a function of the distance $m=\left| {i-j} \right|$. At half
filling, both CFs present a two-site periodicity. The correlation
length $\xi _n$ in Eqs. \eqref{EHM_31} is defined as
\begin{equation}
\xi _n =\left[ {\ln \left( {\frac{1}{\left| p \right|}} \right)}
\right]^{-1}.
 \label{EHM_32}
\end{equation}
When $n<1$ the parameter $\vert p \vert$ is less than 1. Thus, the
CFs \eqref{EHM_31} decrease exponentially, oscillating around the
ergodic values $n^2$ and $D^2$, respectively. On the other hand,
at half filling $\vert p \vert=1$: the correlation length
diverges, a long-range order is established, and both CFs present
a two-site periodicity.

\begin{figure}[b]
\includegraphics[scale=0.2]{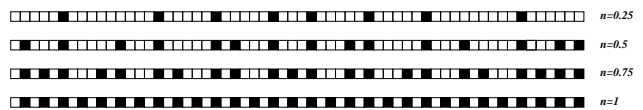}
\caption{\label{fig3} Distribution of the particles along the
chain by varying the particle density at $T=0$ and $U<0$. White
and black squares denote empty and doubly occupied sites,
respectively.}
\end{figure}

\begin{figure}[t]
 \centering
 \subfigure[]
   {\includegraphics[width=1.48in,height=1.10in]{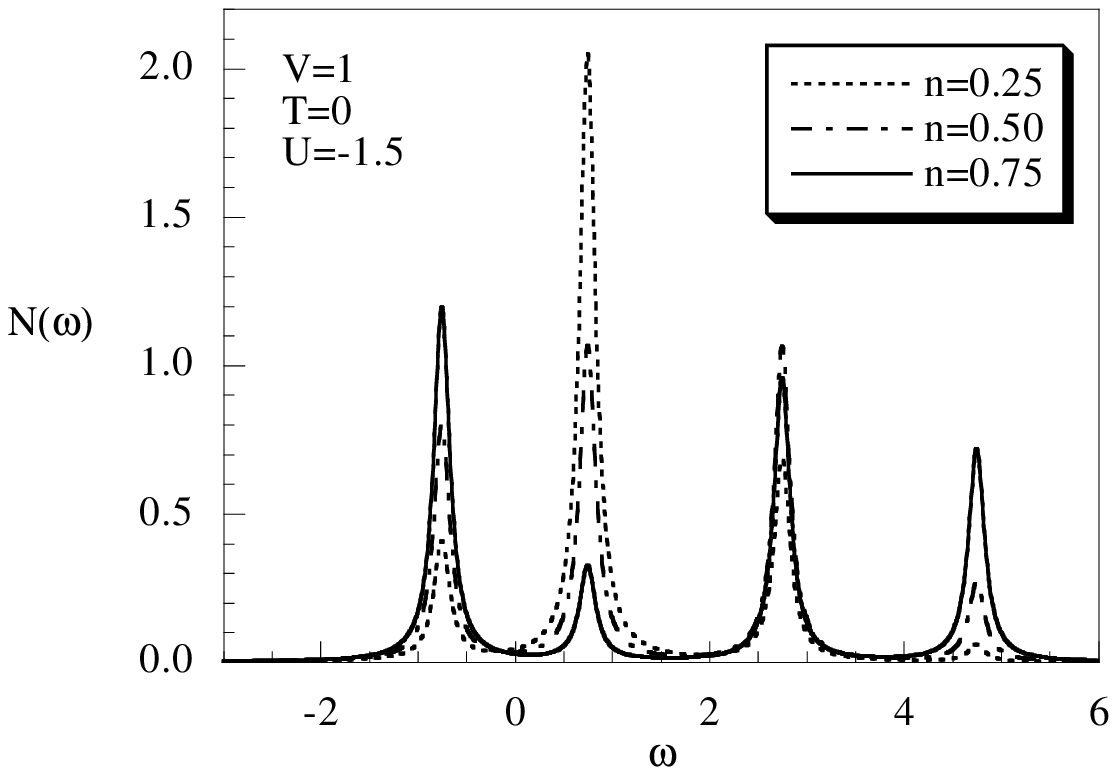}}
 \hspace{1mm}
 \subfigure[]
   {\includegraphics[width=1.48in,height=1.10in]{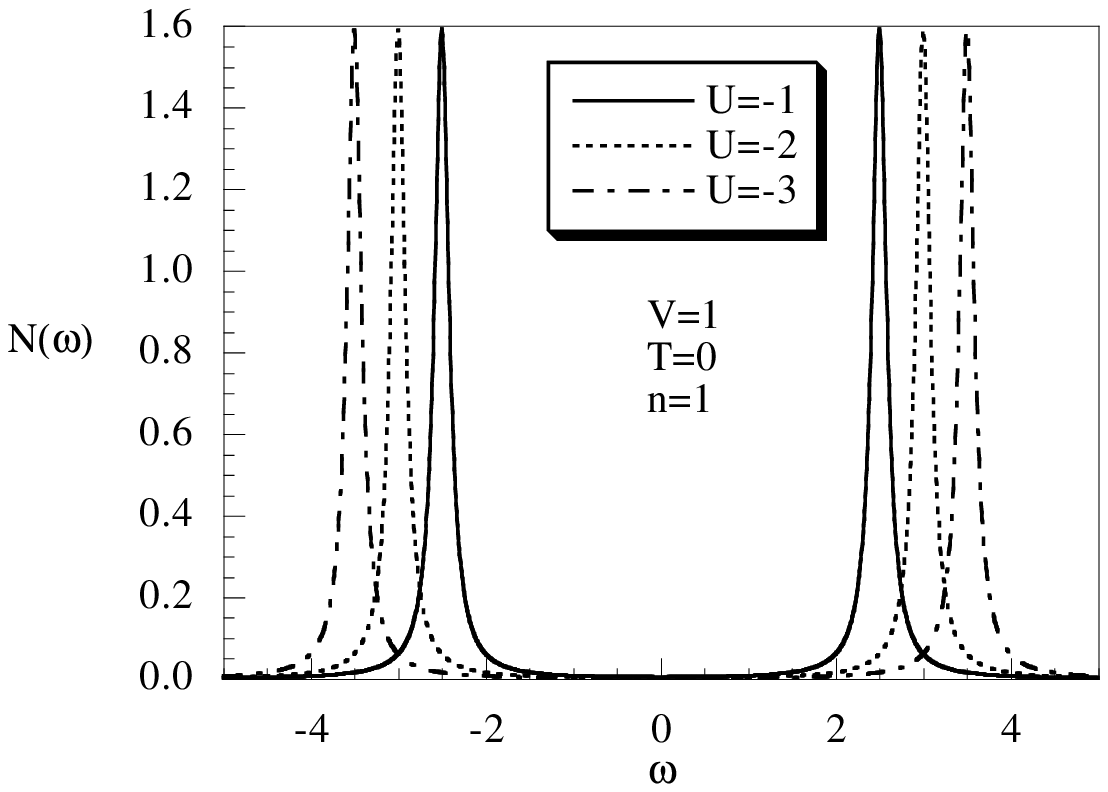}}
 \caption{\label{fig4}(a) The density of states at zero temperature,
$V=1$ and $U=-1.5$ is reported as a function of the frequency for
$n=0.25$, 0.5 and 0.75. (b) The density of states at zero
temperature, $V=1$ and $n=1$ is reported as a function of the
frequency for $U=-1$,$-2$ and $-3$.}
 \end{figure}

The distribution of the particles in the phase ($a$) is shown in
Fig. \ref{fig3}, where we report just one possible configuration.
The configurations in the ground state will have no doubly
occupied sites next to each other. When $n<1$, there is no ordered
pattern in the distribution of the particles, whereas, for $n=1$,
one observes the well-known checkerboard distribution of doubly
occupied sites \cite{Bari71}.
In Figs. \ref{fig4}a-b, we plot the density of states at zero
temperature and attractive $U$, for $n<1$ and $n=1$, respectively.
For $n<1$, the DOS exhibits four peaks at energies
$E_1^{(\eta)}=U/2$, $E_1^{(\xi)}=-U/2$, $E_3^{(\xi)}=-U/2+2V$,
$E_5^{(\xi )} =-U/2+4V$: the electrons can occupy only states with
these associated energies. For $n=1$, the excitation spectrum
contains only two allowed values, leading to a DOS with two peaks
at $E_1^{(\eta)} =U/2-2V$, and $E_5^{(\xi)} =-U/2+2V$. By varying
the particle density, the weights of the peaks change as
\begin{equation*}
\begin{split}
\Sigma _1^{(\eta )} &=\frac{n}{2}
 \quad \quad \quad \quad  \quad  \, \,
\Sigma_1^{(\xi)} =\frac{2(1-n)^2}{2-n}
\\
\Sigma _3^{(\xi )} &=\frac{2n(1-n)}{2-n}
 \quad \quad
\Sigma _5^{(\xi )} =\frac{n^2}{2(2-n)} .
\end{split}
\end{equation*}
The weights relative to the other energies are
zero: $\Sigma _2^{(\xi)}=\Sigma_4^{(\xi)} =\Sigma_2^{(\eta)}
=\Sigma _3^{(\eta)}=\Sigma _4^{(\eta)}=\Sigma_5^{(\eta)} =0$.

{\it Phase (b)}. The phase ($b$) is observed in the region $(0\le
n\le 0.5,\;U>0)$ and is characterized by the following values of
the parameters
\begin{equation*}
\begin{split}
  X_1 &=\frac{n}{1-n}, \\
 X_2 &=D=0,
  \\
\pi ^{(2)}&=\theta ^{(1)}=0 ,\\
\lambda ^{(k)}&=0 ,
\end{split}
\quad
\begin{split}
\kappa ^{(2)} &=\frac{n}{2(1-n)} , \\
\kappa ^{(3)}&=\frac{n(1+2n)}{4(1-n)} , \\
\kappa ^{(4)}&=\frac{n(1+6n)}{8(1-n)},
\end{split}
\quad
\begin{split}
 A&=n(1-n) , \\
 B&=E=C=0  ,\\
 p&=-\frac{n}{1-n}, \\
 q&=0.
\end{split}
\end{equation*}
and
\begin{equation*}
\mu = \left\{
\begin{array}{ll}
 0 &(n<0.5) \\
 U & (n=0.5 \: {\rm and} \:   0<U<2V) \\
2V & (n=0.5 \: {\rm and} \:  U>2V).
\end{array}
\right.
\end{equation*}
 The repulsion between electrons on the same site and on
neighboring sites, leads to a scenario where the electrons will
tend to singly occupy non-neighbor sites, as confirmed by the
expectation values $D=0$ and $\langle n(i)[n^\alpha (i)]^k\rangle
=0$. For particle densities less than quarter filling, i.e.,
$n<0.5$, there is no cost in energy to add one electron, thus
$\mu=0$. At quarter filling, one has to distinguish two cases: (i)
when $0<U<2V$, the chemical potential jumps from the value $\mu
=0$ to the value $\mu =U$; (ii) for $U>2V$, the chemical potential
presents a jump from $\mu =0$ to $\mu =2V$. As a result, for
$n>0.5$, as illustrated in Fig. \ref{fig1}, there are two distinct
phases, ($c$) and ($d$). When $0<U<2V$, by adding one electron the
system enters phase ($c$), characterized by the presence of both
singly and doubly occupied sites. The minimization of the energy
leads to a cost in energy of $\mu =U$ to accommodate an extra
electron. On the other hand, when $U>2V$, by adding one electron,
the system is driven into phase ($d$), characterized by the
presence of neighbor arbitrary-spin singly occupied sites,
requiring thus an energy cost of $\mu =2V$.

\begin{figure}[t]
\includegraphics[scale=0.45]{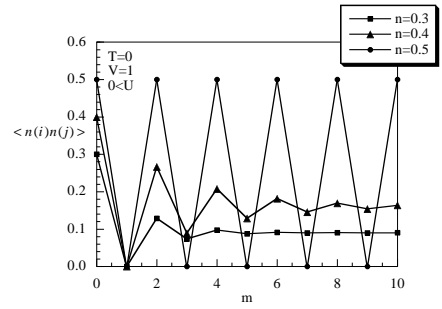}
\caption{\label{fig5} The charge correlation function $\langle
n(i)n(j)\rangle$ for $U>0$ is reported as a function of the
distance $m=\left| {i-j} \right|$ at $T=0$, $V=1$ and $n=0.3$, 0.4
and 0.5.}
\end{figure}

Right at quarter filling, half of the sites are singly occupied:
by varying the particle density, the system undergoes a PT to an
ordered state characterized by a checkerboard distribution of
arbitrary-spin singly occupied sites. This is clearly shown in
Fig. \ref{fig5},
where we plot the charge correlation function $\langle
n(i)n(j)\rangle=n^2+n(1-n)(-1)^m \, e^{-m/\xi_{n}}$ as a function
of the relative distance: when $n<0.5$ one finds that the
parameter $\left| p \right|$ appearing in Eq. \eqref{EHM_28} is
less than 1. As a result, $\langle n(i)n(j)\rangle$ decreases
exponentially, oscillating around the ergodic value $n^2$. At
quarter filling one finds $\left| p \right|=1$: the correlation
length \eqref{EHM_32} diverges and a long-range order is
established, as evidenced by the two-site periodicity of  $\langle
n(i)n(j)\rangle$.

\begin{figure}[!h]
 \centering
 \subfigure[]
   {\includegraphics[width=1.48in,height=1.10in]{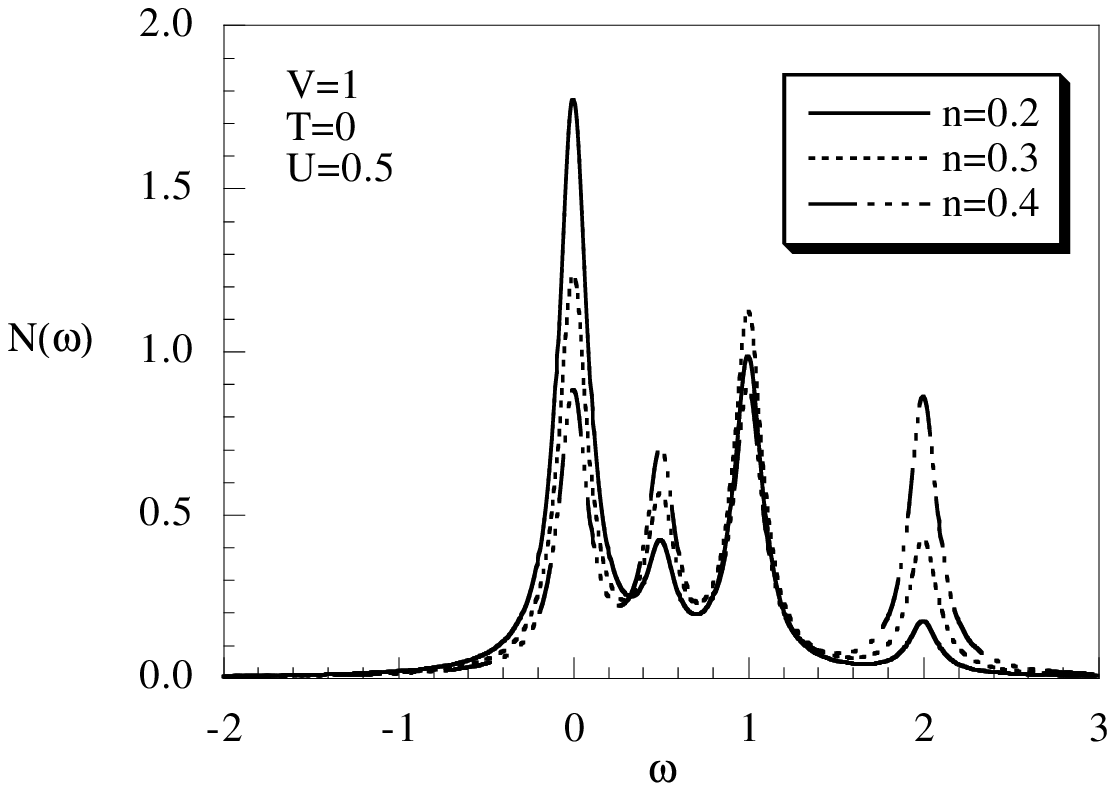}}
 \hspace{0.5mm}
 \subfigure[]
   {\includegraphics[width=1.48in,height=1.10in]{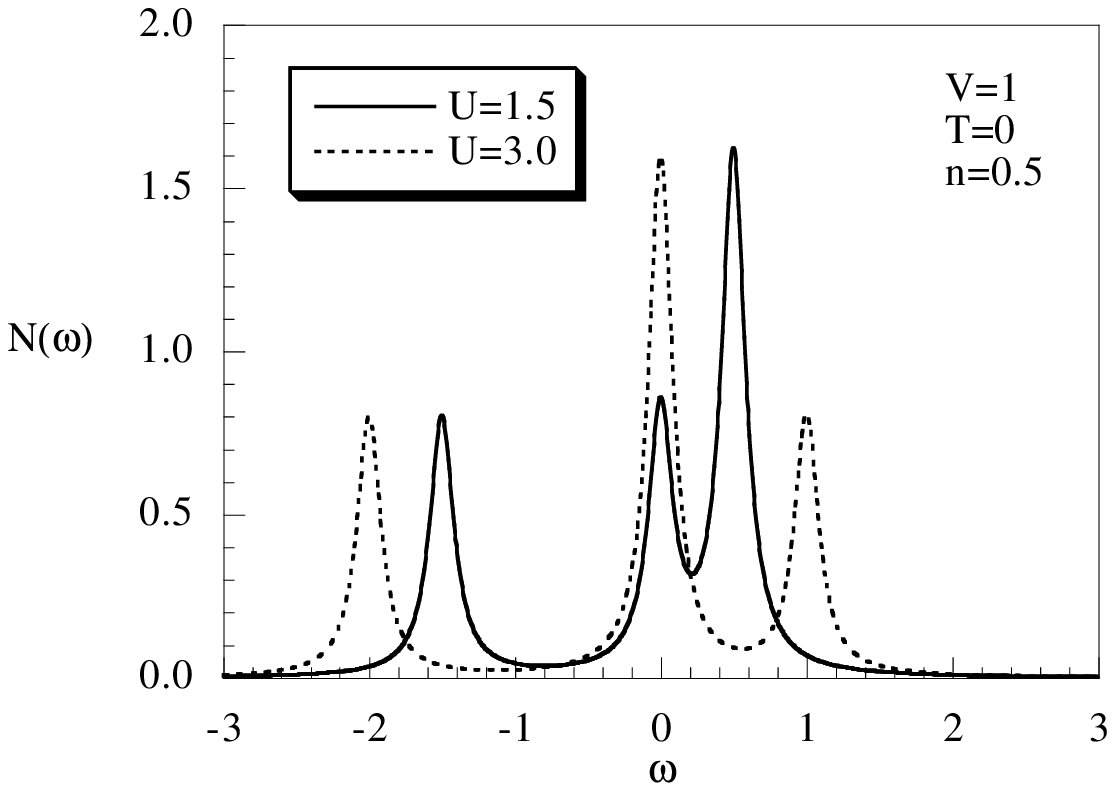}}
 \caption{\label{fig6} The density of states $N(\omega )$ as a function
of the frequency $\omega$  at zero temperature, $V=1$ and  (a)
$U=1/2$ for $n$=0.2, 0.3 and 0.4; (b) $n=0.5$, for $U=3/2$ and
$3$.}
 \end{figure}

The DOS at zero temperature and repulsive $U$ is shown in Fig.
\ref{fig6}a for $n<0.5$. In Fig. \ref{fig6}b we plot the DOS at
$n=0.5$ for $U<2V$ (continuous line) and $U>2V$ (dotted line). For
$n<0.5$, the DOS exhibits four peaks at the energies
$E_1^{(\eta)}=U$, $E_1^{(\xi)}=0$, $E_2^{(\xi)}=V$, $E_3^{(\xi)}
=2V$. For $n=0.5$ and $0<U<2V$, the electrons can occupy only
three states at the energies $E_1^{(\eta)}=0$, $E_1^{(\xi)}=-U$,
$E_3^{(\xi)}=-U+2V$.
For $n=0.5$ and $U>2V$, the DOS still exhibits three peaks but now
at the energies $E_1^{(\eta )}=U-2V$, $E_1^{(\xi)} =-2V$,
$E_3^{(\xi)} =0$. By varying $n$ the weights of the peaks change
as
\begin{equation*}
\begin{split}
\Sigma_1^{(\eta)} &=\frac{n}{2} \quad \quad \quad \quad \quad
\quad \Sigma _1^{(\xi)} =\frac{2-7n(1-n)}{2(1-n)}
\\
\Sigma _2^{(\xi)} &=\frac{2n(2n-1)}{1-n}\quad \quad \Sigma
_3^{(\xi)}=\frac{n^2}{1-n}.
\end{split}
\end{equation*}

The weights relative to the other energies are zero.

{\it Phase (c)} The phase ($c$) is observed in the region ($0.5\le
n\le 1$, $0<U<2V$) and is characterized by the following values of
the parameters:
\begin{equation*}
\begin{split}
 X_1 &=2n ,\\
 X_2 &=2n-1 , \\
 D &=n -1/2 , \\
 \pi ^{(2)}&= n(n-1/2) , \\
\lambda ^{(k)}&=\theta ^{(1)}=0,
\end{split}
\quad
\begin{split}
\kappa ^{(2)} &=\frac{1}{2}(2n^2+3n-1) , \\
\kappa ^{(3)}&=\frac{1}{4}(18n^2+n-3) , \\
\kappa ^{(4)}&=\frac{1}{8}(110n^2-45n-1),
\end{split}
\end{equation*}
and
\begin{equation*}
\begin{split}
 A&=n^2 ,\\
 B&=E=-2n^2+3n-1  ,\\
 C&=\left( {n-1/2} \right)^2 ,
 \end{split}
\quad
\begin{split}
 p&=-1, \\
 q&=0 .
\end{split}
\end{equation*}
The chemical potential takes the value $U$ for $0.5\le n<1$ and
$2V+U/2$ at $n=1$. The repulsive on-site and intersite potentials
force the electrons to prefer seating at different sites and
disfavor the occupation of neighboring sites. However, when the
intersite interaction dominates the on-site one, the minimization
of the energy requires the extra electrons above $n=0.5$ to occupy
nonempty sites. This situation is well described by the
expectation values $D=n-1/2$ and $\langle n(i)[n^\alpha
(i)]^k\rangle =0$. Thus, in the thermodynamic limit, there can be
singly and doubly occupied sites and no neighboring sites are
filled. By increasing $n$, the number of doubly occupied sites
increases linearly. In this region, the correlation length
\eqref{EHM_32} diverges: one finds a charge ordered state
characterized by a checkerboard distribution of alternating empty
and occupied (either singly or doubly) sites. For $n<1$ the energy
necessary to add one electron is $\mu =U$. Upon increasing the
particle density, at half filling the checkerboard distribution is
realized with only doubly occupied sites alternating with empty
ones and, correspondingly, the chemical potential jumps from the
value $\mu =U$ to the value $\mu =2V+U/2$. Since all the nonempty
sites are already doubly occupied, an extra electron would go to
occupy an empty site with a cost in energy of $2V$. The existence
of the ordered states is endorsed by the behavior of the charge
and double occupancy CFs.
\begin{equation*}
\begin{split}
 \langle n(i)n(i+m)\rangle &=n^2[1+(-1)^m] \\
 \langle D(i)D(i+m)\rangle &=D^2[1+(-1)^m] \quad \quad \quad m \ge
 1
 \end{split}
\end{equation*}
\begin{figure}[t]
 \centering
 \subfigure[]
   {\includegraphics[width=1.48in,height=1.10in]{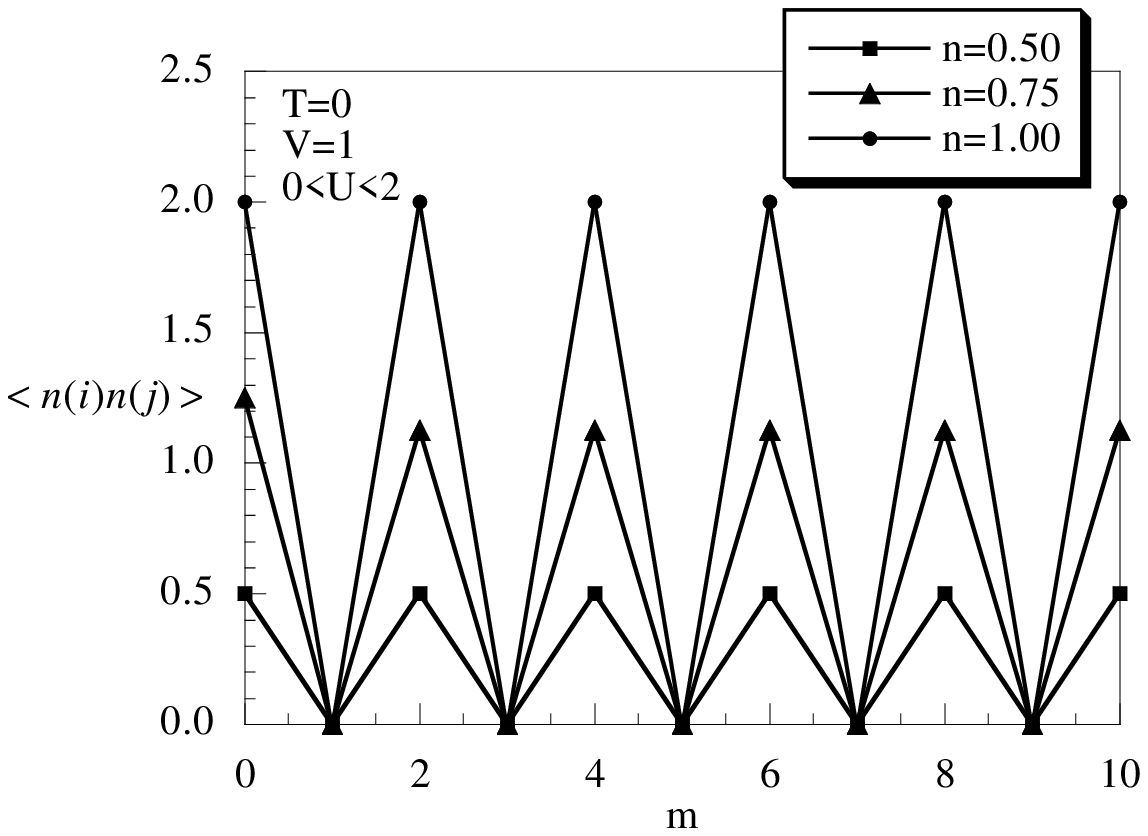}}
 \hspace{1mm}
 \subfigure[]
   {\includegraphics[width=1.48in,height=1.10in]{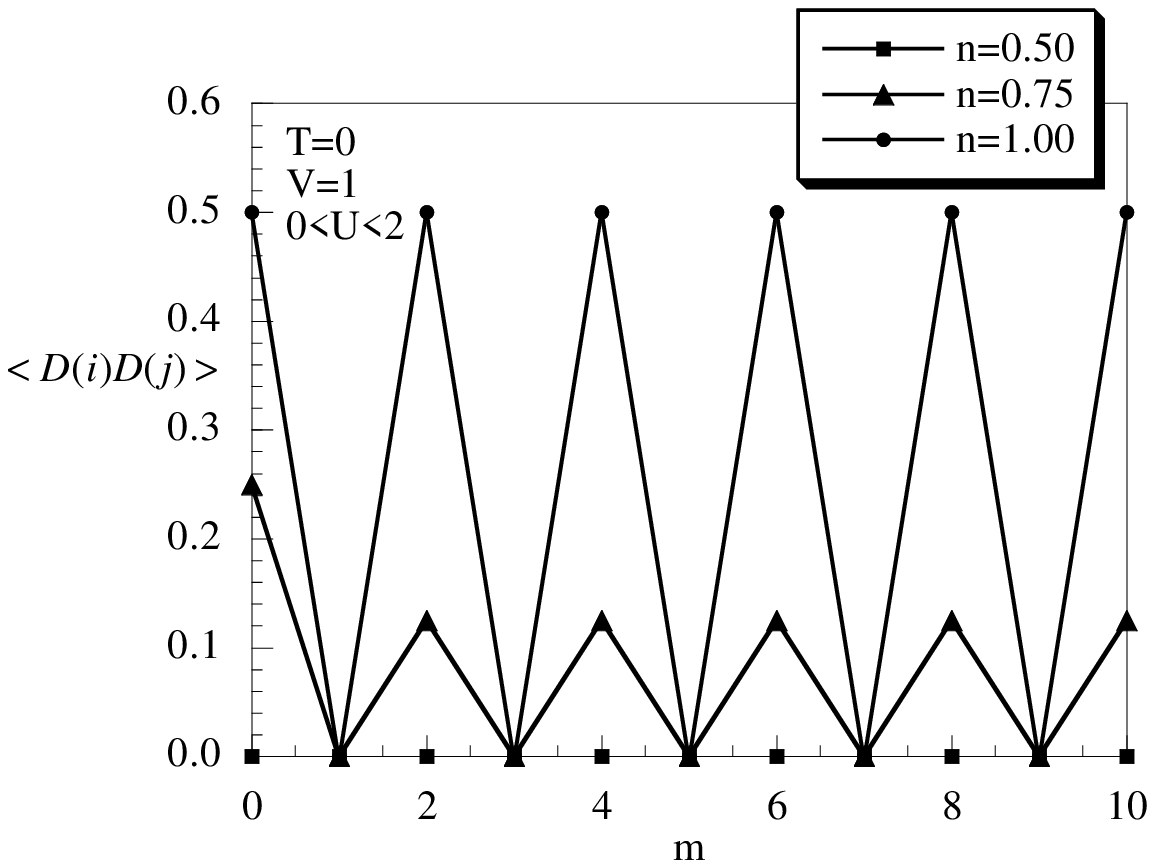}}
 \caption{\label{fig7}The charge (a) and the double occupancy (b)
correlation functions  as functions of the distance $m=\left|
{i-j} \right|$ at $T=0$, $V=1$ and $n=$0.5, 0.75 and 1.}
 \end{figure}
and reported in Figs. \ref{fig7}a and \ref{fig7}b. One clearly
observes a two-site periodicity of $\langle n(i)n(i+m)\rangle$ and
$\langle D(i)D(i+m)\rangle$, leading to an ordered charge
distribution: one every two sites is at least singly occupied. The
double periodicity of the double occupancy CF is, of course, to be
understood as an average over all the possible distributions of
doubly occupied sites that are, of course, energetically
equivalent.

\begin{figure}[b]
\includegraphics[scale=0.2]{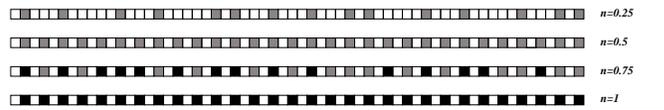}
\caption{\label{fig8} Distribution of the particles  along the
chain by varying the particle density at $T=0$ and $0<U<2V$. White
and black squares denote empty and doubly occupied sites,
respectively. Grey squares denote arbitrary-spin singly occupied
sites.}
\end{figure}

\begin{figure}[ht]
 \centering
 \subfigure[]
   {\includegraphics[width=1.48in,height=1.10in]{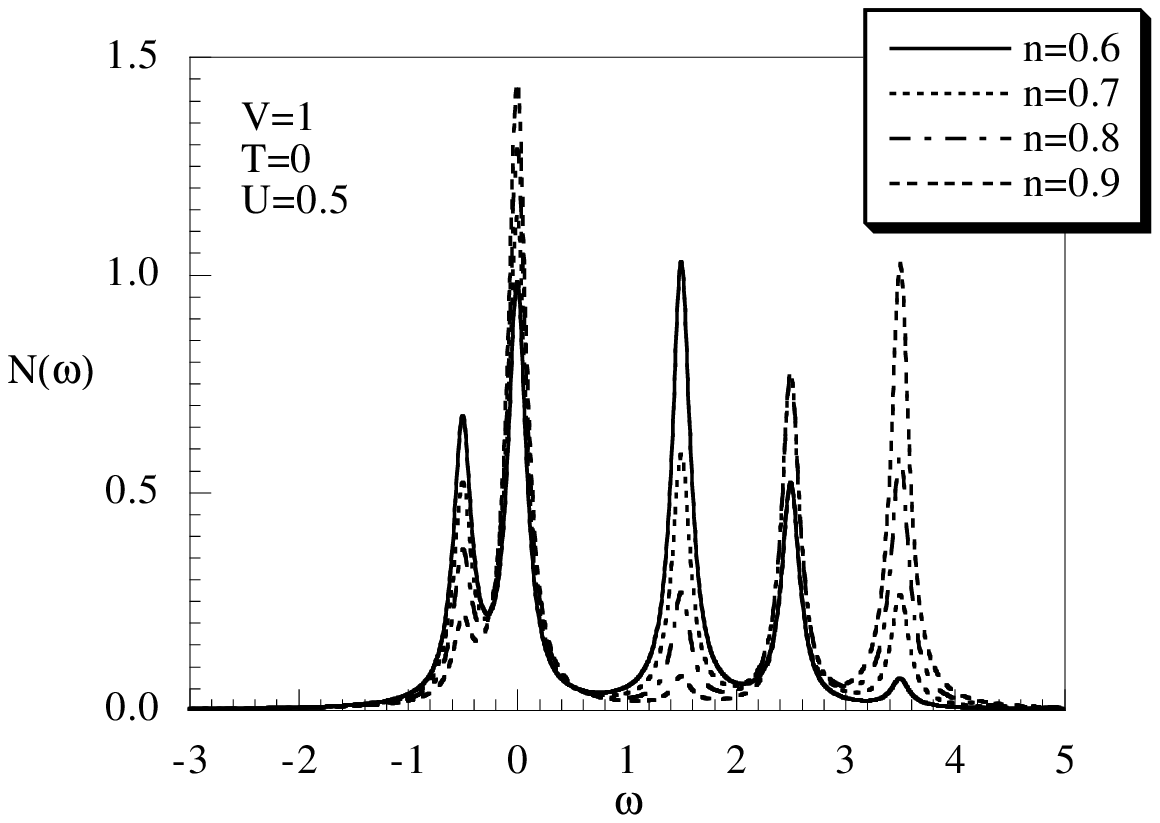}}
 \hspace{1mm}
 \subfigure[]
   {\includegraphics[width=1.48in,height=1.10in]{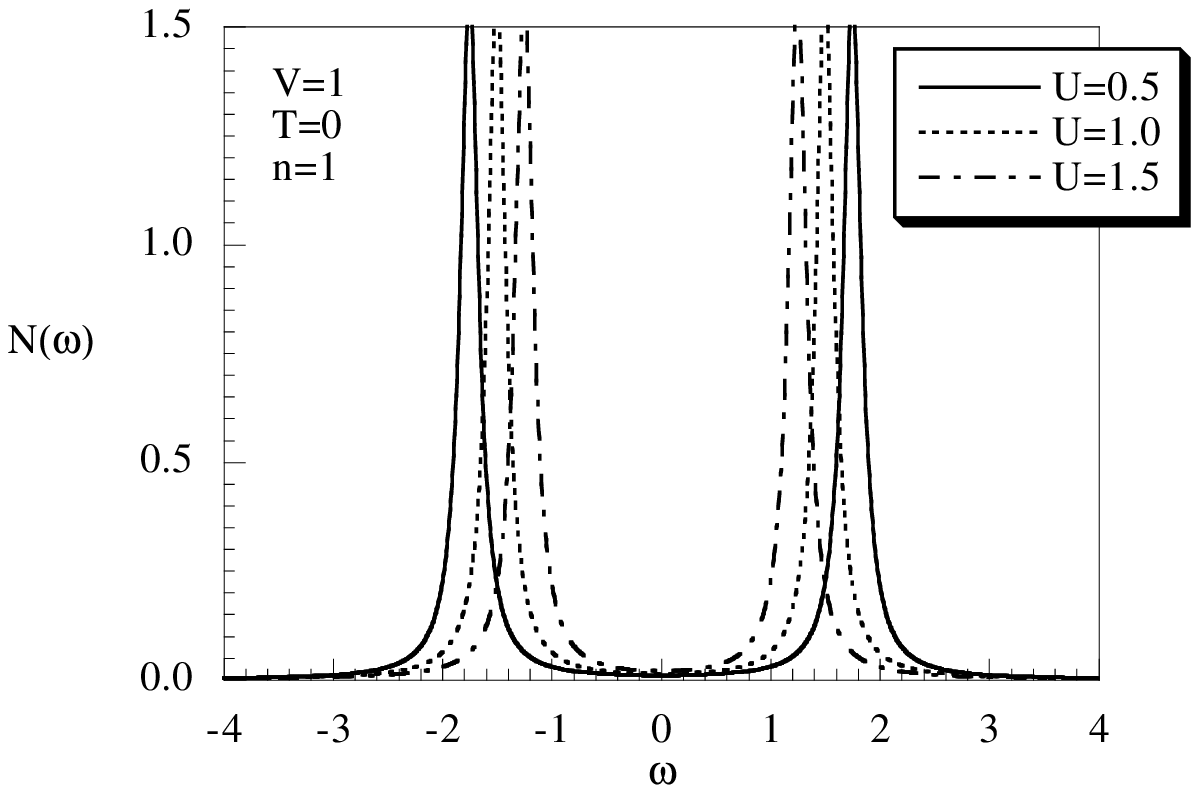}}
 \caption{\label{fig9} The density of states $N(\omega)$ as a function
of the frequency $\omega $  at zero temperature, $V=1$ and  (a)
$U=1/2$ for $n=0.6$, 0.7, 0.8, 0.9; (b) $n=1$, for $U=1/2$, $1$
and $3/2$.}
 \end{figure}

The distribution of the particles in the region ($0\le n\le 1$,
$0<U<2V$) is drawn in Fig. \ref{fig8}, where we represent just one
possible configuration. At quarter filling the ground states of
the Hamiltonian are checkerboard configurations where empty sites
alternate with sites occupied by one particle of arbitrary spin.
By increasing $n$, the charge order persists and one observes an
increase of the number of doubly occupied sites, as illustrated in
Fig. \ref{fig8}. Finally, at half filling, one observes a
checkerboard distribution of doubly occupied sites \cite{Beni74}.
The density of states at zero temperature is plotted in Fig.
\ref{fig9}a for $0.5<n<1$ and $U=V/2$ and in Fig. \ref{fig9}b for
$n=1$ and different values of $U$.
One observes that, when $n<$1, the DOS exhibits five peaks at the
energies $E_1^{(\eta)}=0$, $E_1^{(\xi)} =-U$, $E_3^{(\xi)}=-U+2V$,
$E_4^{(\xi)} =-U+3V$, $E_5^{(\xi)}=-U+4V$. For $n=1$ and $0<U<2V$,
the DOS exhibits only two peaks at the energies $E_1^{(\eta)}
=-U/2-2V$, $E_5^{(\xi)} =-U/2+2V$. By varying $n$ the weights of
the peaks change as
\begin{equation*}
\begin{split}
 \Sigma _1^{(\xi)} &=\frac{1}{2}(1-n)
\quad \quad  \Sigma _5^{(\xi )} =\frac{1}{2}(1-2n)^2
\\
 \Sigma _3^{(\xi)} &=2(1-n)^2
 \quad \quad
  \Sigma _1^{(\eta )} =\frac{n}{2}
  \\
 \Sigma _4^{(\xi)} &=-2+6n-4n^2 .
\end{split}
\end{equation*}
The weights relative to the other energies are
zero.

{\it Phase (d)} The phase ($d$) is observed in the region ($0.5\le
n\le 1$, $U>2V$) and is characterized by the following values of
the parameters,
\begin{figure}[b]
\includegraphics[scale=0.45]{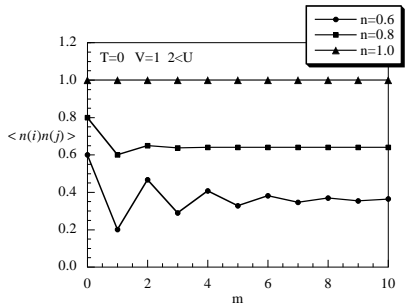}
\caption{\label{fig10} The charge correlation function $\langle
n(i)n(j) \rangle$ for $U>2V$ as a function of the distance
$m=\left| {i-j} \right|$ at $T=0$, $V=1$ and $n=0.6$, 0.8 and 1.}
\end{figure}

\noindent
 for $0.5\le n<1$
\begin{equation*}
\begin{split}
 \mu &=2V ,\\
  X_1 &=1 ,\\
 X_2 &=0 ,\\
 D &=0, \\
\kappa ^{(2)} & > \kappa ^{(3)}> \kappa ^{(4)},
\end{split}
\quad
\begin{split}
\lambda ^{(1)} & > \lambda ^{(2)}>
 \lambda ^{(3)}> \lambda ^{(4)},
 \\
 A&=(1-n)n ,\\
 B&=C=E=0  , \\
 p&=-(1-n)/n , \\
 q&=0 ,
\end{split}
\end{equation*}
and for $n=1$,
\begin{equation*}
\begin{split}
 \mu &=2V+U/2\\
 X_1 &=1 ,\\
 X_2 &=0, \\
 D &=0,
\end{split}
\quad
\begin{split}
\kappa ^{(k)} &=1, \\
\pi ^{(2)}&= 0, \\
\lambda ^{(k)}&=1/2, \\
\theta ^{(1)}&=0,
\end{split}
\quad
\begin{split}
 A&=B=0 ,\\
 C&=E=0,\\
 p&\neq 0 ,\\
 q&\neq 0 .
\end{split}
\end{equation*}
In this region the on-site interaction dominates the intersite
one. The minimization of the energy requires the electrons not to
be paired and allows for the occupation of neighboring sites:
$D=0$ and $\langle n(i)[n^\alpha (i)]^k \rangle \ne 0$. For $n<1$
the energy necessary to add one electron is $\mu =2V$. At half
filling the chemical potential jumps from the value $\mu =2V$ to
$\mu =2V+U/2$. This is because an extra electron would go to
occupy a singly occupied site, requiring thus an energy equal to
$U$. The behavior of the charge correlation function is reported
in Fig. \ref{fig10} as a function of the relative distance. One
observes that increasing the particle density, $\langle n(i)
n(j)\rangle$ becomes more and more uniform losing the two-site
periodicity peculiar of the region $0<U<2V$. At half filling the
system presents an homogeneous distribution, with all singly
occupied sites \cite{Beni74}.

\begin{figure}[t]
 \centering
 \subfigure[]
   {\includegraphics[width=1.48in,height=1.10in]{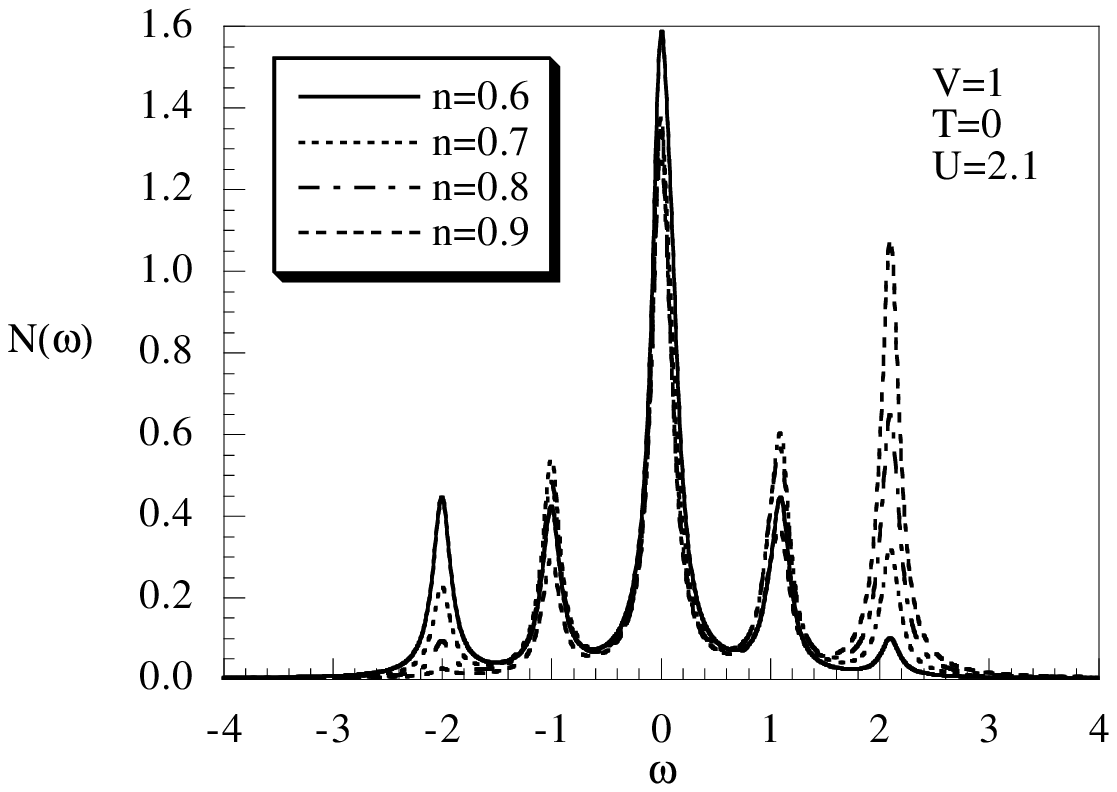}}
 \hspace{1mm}
 \subfigure[]
   {\includegraphics[width=1.48in,height=1.10in]{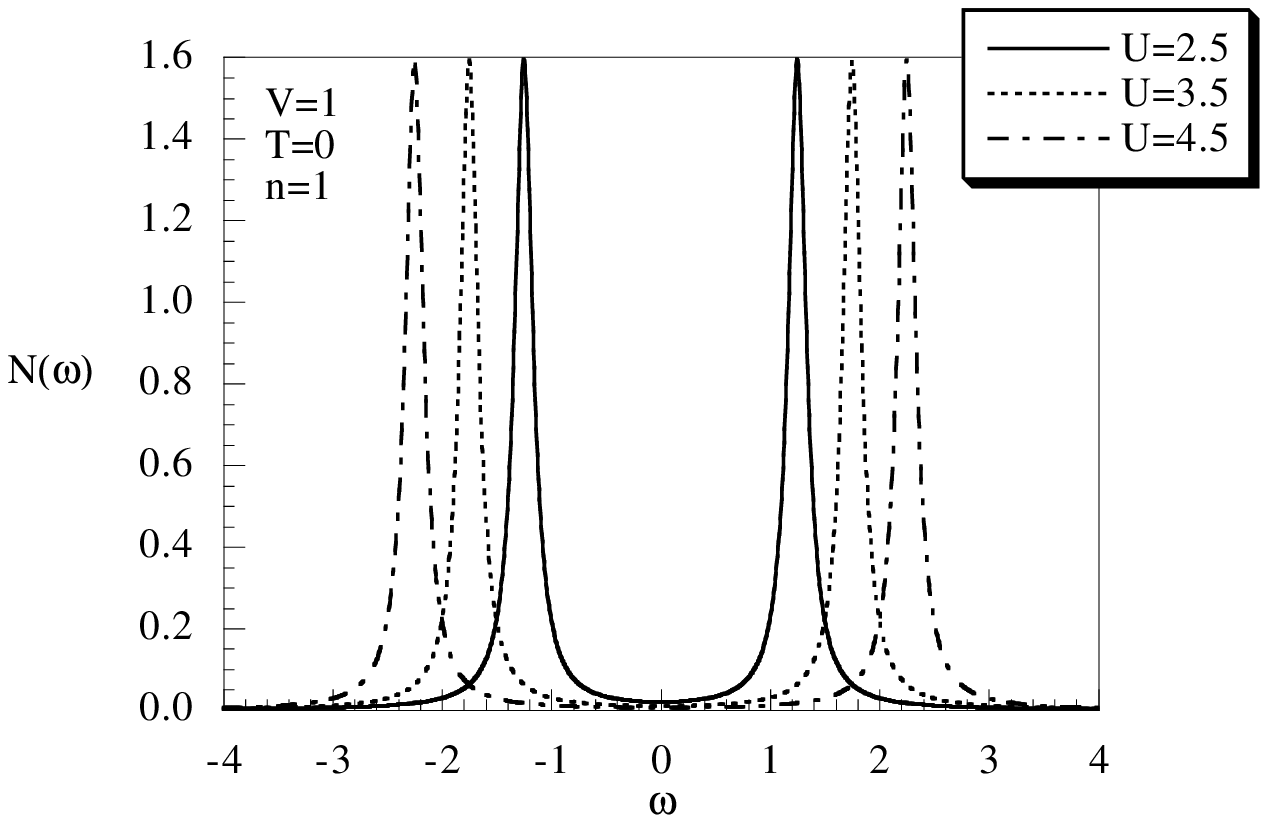}}
 \caption{\label{fig11}
The density of states $N(\omega)$ as a function of the frequency
$\omega$  at zero temperature, $V=1$ and (a) $U=2.1$ for $n=0.6$,
0.7, 0.8, 0.9; (b) $n=1$, for $U=5/2$, $7/2$ and $U=9/2$.}
 \end{figure}
The density of states at zero temperature and $U>2V$ is plotted in
Fig. \ref{fig11}a for $0.5<n<1$ and in Fig. \ref{fig11}b for $n=1$
and different values of $U$ ($U>2V$). For $n<$1, the DOS exhibits
five peaks at the energies $E_1^{(\xi)}=-2V$, $E_2^{(\xi)}=-V$,
$E_3^{(\xi)}=0$, $E_2^{(\eta)}=U-V$, $E_3^{(\eta)} =U$. For $n=1$,
the electrons can occupy two states at the energies
$E_3^{(\xi)}=-U/2$, $E_3^{(\eta)}=U/2$.
\begin{figure}[b]
\includegraphics[scale=0.2]{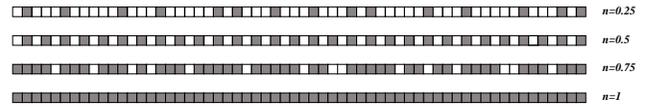}
\caption{\label{fig12} Distribution of the particles  along the
chain by varying the particle density at $T=0$ and $U>2V$. White
and grey squares denote empty and arbitrary spin singly occupied
sites, respectively.}
\end{figure}
The distribution of the particles in the region $(0\le n\le 1$,
$U>2V)$ is presented in Fig. \ref{fig12}, where just one possible
configuration is shown. As one can clearly see, in region ($d$)
there are no charge ordered states but at $n=0.5$.

\begin{figure}[t]
 \centering \vspace{-4mm}
 \subfigure[]
   {\includegraphics[width=1.48in,height=1.10in]{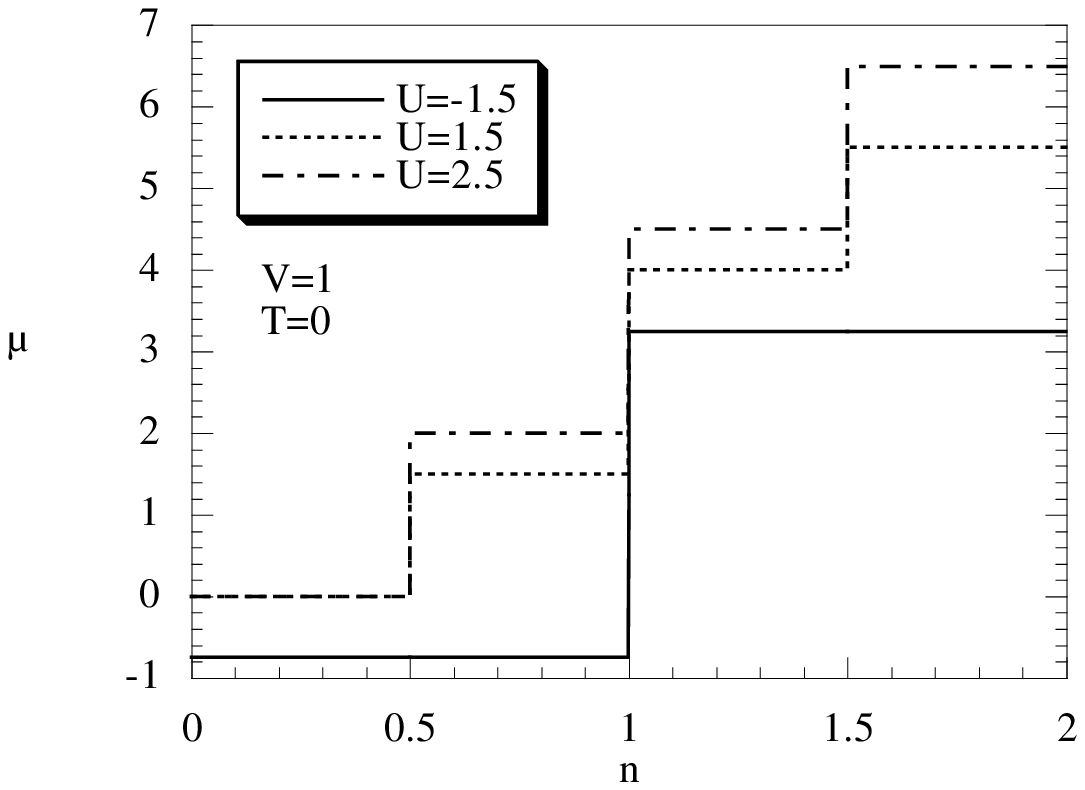}}
 \hspace{4mm}
 \subfigure[]
   {\includegraphics[width=1.48in,height=1.10in]{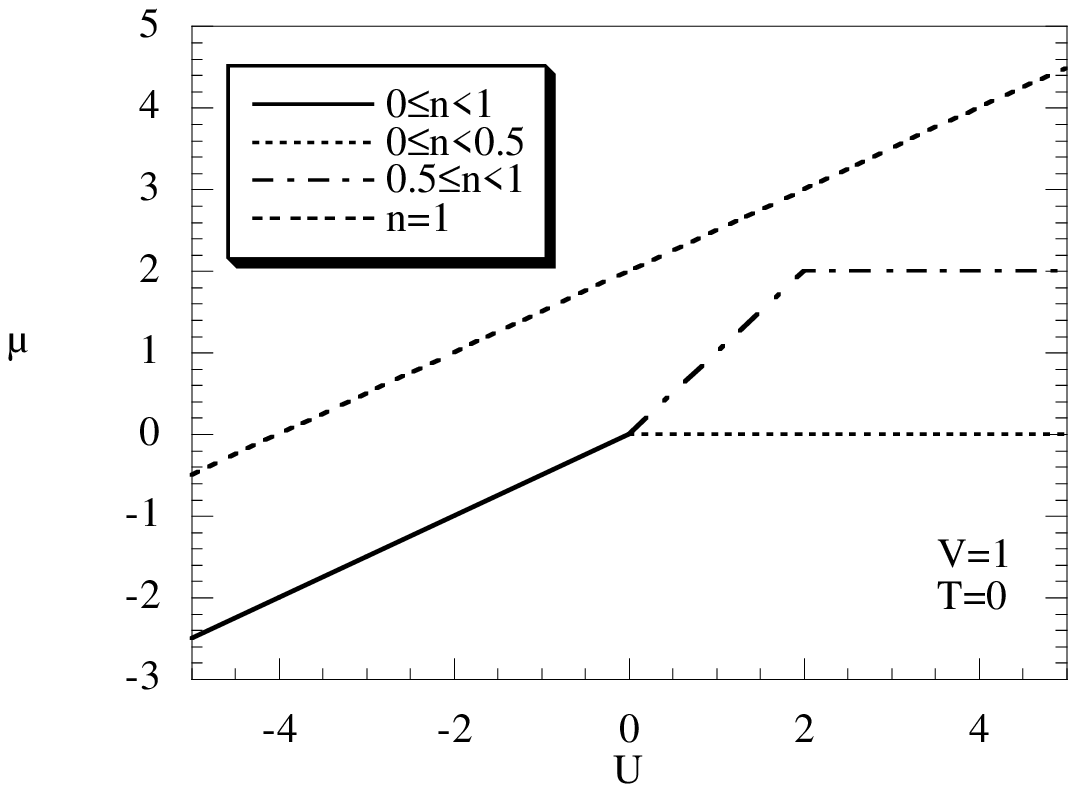}}
\vspace{-4mm}
 \caption{\label{fig13}(a) The chemical potential $\mu$  as
a function of the particle density for $V=1$, $T=0$ and $U=-1.5$,
$1.5$ and $2.5$. (b) The chemical potential as a function of the
on-site potential $U$ for $V=1$, $T=0$ in the region $0\le n\le
1$.}
\end{figure}

\begin{figure}[!h]
  \vspace{-4mm}
 \centering
 \subfigure[]
   {\includegraphics[width=1.48in,height=1.10in]{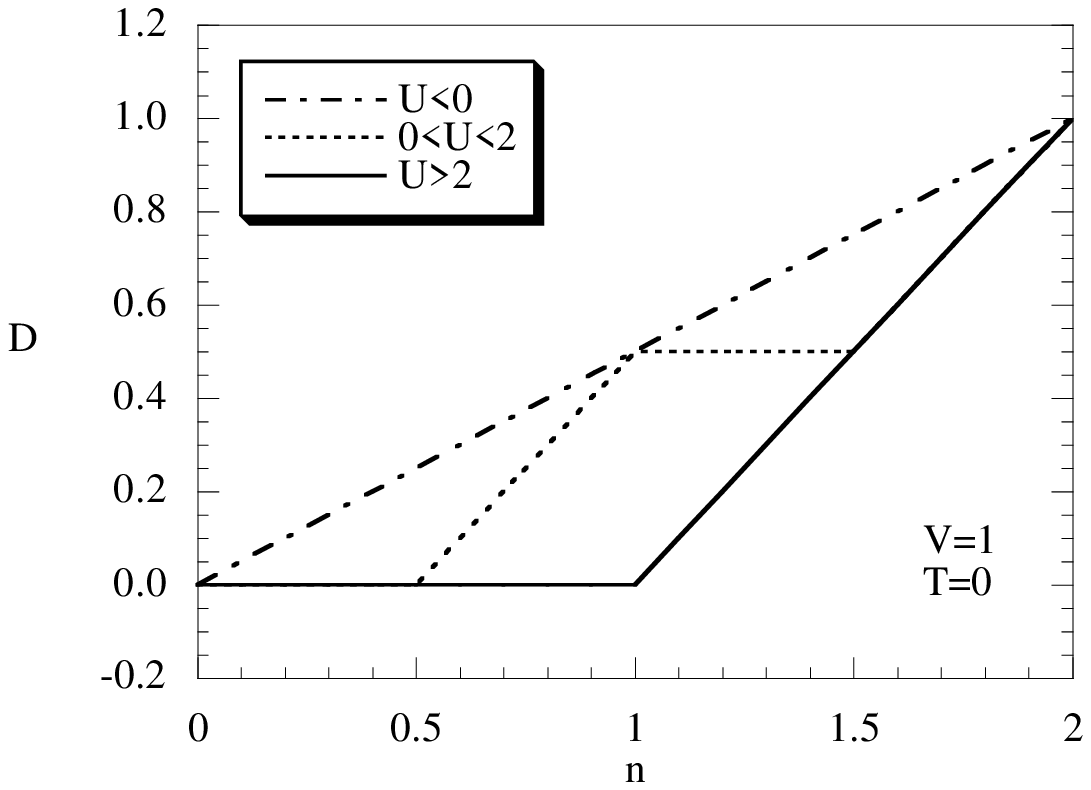}}
 \hspace{1mm}
 \subfigure[]
   {\includegraphics[width=1.48in,height=1.10in]{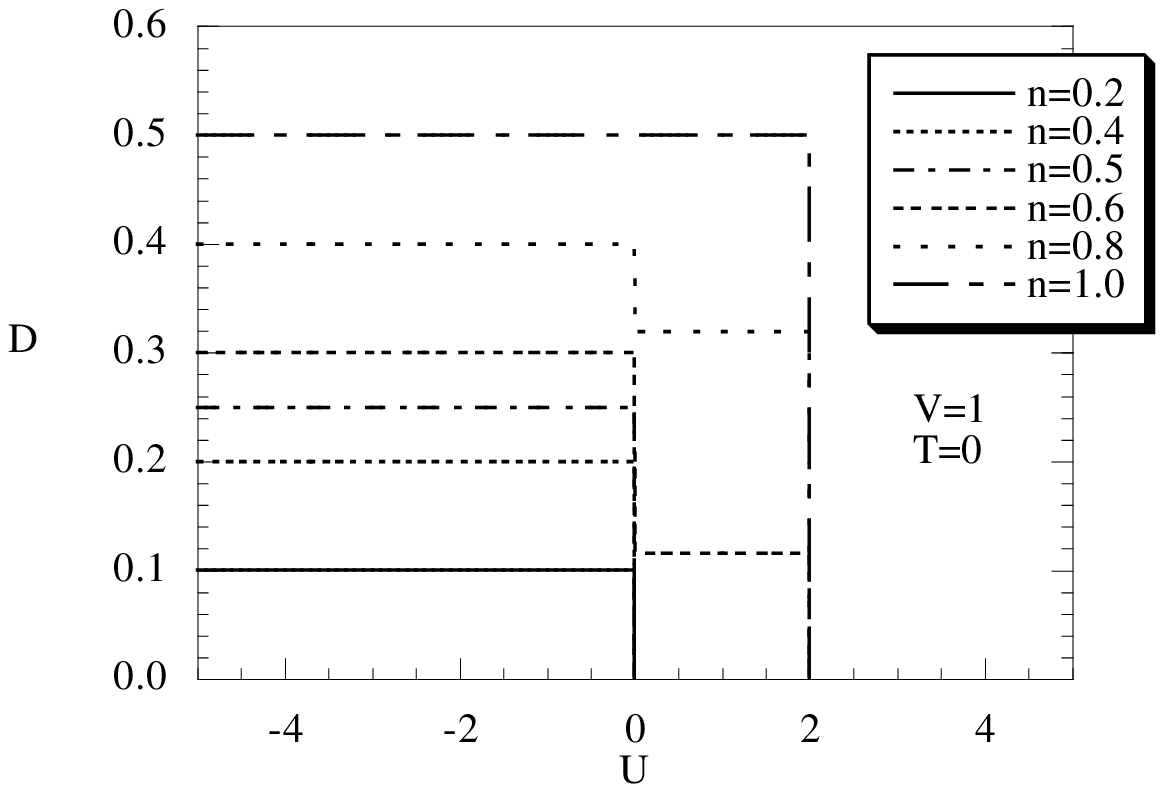}}
\vspace{-4mm}
 \caption{\label{fig14}(a) The double occupancy as a
function of the particle density for $V=1$, $T=0$,  $U<0$
(dot-dashed line), $0<U<2V$ (dotted line) and $U>2V$ (continuous
line). (b) The double occupancy as a function of the on-site
potential $U$ for $V=1$, $T=0$ and various values of $n$.}
 \end{figure}

One can now investigate some thermodynamic quantities, whose
behavior can also be relevant to ascertain some of the results
previously obtained. In particular, we shall investigate the
behavior of the chemical potential and of the double occupancy as
a function of the particle density and of the on-site potential.
In the former case, we shall consider the full range $0 \le n \le
2$ to better recognize the peculiar behaviors of $\mu$ and $D$. In
Fig. \ref{fig13}a we plot the chemical potential as a function of
$n$ for different values of $U$: $\mu$ exhibits plateaus and
discontinuities. In particular, (i) for $U<0$, $\mu$ takes the
constant value $\mu=U/2$ in the region $0<n<1$, corresponding to
phase ($a$). At $n=1$ there is a discontinuity corresponding to
the PT to a checkerboard distribution of doubly occupied sites;
(ii) for $0<U<2V$, $\mu$ takes the constant value $\mu=0$ in the
range $0<n<0.5$, corresponding to phase ($b$). At $n=0.5$ there is
a discontinuity corresponding to the phase transition to the
checkerboard distribution of singly occupied sites. In the range
$0.5<n<1$, $\mu$ takes the constant value $\mu=U$, corresponding
to phase ($c$). At $n=1$ there is a discontinuity corresponding to
the PT to the checkerboard distribution of doubly occupied sites;
(iii) for $U>2V$, $\mu$ takes the constant value $\mu=2V$ in the
region $0.5<n<1$, corresponding to phase ($d$). At $n=1$ there is
a discontinuity corresponding to the PT to the homogeneous
distribution of singly occupied sites. When plotted as a function
of $U$ [see Fig. \ref{fig13}b], the chemical potential shows a
linear behavior for $n=1$, as required by the particle-hole
symmetry. For all other values of the filling, $\mu$ presents a
linear behavior for attractive on-site interaction. Then, for
positive $U$, $\mu$ presents a plateau ($\mu=0$) when $n<0.5$,
whereas, for $0.5 \le n<1$, it increases linearly with $U$ up to
$U=2V$, where it takes the constant value $\mu=2V$.
As it is evident in Fig. \ref{fig14}a, the double occupancy shows
three different behaviors according to the value of $U$. For $0
\le n \le 1$, the double occupancy is always zero when $U>2V$, it
shows a linear behavior for negative values of $U$ and presents a
discontinuity at quarter filling, when $0<U<2V$. In Fig.
\ref{fig14}b we plot the double occupancy as a function of the
on-site potential $U$ for several values of $n$: one observes a
one-step or a two-step behavior depending on the particle density.

In closing this section, it is worthwhile to mention that at zero
temperature all the phases (with the exception of the case $n=1$
and $U<2V$) exhibit a macroscopic degeneracy growing exponentially
with the volume of the lattice.

\section{finite temperature results}
\label{sec_IV}

In this section we shall investigate the finite temperature
properties of the AL-EHM. We study the behavior of the system as a
function of the parameters $n$, $T$, and $U$ and again we take
$V=1$ as the unit of energy. We shall focus our attention on
several thermodynamic quantities: namely, the chemical potential
$\mu$, the double occupancy $D$, the nearest-neighbor charge
correlation $\chi^\alpha =2\lambda ^{(1)}=\langle n(i)n^\alpha
(i)\rangle$, the thermal compressibility $\kappa ^{(T)}= (1/n^2)
\cdot dn/d\mu $, the density of states, the charge $\langle n(i)
n(j)\rangle$ and double occupancy $\langle D(i)D(j)\rangle$
correlation functions, the charge $\chi_c$ and the spin $\chi_s$
susceptibilities, the internal energy $E=UD+V\chi^\alpha$, the
specific heat $C=dE/dT$, and the entropy $S$. Under the
transformation $n\to 2-n$, these quantities transform as:
\begin{equation}
\label{EHM_33}
\begin{split}
 \mu (2-n)&=4V+U-\mu (n) \\
 D(2-n)&=1-n+D(n) \\
 \chi^\alpha (2-n)&=4(1-n)+\chi^\alpha (n) \\
 \chi_c(2-n) &=  \chi_c(n)   \\
\chi_s(2-n) &=  \chi_s(n)   \\
 E(2-n)&=(U+4V)(1-n)+E(n) \\
 C(2-n)&=C(n).
 \end{split}
\end{equation}
Therefore, unless otherwise stated, we shall limit the analysis to
the range $0\le n\le 1$. In order to characterize the various
features of the thermodynamic quantities, in the following we
shall distinguish three intervals along the $U$ line (at the
borders of which zero temperature transitions occur): namely,
$U<0$, $0<U<2V$, and $U>2V$.

\subsection{The chemical potential}

\begin{figure}[t]
 \centering
 \subfigure[]
   {\includegraphics[width=1.48in,height=1.10in]{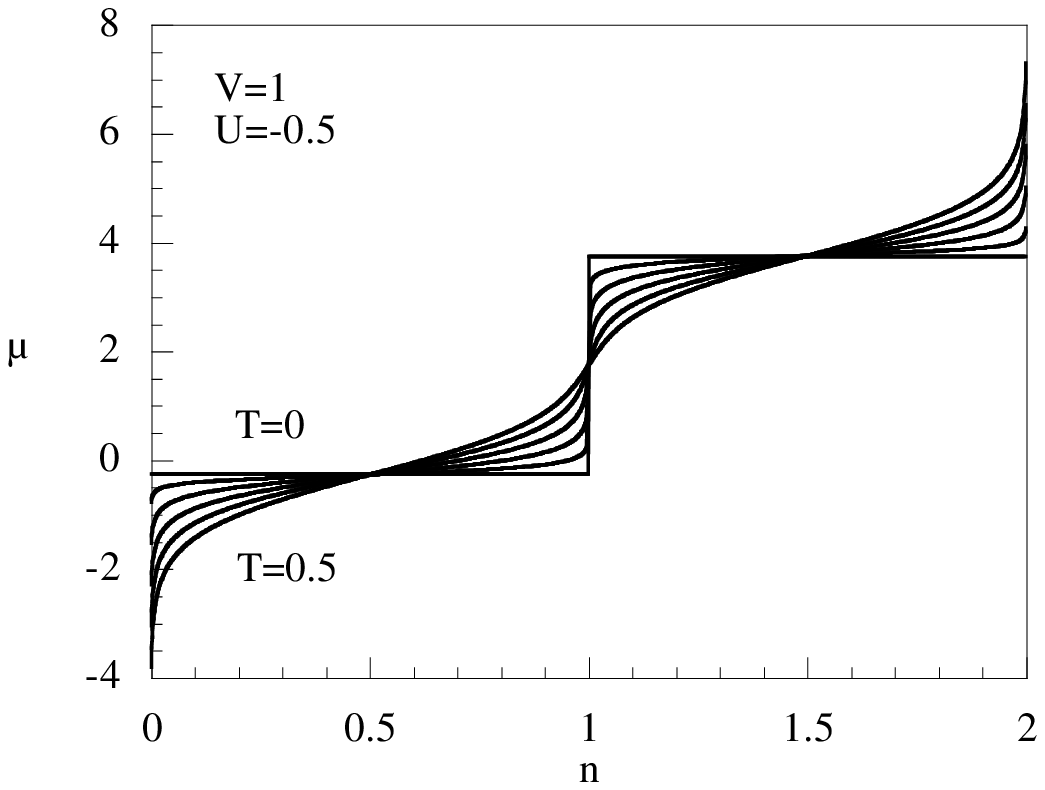}}
 \hspace{1mm}
 \subfigure[]
   {\includegraphics[width=1.48in,height=1.10in]{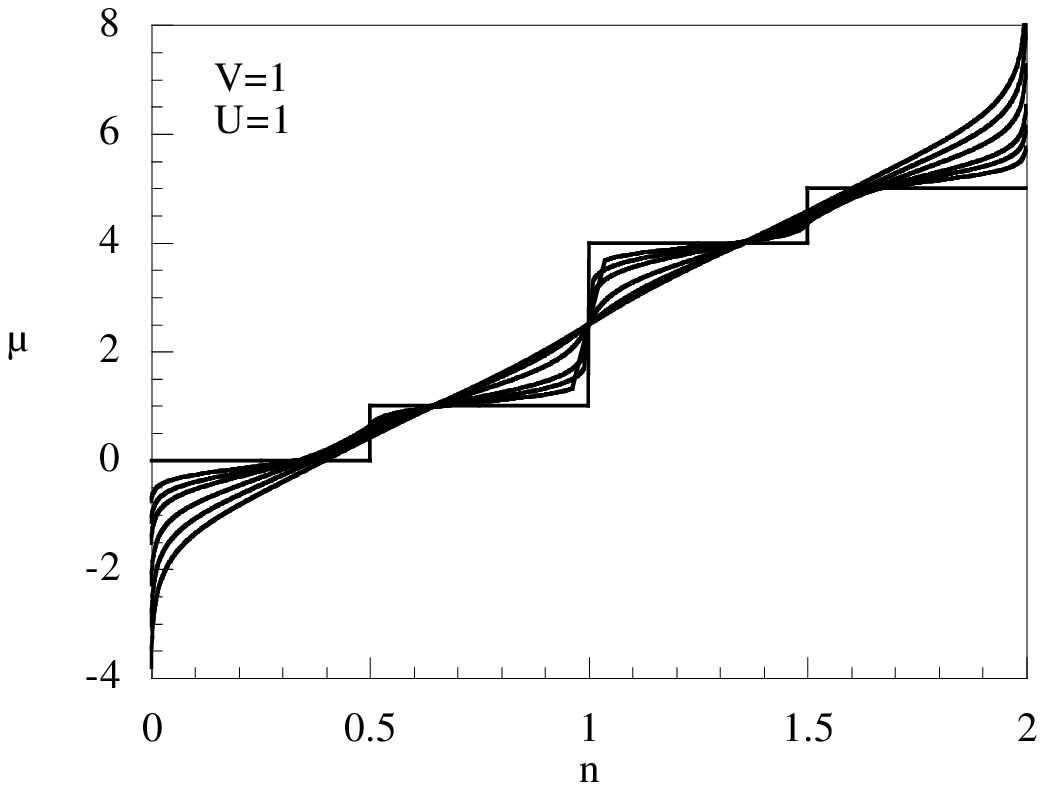}}
   \hspace{1mm}
 \subfigure[]
   {\includegraphics[width=1.48in,height=1.10in]{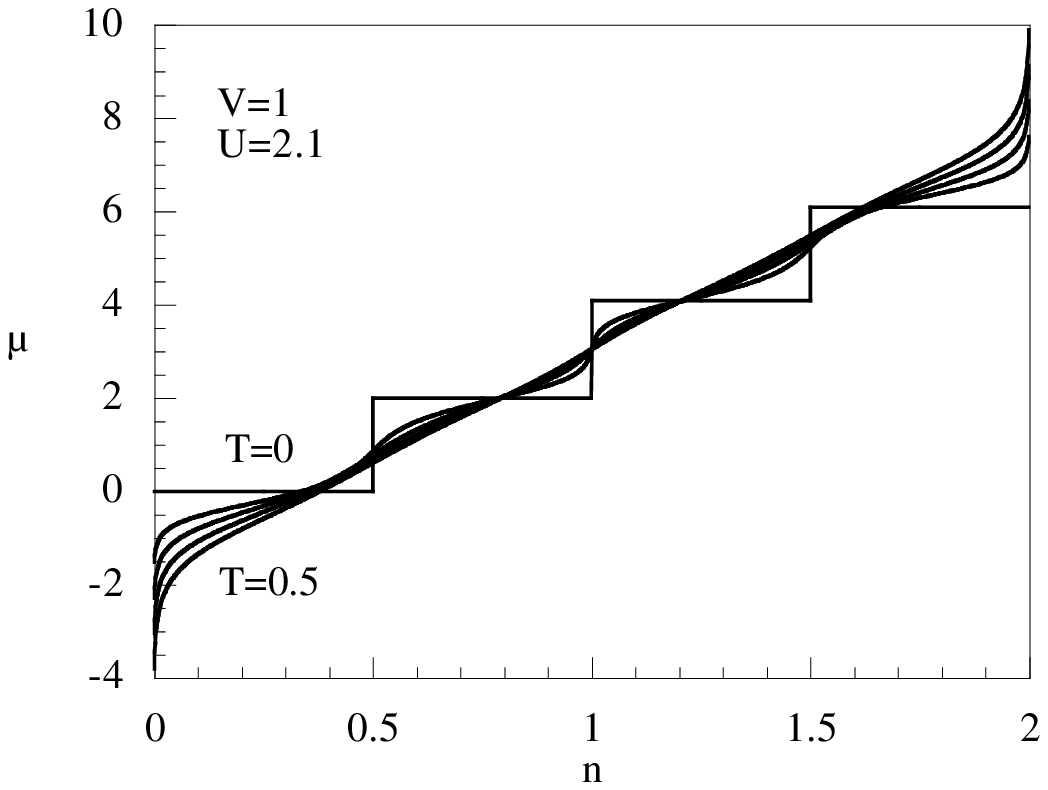}}
 \caption{\label{fig15} The chemical potential $\mu$ as a function of
the particle density $n$ for various temperatures and (a) $U<0$,
(b) $0<U<2V$, (c) $U>2V$.}
 \end{figure}

The behavior of the chemical potential as a function of the
particle density $n$ is reported in Figs. \ref{fig15}a-c as the
temperature varies, for the three regions $U<0$, $0<U<2V$ and
$U>2V$, respectively. One can immediately see that $\mu$ is always
an increasing function of $n$, hinting at a thermodynamically
stable system for all ranges of the parameters $n$, $T$, and $U$.
We consider the full range $0 \le n\le 2$ to better appreciate the
jumps in $\mu$ as the particle density is varied. An interesting
feature is the presence of crossing points in the chemical
potential curves, when plotted for different temperatures and
$U<2V$. More precisely, one observes crossing points at the values
$n=0.5$, $n=1$, and $n=1.5$, where PTs are observed at $T=0$. In
between these values, for temperatures close to zero, the chemical
potential presents plateaus. In Sec. \ref{spec_heat} we shall see
that nearly universal crossing points are also exhibited by the
specific heat. The value $n=0.5$ is a turning point for the
derivative $d\mu /dT$: for $n<0.5$ ($n>0.5$), $\mu$ is a
decreasing (increasing) function of $T$.
\begin{figure}[th]
\centerline{\includegraphics[scale=0.45]{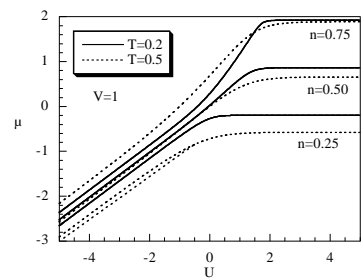}}
\caption{\label{fig16} The chemical potential $\mu$ as a function
of the on-site interaction $U$ for $n=0.25$, 0.5, 0.75 and for
$T=0.2$ (solid line) and $T=0.5$ (dotted line).}
\end{figure}
In Fig. \ref{fig16} we show the chemical potential as a function
of the on-site interaction $U$ for $n=0.25$, 0.5, 0.75 and for two
temperatures. For negative $U$, $\mu $ increases by increasing $U$
following approximately a linear law ($\mu =U/2$ for $T=0$). When
$U>0$, $\mu $ deviates from the linear behavior and, for $U\approx
2V$, tends to a constant value depending both on $n$ and $T$. The
latter behavior has been observed also for the 2D Hubbard model
(see Ref. \cite{mancini_95} and references therein).

\subsection{The double occupancy}

\begin{figure}[t]
 \centering
 \subfigure[]
   {\includegraphics[width=1.48in,height=1.10in]{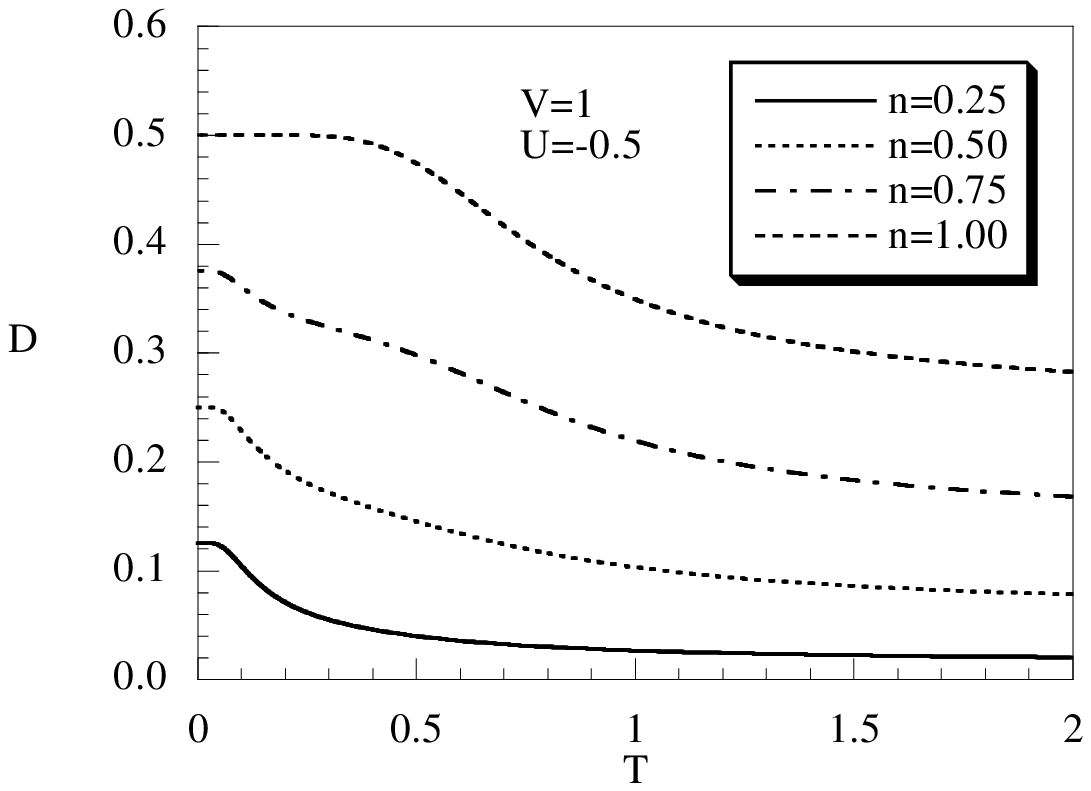}}
 \hspace{1mm}
 \subfigure[]
   {\includegraphics[width=1.48in,height=1.10in]{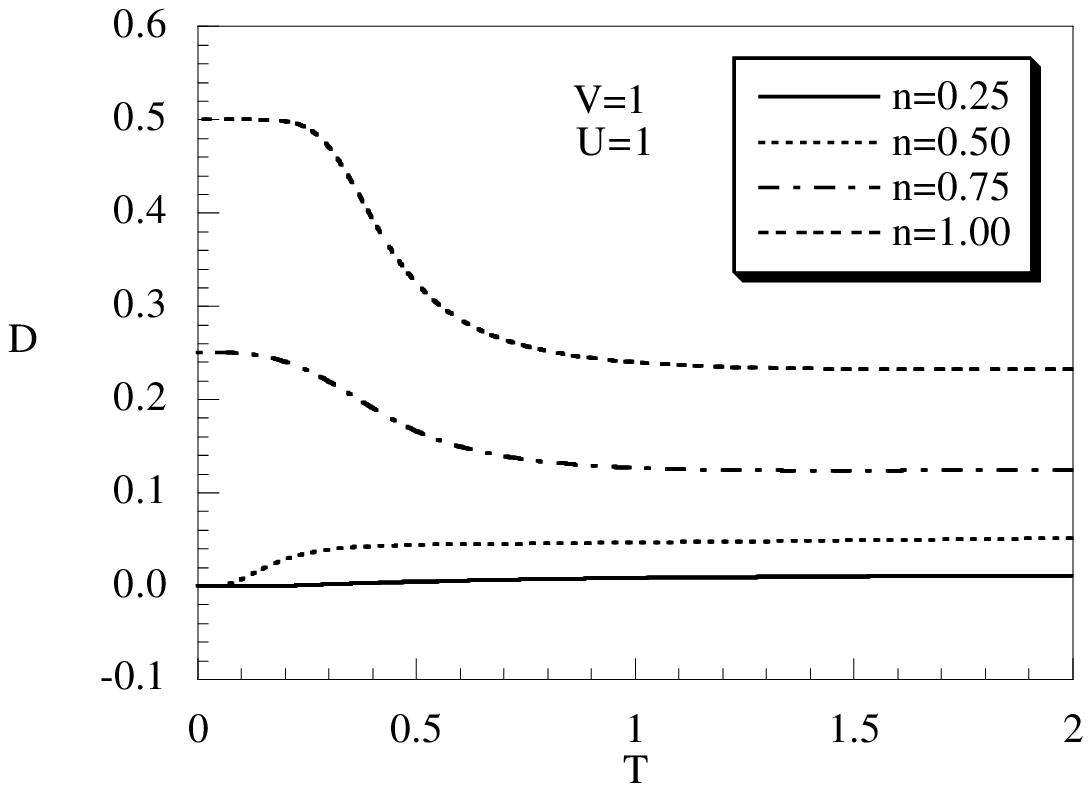}}
   \hspace{1mm}
 \subfigure[]
   {\includegraphics[width=1.48in,height=1.10in]{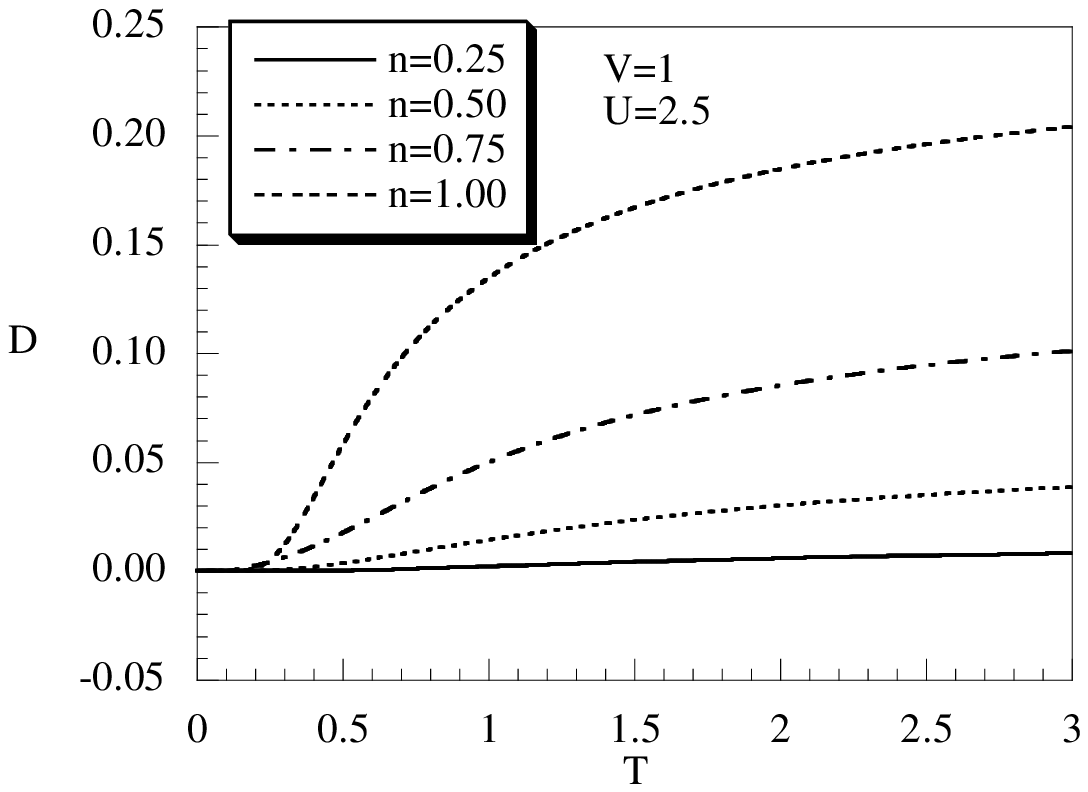}}
 \caption{\label{fig17} The double occupancy as a function of
the temperature for different values of the particle density and
(a) $U=-1/2$, (b) $U=1$, (c) $U=5/2$.}
 \end{figure}

In Figs. \ref{fig17}a-c  the temperature dependence of the double
occupancy $D$ is shown  for the three regions $U<0$, $0<U<2V$ and
$U>2V$, respectively. In the first region (attractive $U$), $D$ is
always a decreasing function of the temperature since the thermal
excitations tend to break the doublons, i.e., the charge carriers
of doubly occupied sites.
In the second region ($0<U<2V$), the double occupancy is an
increasing (decreasing) function for $n<0.5$ ($n>0.5$). This is
easily understandable if one thinks of the zero-temperature
scenario: for filling less than one-quarter, the thermal
excitations may favor the formation of doublons, against the
repulsive interaction $U$. For filling greater than one-quarter,
the thermal excitations may favor the formation of neighboring
singly occupied sites, against the repulsive interaction $V$. In
the third region ($U>2V$), the double occupancy is always an
increasing function of $T$, since the thermal excitations favor
the formation of doublons, against the repulsive interaction $U$.
\begin{figure}[ht]
\centerline{\includegraphics[scale=0.45]{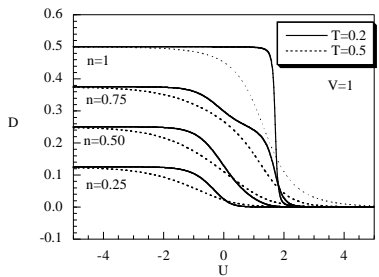}} \caption{
\label{fig18}The double occupancy as a function of the on-site
interaction $U$ for $n=0.25$, 0.5, 0.75, 1 and for two different
temperatures.}
\end{figure}
A similar behavior of the double occupancy by varying the
temperature has been found in a strong coupling analysis of the
EHM at half filling \cite{Glocke07}.

It is interesting to plot the double occupancy as a function of
the on-site potential $U$. As expected, at all temperatures, $D$
is always a decreasing function of $U$. As one can see from Fig.
\ref{fig18}, the discontinuities observed at zero temperature for
$U=0$ and $U=2V$ are more and more smoothed as the temperature is
increased. In the limit $U\to -\infty$ the double occupancy tends
to the constant value $D=n/2$, independently of the temperature:
all the electrons are paired. In the opposite limit, namely $U\to
\infty$, the on-site repulsion prevents the double occupations of
the sites.

\subsection{The nearest-neighbor charge correlation function}

\begin{figure}[ht]
 \centering
 \subfigure[]
   {\includegraphics[width=1.48in,height=1.10in]{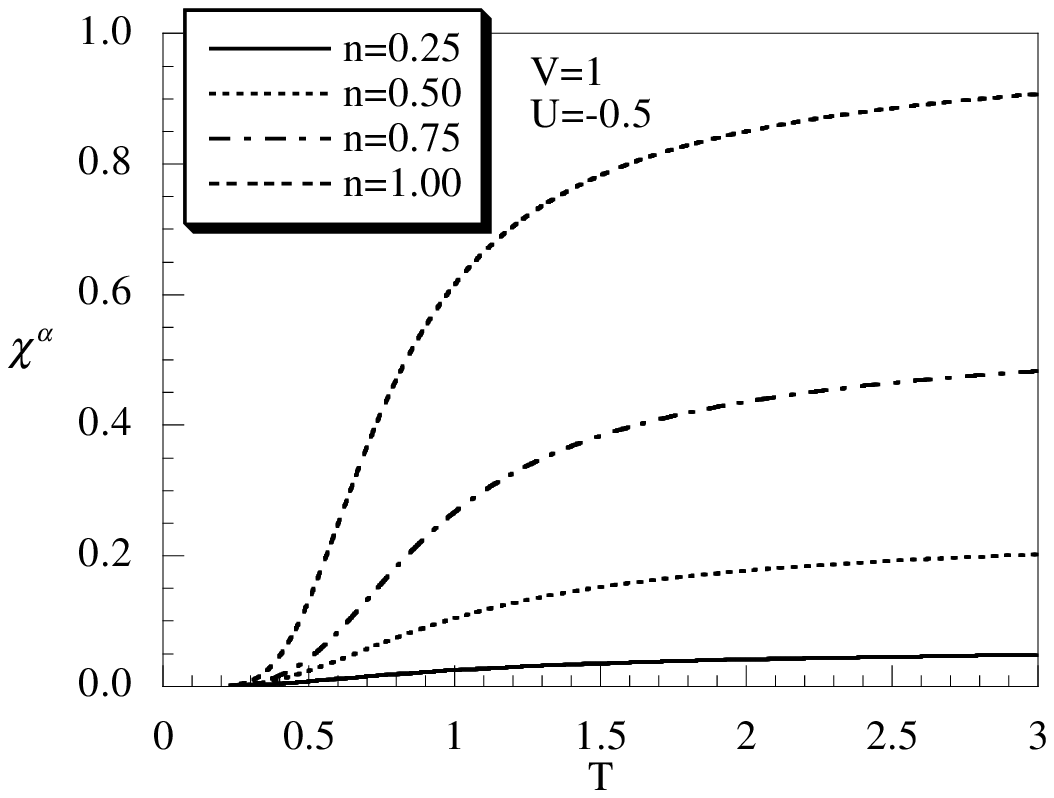}}
 \hspace{1mm}
 \subfigure[]
   {\includegraphics[width=1.48in,height=1.10in]{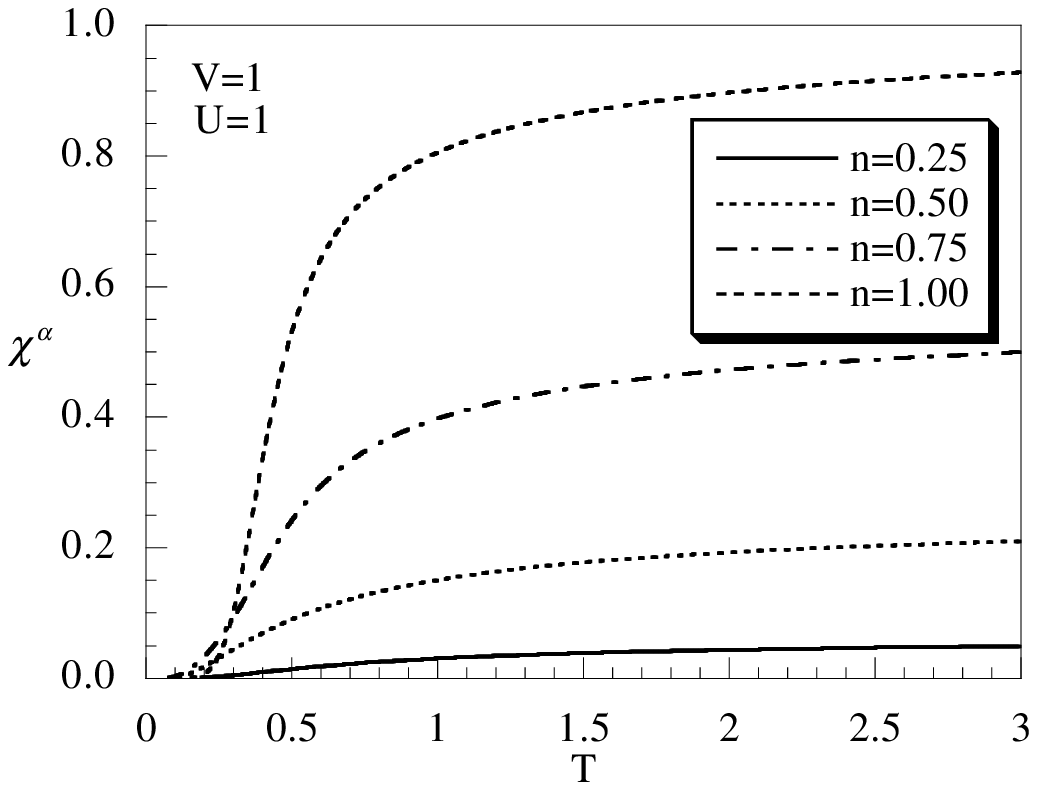}}
  \hspace{1mm}
 \subfigure[]
   {\includegraphics[width=1.48in,height=1.10in]{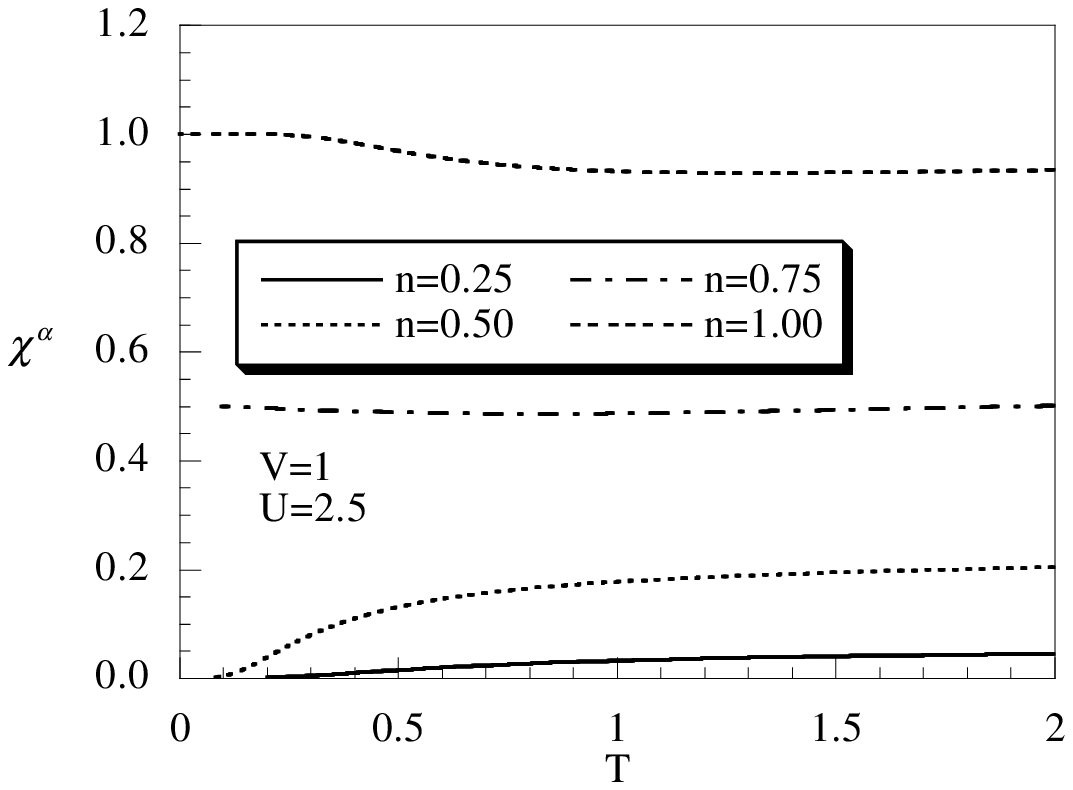}}
 \caption{\label{fig19} The nearest-neighbor charge correlation function
$\chi ^\alpha =\langle n(i) n^\alpha (i)\rangle$ as a function of
the temperature for different values of the particle density and
(a) $U=-1/2$, (b) $U=1$, (c) $U=5/2$.}
 \end{figure}

As discussed in Sec. \ref{sec_III}, there are two interesting
quantities which mainly characterizes the state of the system: one
is the double occupancy, discussed in the previous subsection, the
other is the nearest-neighbor charge correlation function $\chi
^\alpha =\langle n(i) n^\alpha (i) \rangle=2\lambda ^{(1)}$. The
behavior of $\chi^\alpha$ as a function of the temperature is
shown in Figs. \ref{fig19}a-c, for the three regions $U<0$,
$0<U<2V$ and $U>2V$, respectively. As one would have expected, the
behavior of $\chi^\alpha$ is opposite to the one of $D$ in some
regions of the parameters: for instance, for attractive on-site
interactions, thermal excitations tend to break the doublons and
to consequently favor the occupation of neighboring sites.
For $U<2V$, $\chi^\alpha$ is always an increasing function of $T$
and vanishes in the limit $T\to 0$, where occupation of neighbor
sites is clearly disfavored. As one can see in Figs.
\ref{fig19}a-c, the thermal excitations may favor the occupation
of neighboring sites and, in the limit $T\to \infty$,
$\chi^\alpha$ tends to the value $n^2$ for all values of $U$ and
$n$. When the on-site potential is greater than $2V$, the behavior
of $\chi^\alpha$ depends on the filling: it is an increasing
(decreasing) function for $n<0.5$ ($n>0.5$) at low temperatures.
This different behavior can be understood if one looks at Fig.
\ref{fig12}: it is clear that for lower fillings thermal
excitations are more effective in augmenting the nearest-neighbor
charge correlations with respect to higher fillings, where
electrons are already clustered together.

\begin{figure}[ht]
\centerline{\includegraphics[scale=0.45]{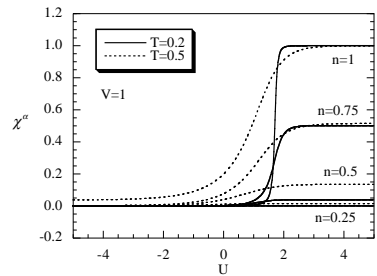}}
 \caption{ \label{fig20} The nearest-neighbor charge correlation
function $\chi^\alpha =\langle n(i)n^\alpha (i)\rangle$ as a
function of the on-site interaction $U$ for $n=0.25$, 0.5, 0.75, 1
and for $T=0.2$ (solid lines) and $T=0.5$ (dotted lines).}
\end{figure}

In Fig. \ref{fig20} we plot $\chi^\alpha$ as a function of the
on-site potential: one observes that $\chi^\alpha$ is always an
increasing function of $U$, and tends to a constant value (which
depends on $n$), independently of the temperature, in the limit of
$U\to +\infty$.

\subsection{The compressibility}

In this subsection we study the thermodynamic compressibility,
defined as $\kappa =\left(\partial n /\partial \mu \right) /n^2$.
 \begin{figure}[t]
 \centering
 \subfigure[]
   {\includegraphics[width=1.48in,height=1.10in]{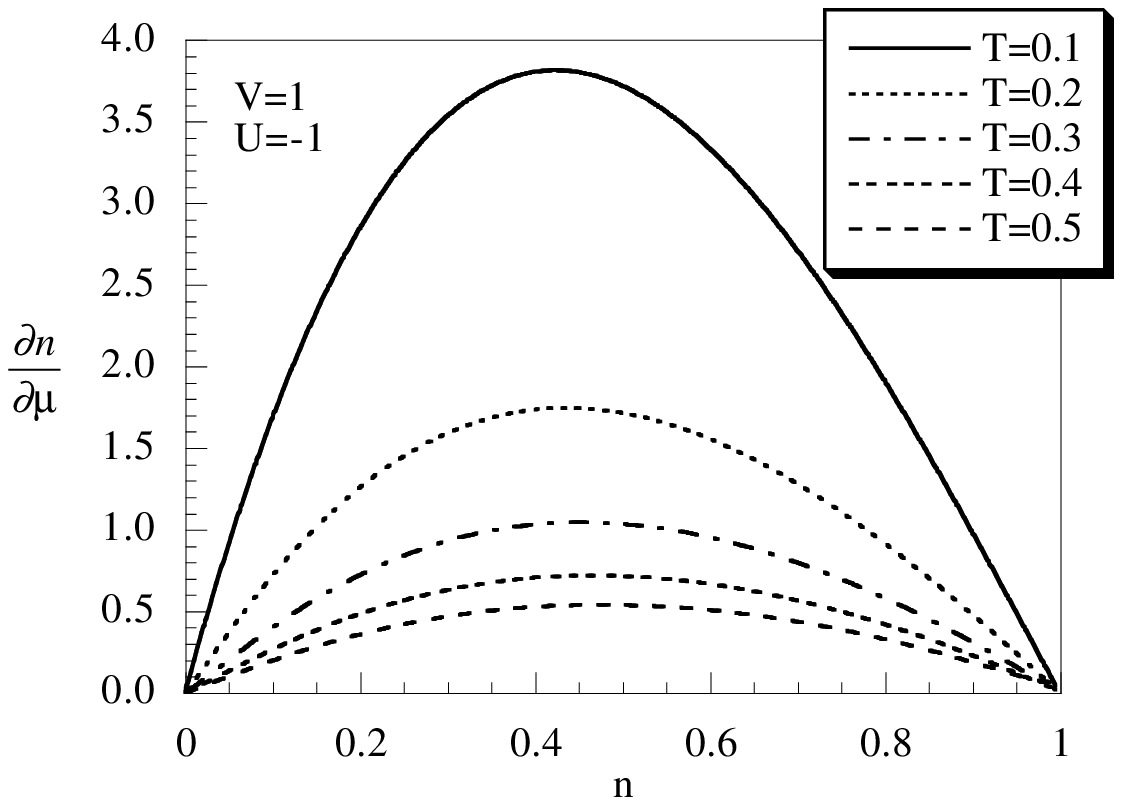}}
 \hspace{1mm}
 \subfigure[]
   {\includegraphics[width=1.48in,height=1.10in]{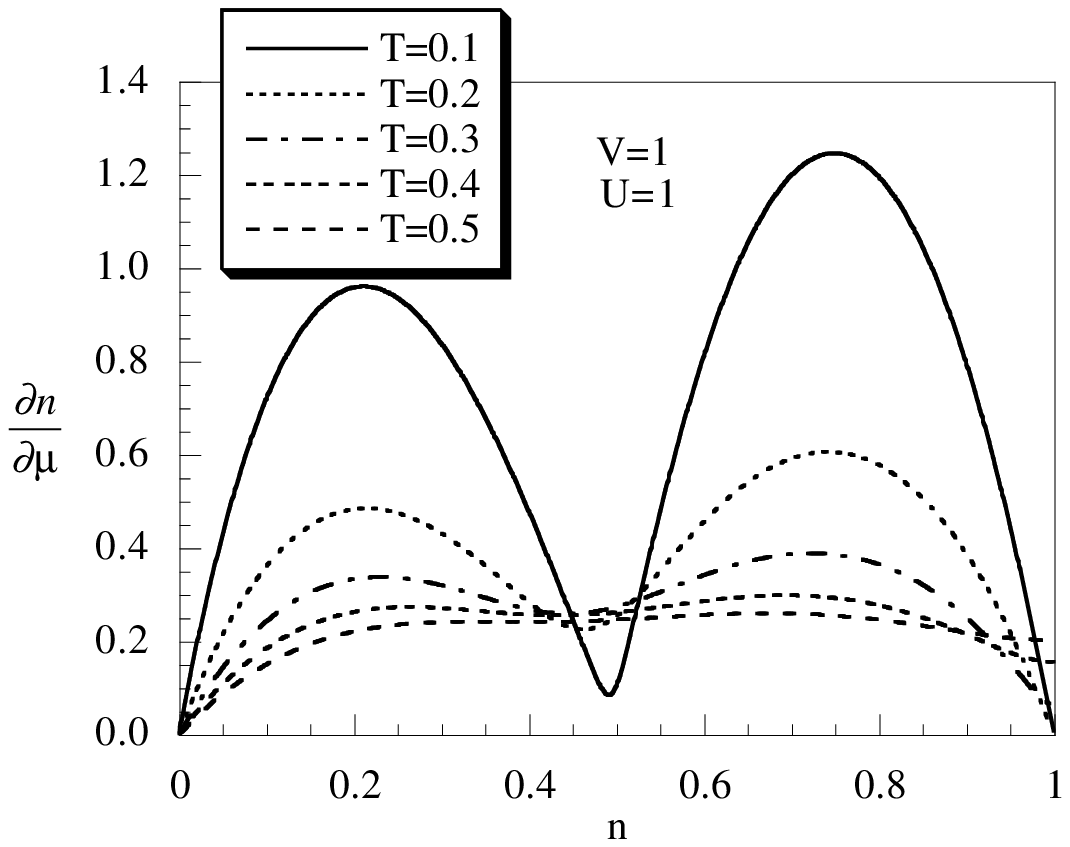}}
  \hspace{1mm}
 \subfigure[]
   {\includegraphics[width=1.48in,height=1.10in]{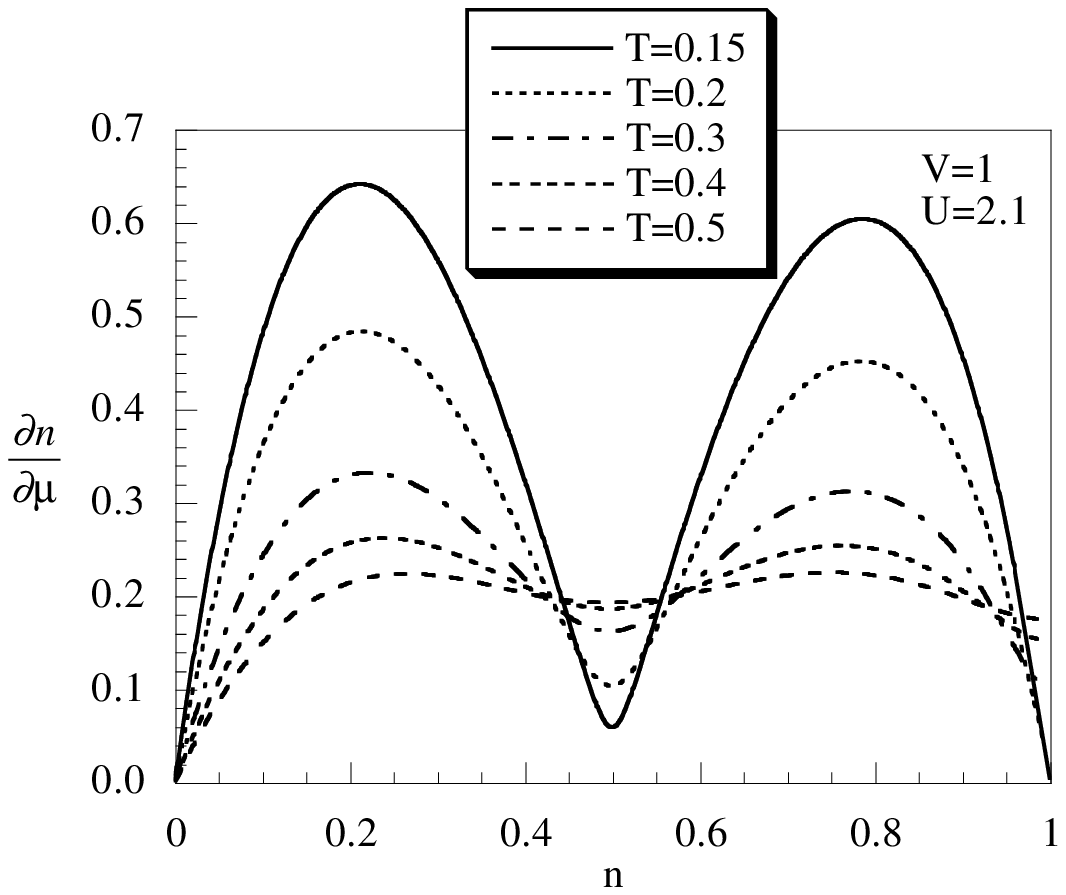}}
 \caption{\label{fig21} $\partial n/\partial \mu $ as
a function of the particle density $n$ for various temperatures
and (a) $U<0$, (b) $0<U<2V$, (c) $U>2V$.}
 \end{figure}
In Figs. \ref{fig21}a-c we plot $\kappa$ as a function of the
filling $n$, for three representative values of $U$ ($U=-V$, $U=V$
and $U=2.1V$, respectively) and for different temperatures. To
avoid divergences in the limit $n \to 0$, we plot only the
derivative $\left(\partial n /\partial \mu \right)$. In the first
region ($U<0$, phase ($a$)), the compressibility increases by
increasing $n$ up to a maximum in correspondence of
quarter-filling; then it decreases exhibiting a minimum at
half-filling, where a charge ordered state is established at $T=0$
(see Fig. \ref{fig3}). In the second region ($0<U<2V$, phases
($b$) and ($c$)), $\kappa$ has a double peak structure with two
maxima at $n=0.25$ and $n=0.75$ and two minima at $n=0.5$ and
$n=1$; the latter points correspond to the situation where a
charge ordered state (checkerboard distribution for singly and
doubly occupied sites, respectively) is observed at zero
temperature (see Fig. \ref{fig8}).
 \begin{figure}[ht]
 \centering
 \subfigure[]
   {\includegraphics[width=1.48in,height=1.10in]{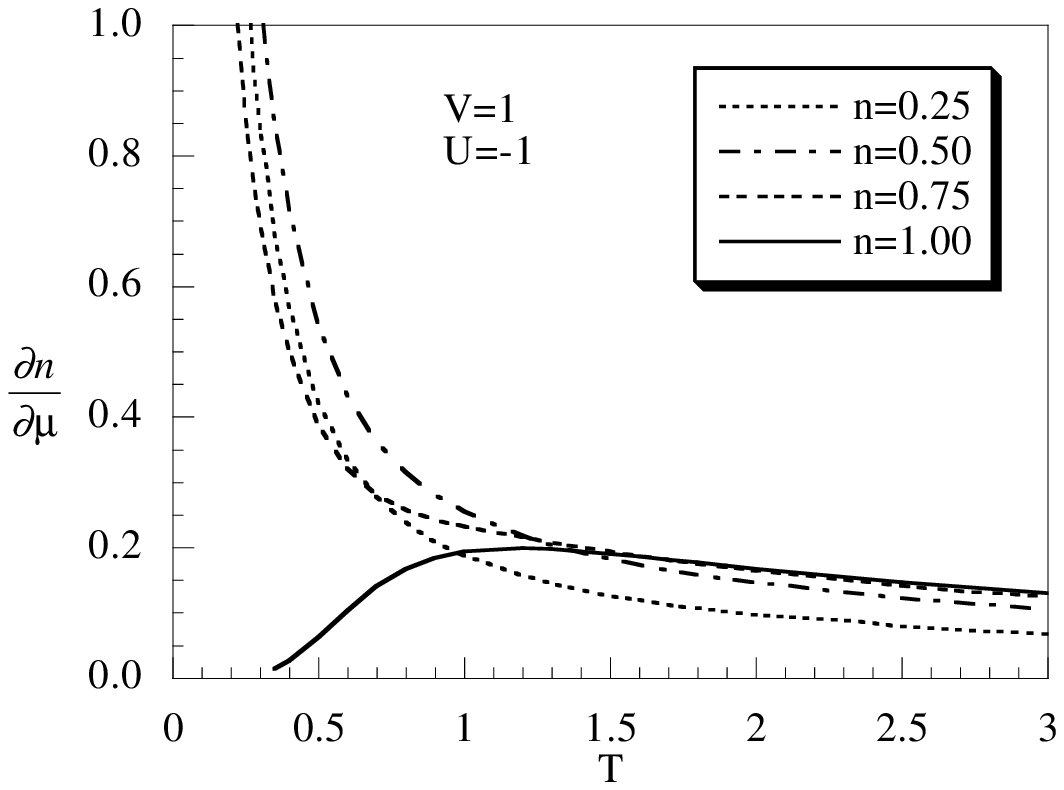}}
 \hspace{1mm}
 \subfigure[]
   {\includegraphics[width=1.48in,height=1.10in]{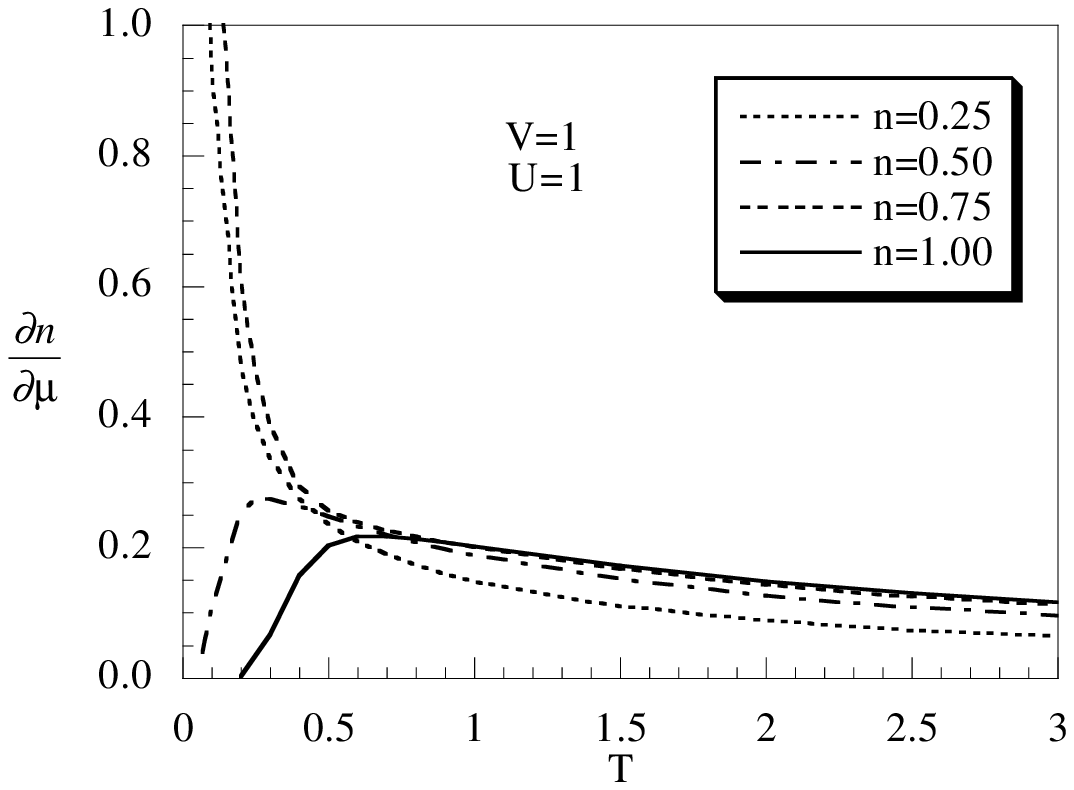}}
  \hspace{1mm}
 \subfigure[]
   {\includegraphics[width=1.48in,height=1.10in]{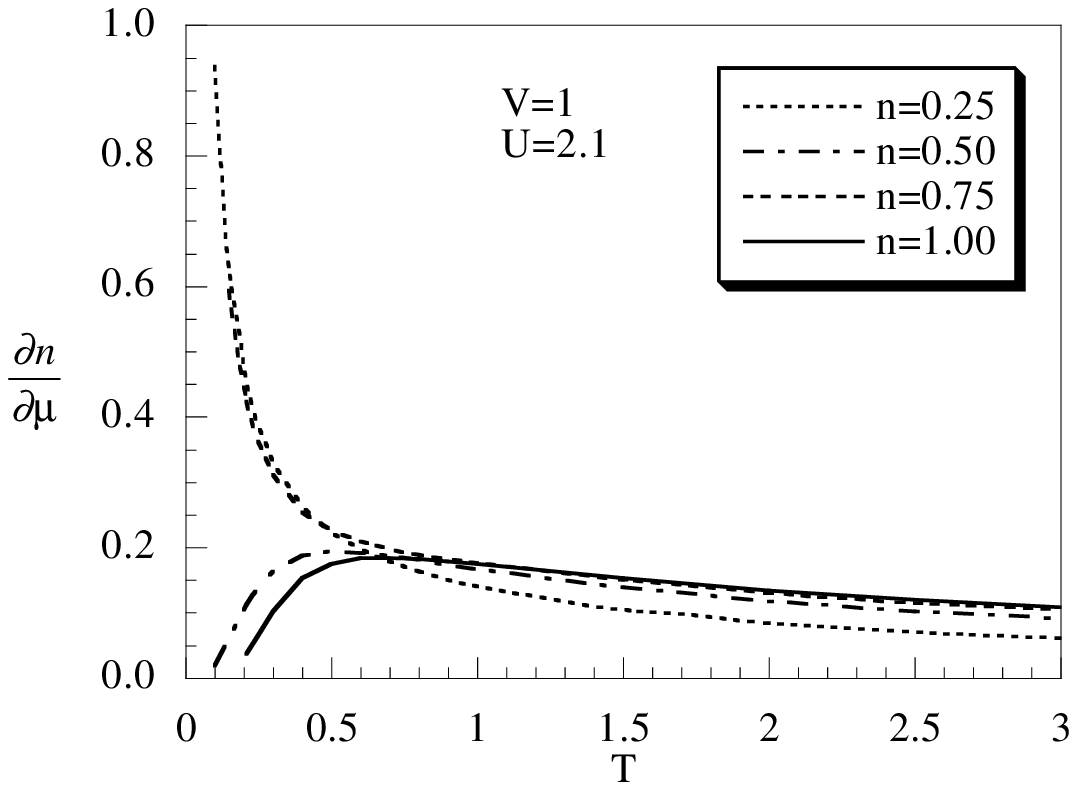}}
 \caption{\label{fig22} $\partial n/\partial \mu $ as
a function of the temperature for various fillings and (a) $U<0$,
(b) $0<U<2V$, (c) $U>2V$.}
 \end{figure}
In the third region ($U>2V$, phases ($b$) and ($d$)), $\kappa$ has
a behavior similar to the one exhibited in the second region, with
the difference that the two peaks have now the same height. This
difference is due to the fact that in the phase $(d)$, for all
values of the filling, there are only singly occupied sites (see
Fig. \ref{fig12}). In Figs. \ref{fig22}a-c we plot $\left(\partial
n /\partial \mu \right)$ as a function of the temperature, again
in the three regions of $U$ ($U=-V$, $U=V$ and $U=2.1V$,
respectively) and for different values of the filling ($n=0.25$,
0.5, 0.75, 1). For all values of $U$ and $n$, in the limit of
large temperatures, the compressibility vanishes following the law
$\kappa \propto T^{-1}$, but becomes strongly filling dependent as
$T$ is reduced. In particular, in the limit $T \to 0$ and in the
regions where no CO state is observed, $\kappa$ increases by
lowering $T$ diverging at $T=0$. On the other hand, in the regions
where a charge ordered state is present, the compressibility
exhibits a maximum at a certain temperature, then, further
decreasing $T$, $\kappa$ vanishes.

\subsection{Correlation functions at finite temperature}

 \begin{figure}[t]
 \centering
 \subfigure[]
   {\includegraphics[width=1.48in,height=1.10in]{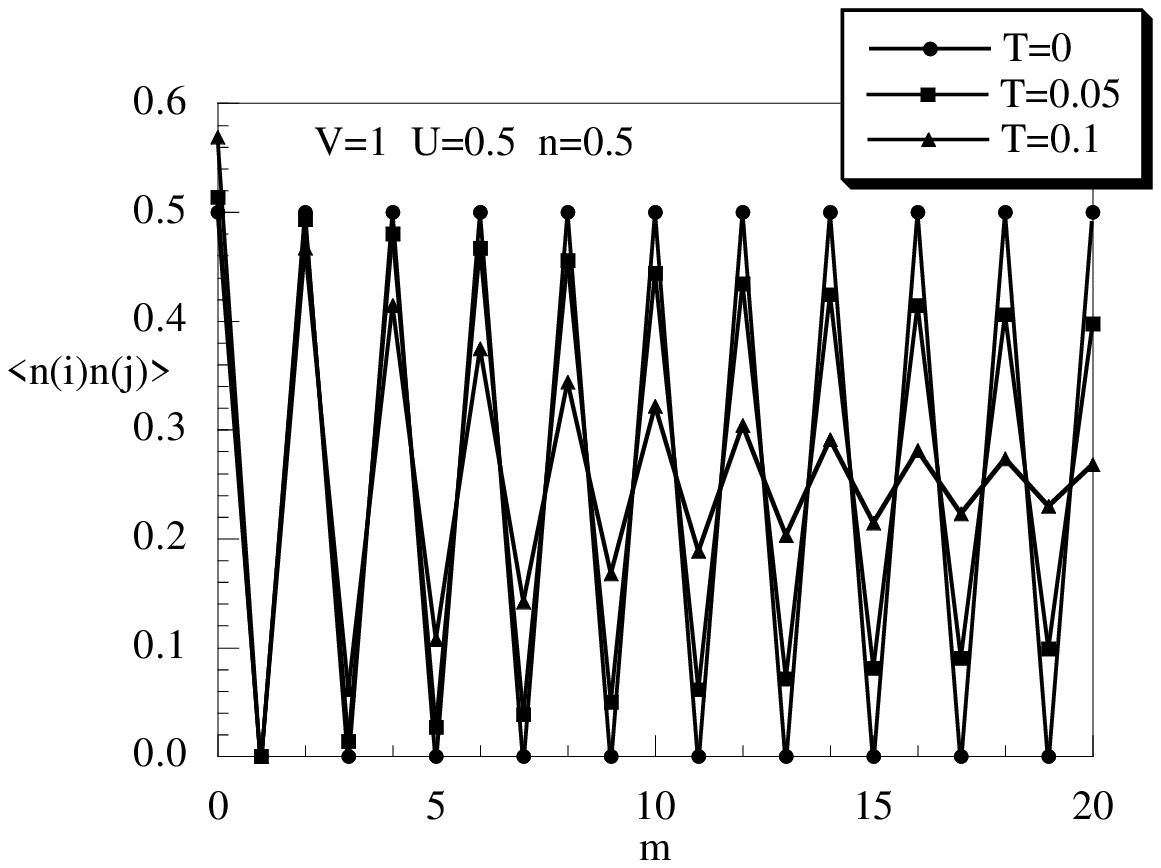}}
 \subfigure[]
   {\includegraphics[width=1.48in,height=1.10in]{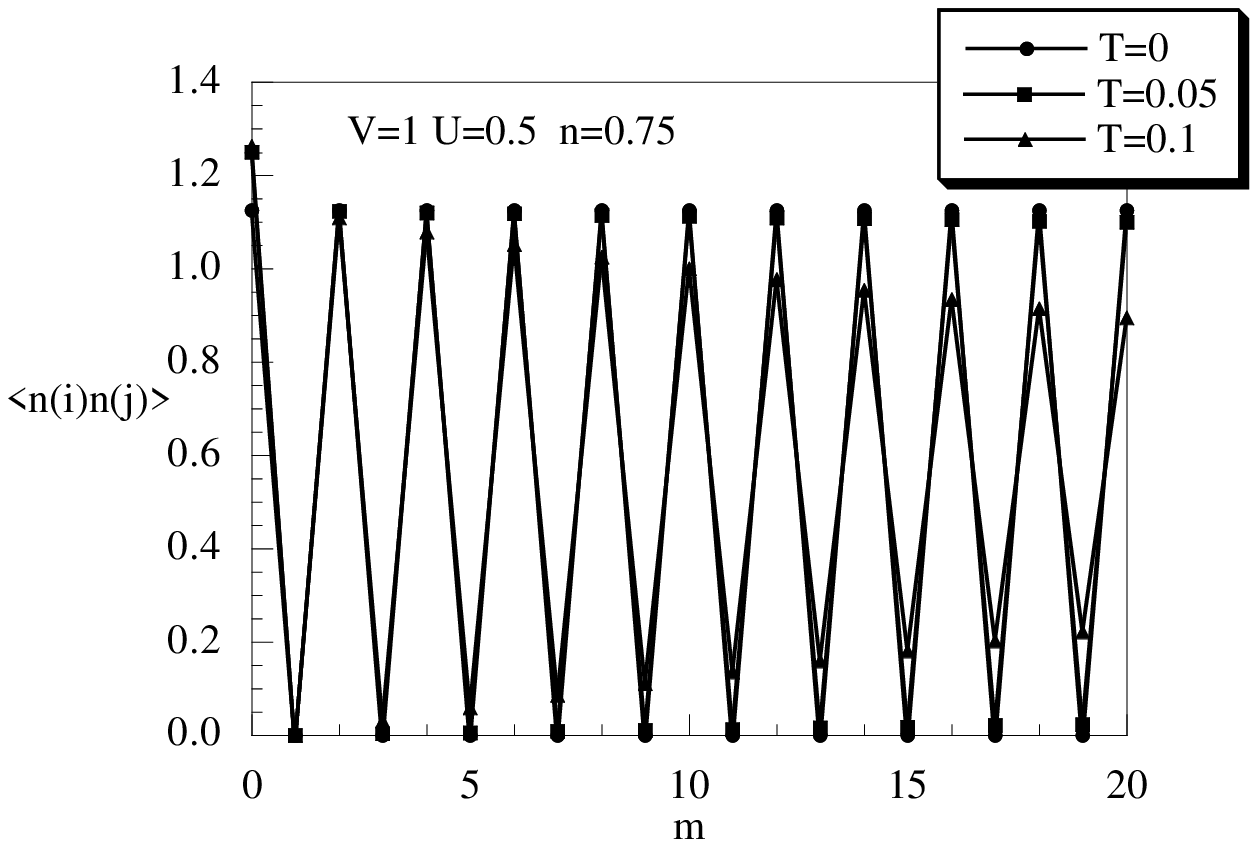}}
 \subfigure[]
   {\includegraphics[width=1.48in,height=1.10in]{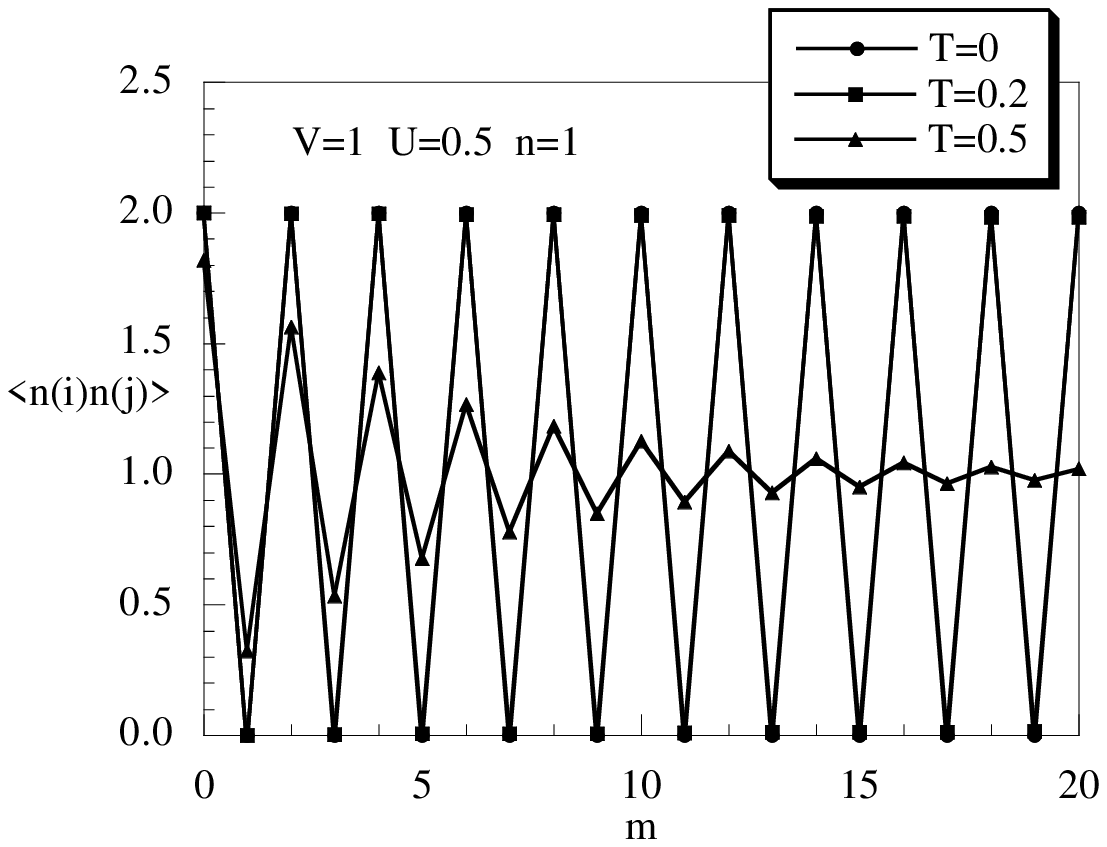}}
    \subfigure[]
   {\includegraphics[width=1.48in,height=1.10in]{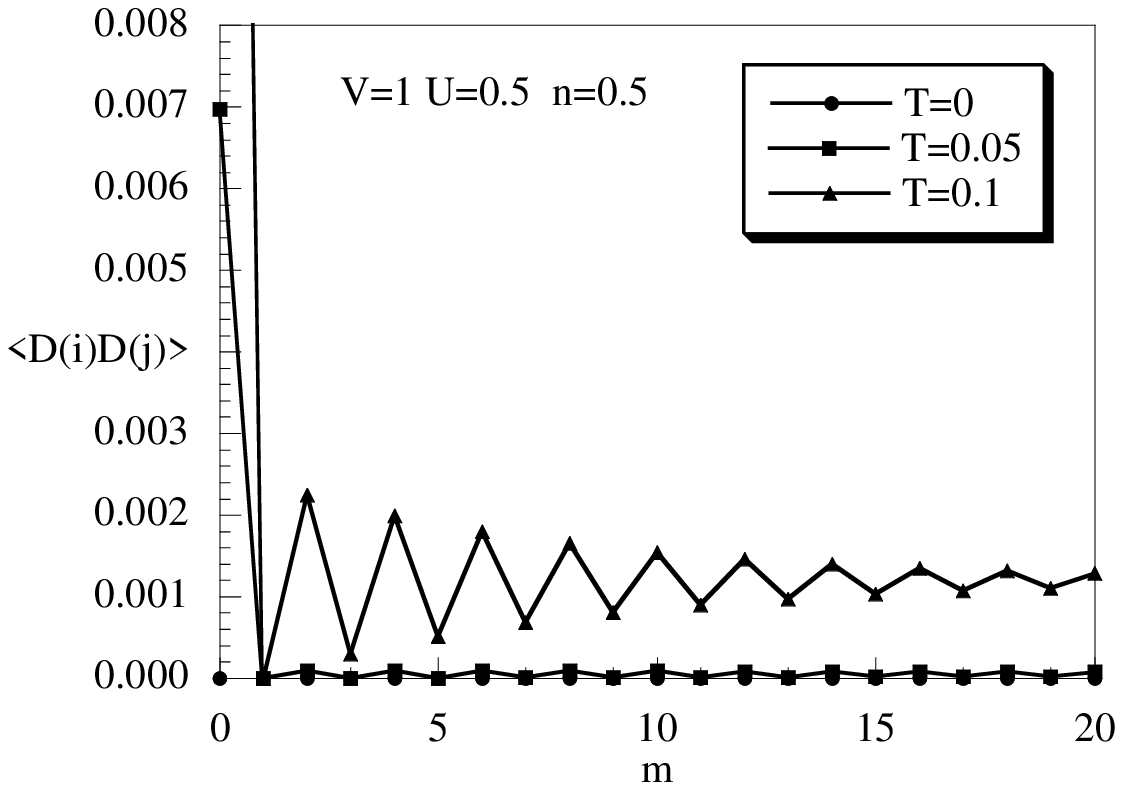}}
    \subfigure[]
   {\includegraphics[width=1.48in,height=1.10in]{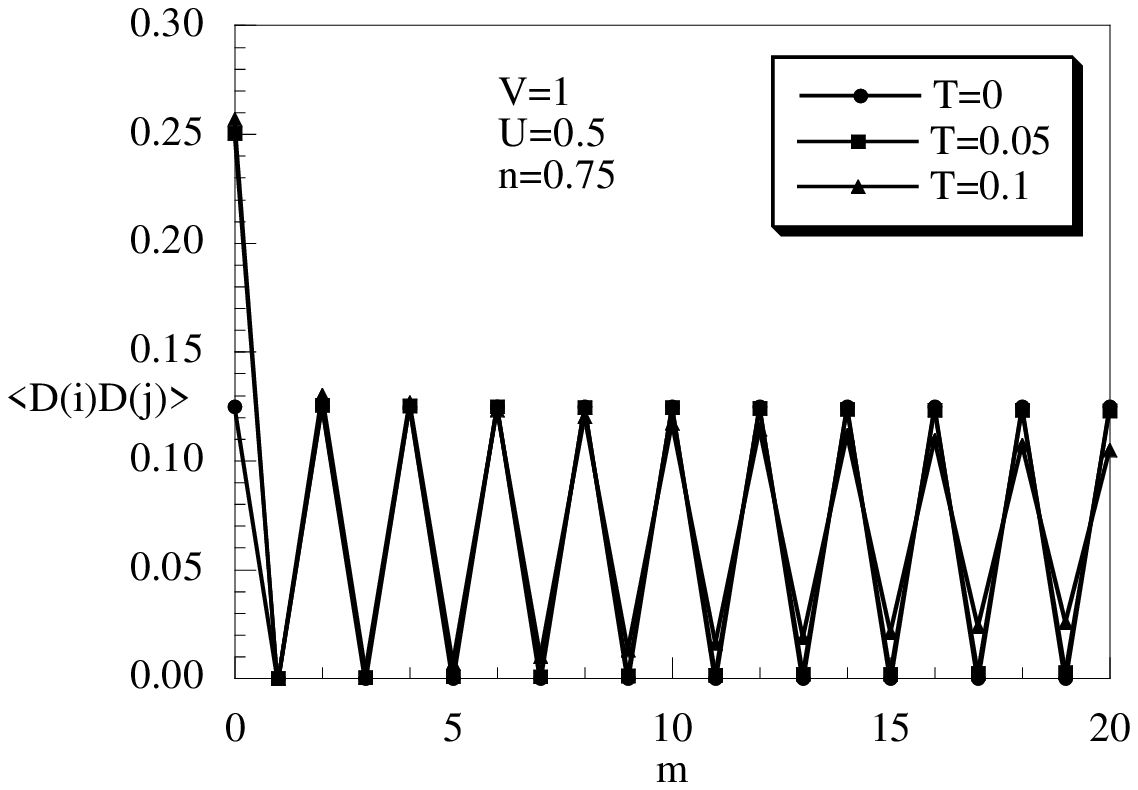}}
    \subfigure[]
   {\includegraphics[width=1.48in,height=1.10in]{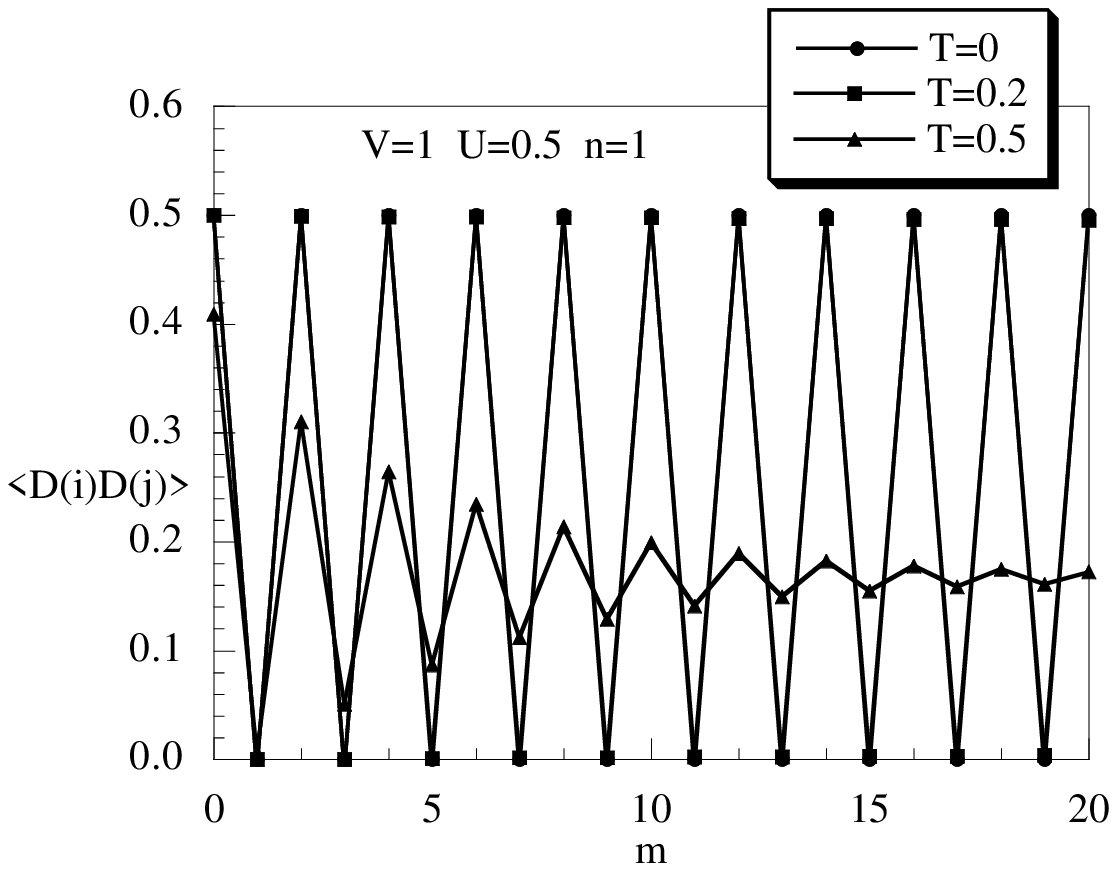}}
 \caption{\label{fig23} The charge (a), (b), (c) and the double occupancy
 (d), (e), (f) correlation functions as a function of the distance $m=\left|
{i-j} \right|$ for pertinent values of the temperature, $V=1$ and
$U=1/2$. The value of the filling grows from left to right.}
 \end{figure}
At low but finite temperatures, the charge and double occupancy
correlation functions $\langle n(i)n(j)\rangle$ and $\langle
D(i)D(j)\rangle$ still show a finite-range order evidenced by a
two-site periodicity. At quarter filling, one clearly observes in
Figs. \ref{fig23}a and \ref{fig23}d that one every two sites is
singly occupied: the two-site periodicity of the charge CF
decreases with increasing $T$ and distance $m$. The double
occupancy CF is still very close to zero hinting at the absence of
doubly occupied sites also at low temperatures. Upon increasing
the filling, for instance at $n=0.75$ (see Figs. \ref{fig23}b and
\ref{fig23}e), both $\langle n(i)n(j)\rangle$ and $\langle
D(i)D(j)\rangle$ basically retain the same zero-temperature
behavior also at low temperatures. At half filling the double
periodicity of the CFs persists even at higher temperatures, as
evidenced in Figs. \ref{fig23}c and \ref{fig23}f.

\subsection{Charge and spin susceptibilities}

Another important quantity useful for studying the critical
behavior of the system is the susceptibility. In this subsection
we shall compute the charge and spin susceptibilities.

\subsubsection{Charge susceptibility}

Let $\Lambda(k)$  be the Fourier transform of the charge
correlation function
\begin{equation*}
\langle n(j) n(l)    \rangle = \frac{a}{2\pi}\int_{-\pi /a}^{\pi
/a} dk \, e^{ik(j-l)} \Lambda(k) .
\end{equation*}
From Eq.
\eqref{EHM_28} it is easy to derive
\begin{equation*}
\begin{split}
\Lambda (k) &= \frac{2\pi }{a}n^2\delta (k) + \frac{A(1 - p^2)}{1
+ p^2 - 2p\cos (ka)}
\\
&+ \frac{B(1 - q^2)}{1 + q^2 - 2q\cos (ka)} .
\end{split}
\end{equation*}
Then, it immediately follows that the charge static susceptibility
$\chi _c$ is given by
\begin{equation}
\label{EHM_34} \chi _c = \Lambda (0) = Nn^2 + \frac{A(1 + p)}{(1 -
p)} + \frac{B(1 + q)}{(1 - q)},
\end{equation}
where $N$ is the number of sites. It is worthwhile to recall that
$\chi_c$ can be also computed by means of the thermodynamics
through the formula
\begin{equation}
\label{EHM_35}
 \chi_c = N n^2 + \frac{1}{\beta }\frac{\partial
n}{\partial \mu } .
\end{equation}
The difference between the above two expressions for $\chi _c$
lies in the fact that \eqref{EHM_34} requires the knowledge of a
two-point electronic Green's function, while \eqref{EHM_35}
requires only the knowledge of a one-point electronic Green's
function. In condensed matter theory there are many examples where
a physical quantity can be computed in different ways by using GFs
of different hierarchy (see, for example, Sec. 1.5 in Ref.
\cite{manciniavella}).
\begin{figure}[t]
 \centering
 \subfigure[]
    {\includegraphics[width=1.48in,height=1.10in]{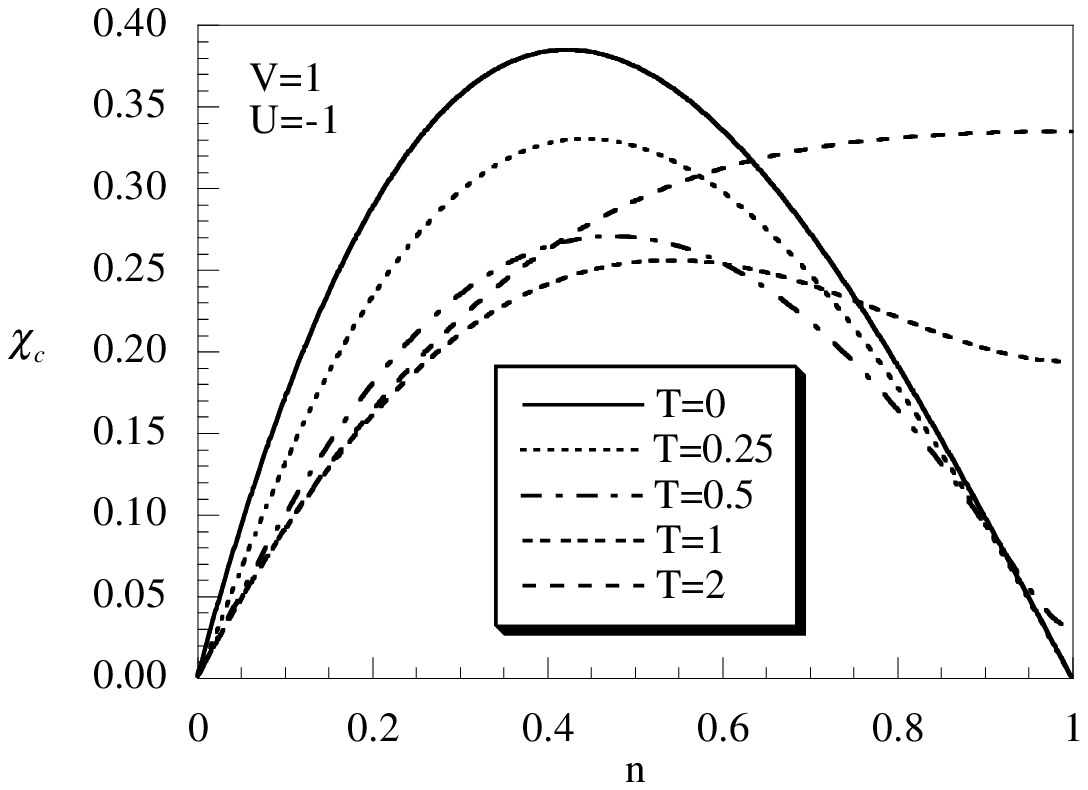}}
 \hspace{1mm}
 \subfigure[]
   {\includegraphics[width=1.48in,height=1.10in]{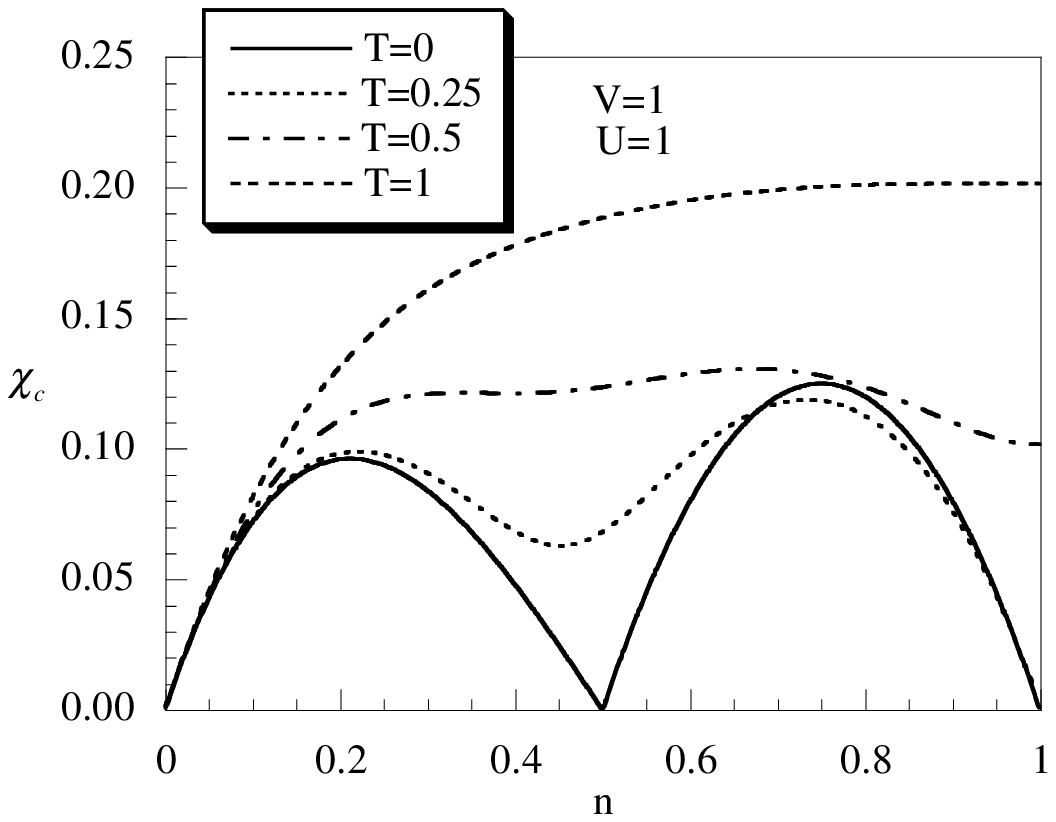}}
\hspace{1mm}
 \subfigure[]
   {\includegraphics[width=1.48in,height=1.10in]{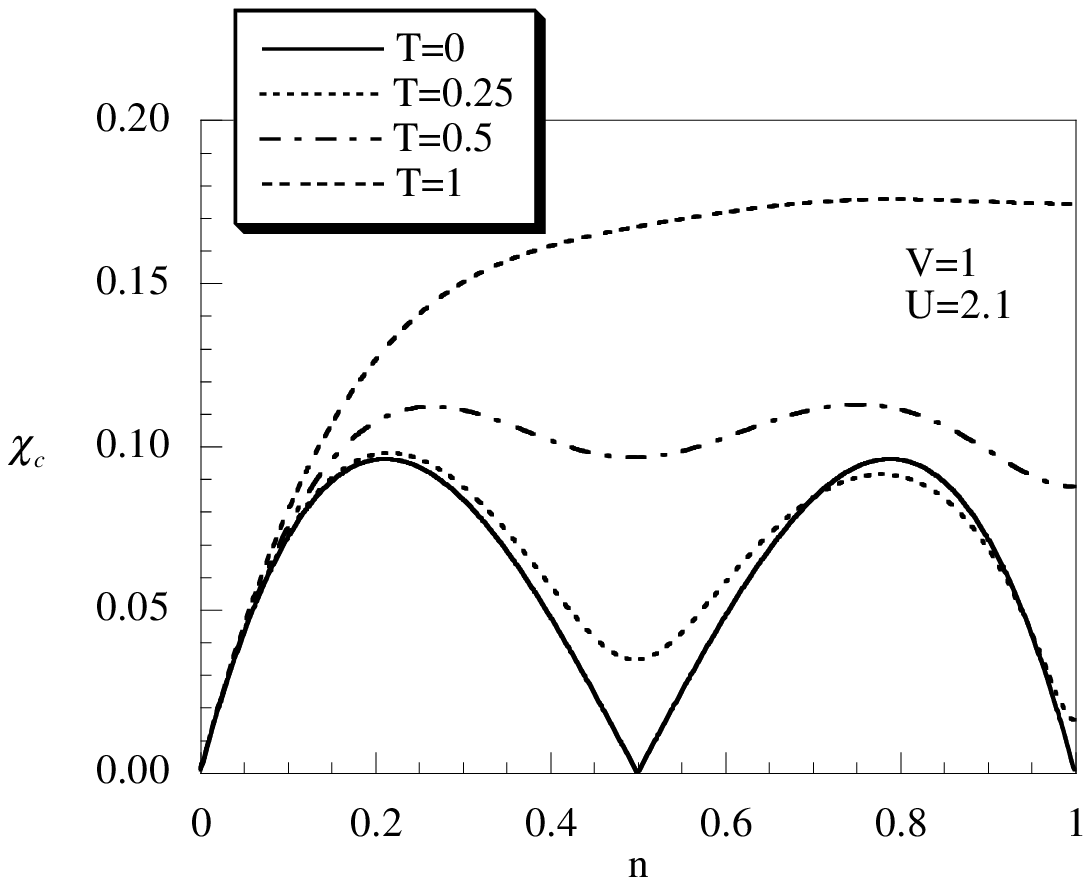}}
 \caption{\label{fig24} The charge susceptibility as a
function of the particle density for $V=1$ and different values of
the temperature, and  for (a) $U=-1$ ; (b) $U=1$; (c) $U=2.1$.}
 \end{figure}
Of course the result should not depend on which path of
calculation one has followed. This is true if one is capable to
exactly solve the model. However, in many practical situations
this is not possible and one needs to use some approximations. As
a consequence, one may obtain different results, because
approximation methods usually perform in different ways according
to the rank of the GFs under analysis. Actually, by comparing
different expressions of the same quantity one may envisage the
effectiveness of an approximation. For the present model, it is
not difficult to show that Eqs. \eqref{EHM_34} and \eqref{EHM_35}
give the same result since the model has been exactly solved. In
the following, we will use one or the other according the
convenience of numerical calculations. At $T=0$, the charge
susceptibility can be analytically computed since one knows all
the coefficients $A$, $B$, $p$, and $q$, as they are given in
Appendix \ref{sec_III}. In particular, one can compute $\chi_c$ in
each region of the phase diagram:
\begin{equation*}
\begin{split}
\chi_c^{(a)} &= n(1 - n)(2 - n) ,\\
\chi_c^{(b)} &= n(1 - n)(1 - 2n), \\
\chi_c^{(c)} &= (1 - n)(2n - 1) ,\\
\chi_c^{(d)} &= n(1 - n)(2n - 1).
\end{split}
\end{equation*}

The superscript in the above expressions labels the different
phases at $T=0$. In Figs. \ref{fig24}a-c we plot $\chi_c$ as a
function of the filling $n$ for three fixed values of $U$ ($U=-V$,
$U=V$ and $U=2.1V$, respectively) and for different temperatures.
One immediately sees that, at $T=0$, the charge susceptibility is
finite except at $n=0.5$ and $n=1$, where, at fixed on-site
potential, PTs occur. In the first region [$U<0$, phase ($a$)],
the charge susceptibility increases by increasing $n$ and has a
maximum in correspondence of quarter filling. Further increasing
$n$, $\chi _c$ decreases and vanishes at half filling, where a
charge ordered state is established at $T=0$ (see Fig.
\ref{fig3}). In the second region [$0<U<2V$, phases ($b$) and
($c$)], for low temperatures, $\chi_c$ has a double peak structure
with two maxima around $n=0.25$ and $n=0.75$ and two minima around
$n=0.5$ and $n=1$; the latter points correspond to the situation
where a charge ordered state (checkerboard distribution of singly
and doubly occupied sites, respectively) is observed at zero
temperature (see Fig. \ref{fig8}). In the third region [$2V<U$,
phases ($b$) and ($d$)], $\chi_c$ shows a behavior similar to the
one exhibited in the second region, the difference being that the
two peaks have the same height. This difference is due to the fact
that in the phase ($d$) one has only singly occupied sites for all
values of the filling (see Fig. \ref{fig12}). For high
temperatures, $\chi_c$ is always an increasing function of $n$.
\begin{figure}[t]
 \centering
 \subfigure[]
   {\includegraphics[width=1.48in,height=1.10in]{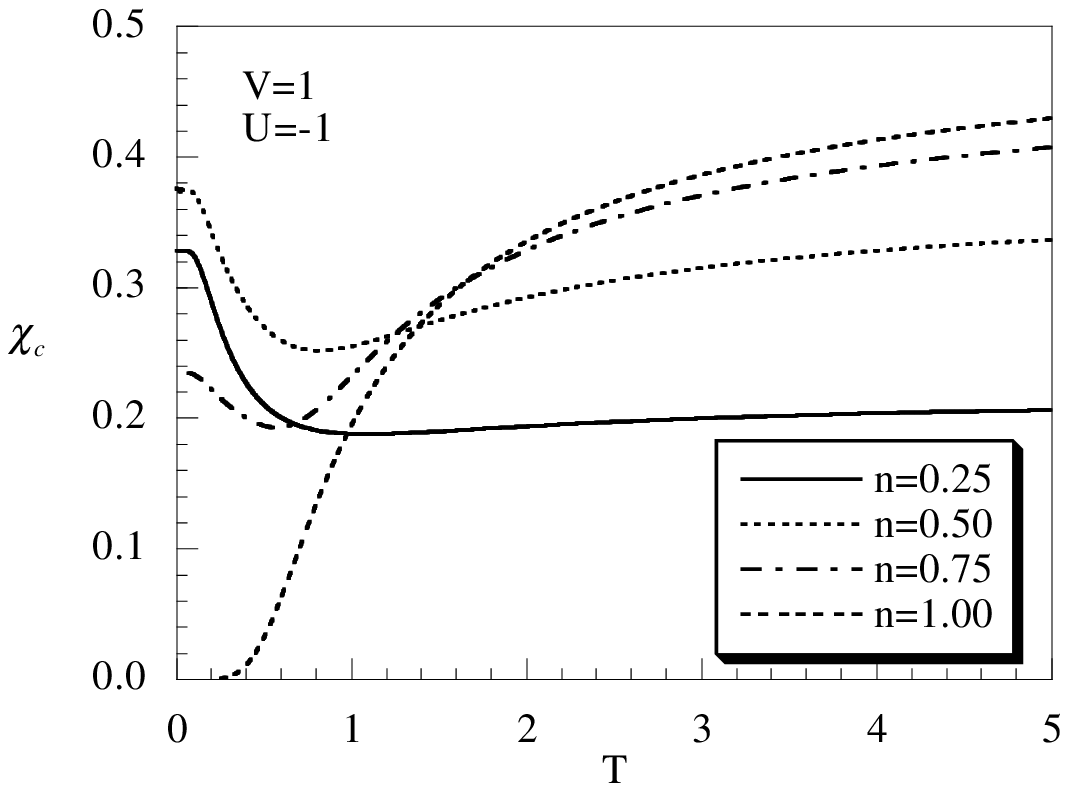}}
 \hspace{1mm}
 \subfigure[]
   {\includegraphics[width=1.48in,height=1.10in]{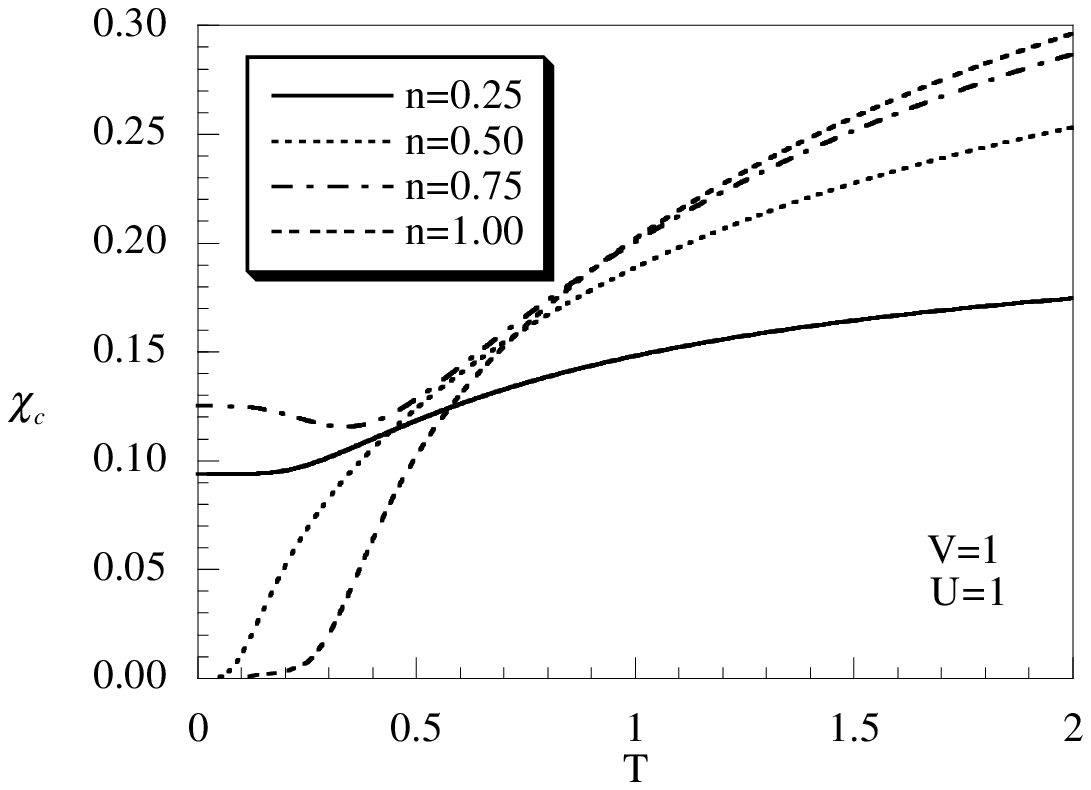}}
\hspace{1mm}
 \subfigure[]
   {\includegraphics[width=1.48in,height=1.10in]{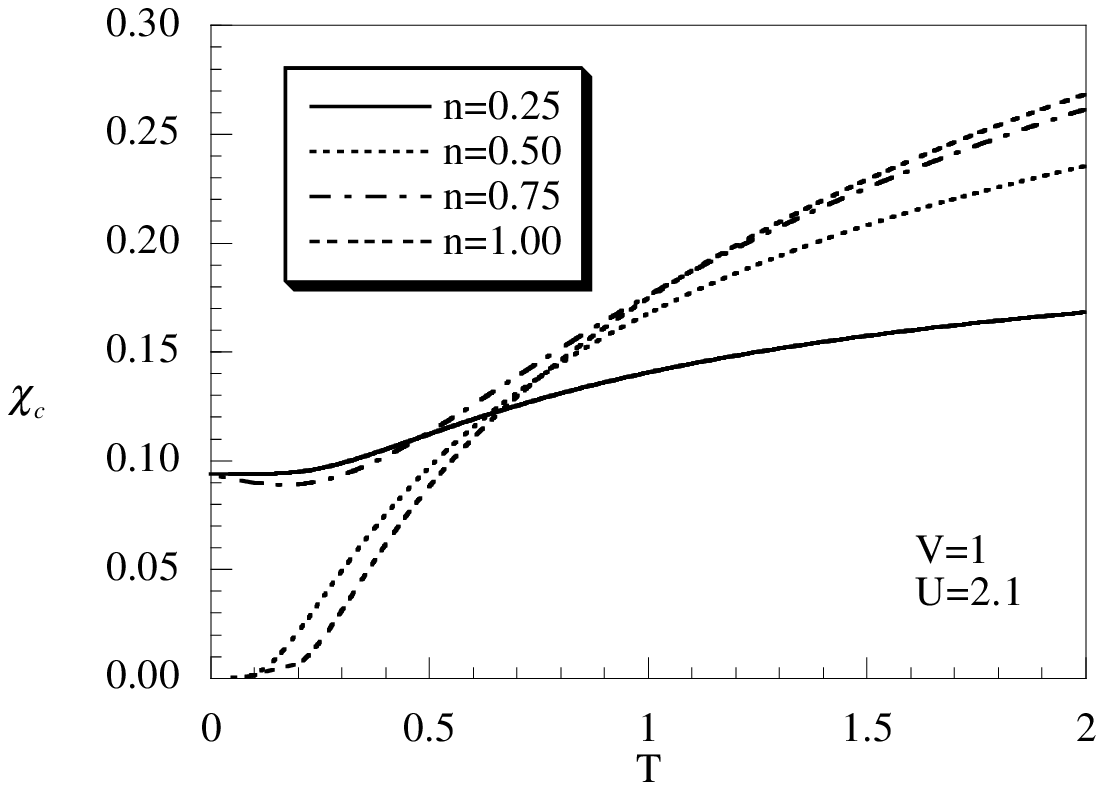}}
 \caption{\label{fig25} The charge susceptibility as a
function of the temperature for $V=1$, $n=0.25$, 0.5, 0.75, 1 and
 for (a) $U=-1$ ; (b) $U=1$; (c) $U=2.1$.}
 \end{figure}
In Figs. \ref{fig25}a-c we plot $\chi _c$ as a function of the
temperature, again in the three regions of $U$ ($U=-V$, $U=V$ and
$U=2.1V$, respectively) and for different values of the filling
($n=0.25$, 0.5, 0.75, 1). For all values of $U$ and for large
temperatures, the charge susceptibility increases and tends to a
constant value in the limit of $T$ infinite. At low temperatures,
one observes a different behavior: at $T=0$, $\chi_c$ is finite
for values of the particle density corresponding to nonordered
states. Further increasing $T$, $\chi_c$ decreases, it reaches a
minimum and then increases again. On the other hand, $\chi _c$
decreases with $T$ and vanishes at $T=0$ for fillings
corresponding to charge ordered states. In the limit of high
temperatures, the charge susceptibility tends to a constant value
which does not depend on $U$ but only on $n$ according to the law
\begin{equation}
\lim_{T \to \infty } \chi_c =\alpha(n),
 \label{EHM_36}
\end{equation}
where $\alpha(n)=n(2 - n)/2$.

\subsubsection{Spin susceptibility}

The spin magnetic susceptibility $\chi_s$ can be computed by
introducing an external magnetic field $h$, taking the derivative
of the magnetization $m = \langle n_\uparrow (i) - n_\downarrow
(i)\rangle$ with respect to $h$ and letting $h$ going to zero:
\begin{equation*}
\chi_s= \left( {\frac{\partial m}{\partial h}}
\right)_{h = 0}.
\end{equation*}
The addition of a homogeneous magnetic field does not dramatically
modify the framework of calculation given in Sec. \ref{sec_II},
once one has taken into account the breakdown of the spin
rotational invariance. Details of the calculations will be
presented elsewhere \cite{mancini08}: here we shall present only
some results for the spin susceptibility $\chi_s$ relevant to our
discussion.
\begin{figure}[t]
 \centering
 \subfigure[]
   {\includegraphics[width=1.48in,height=1.10in]{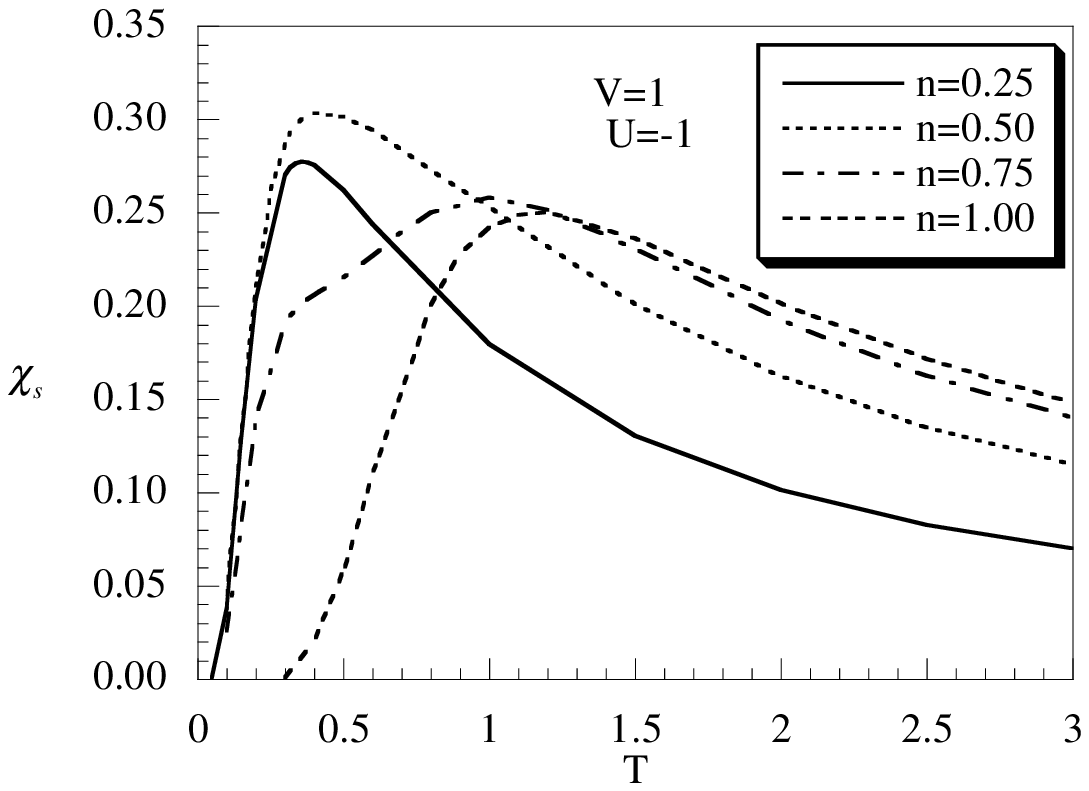}}
 \hspace{1mm}
 \subfigure[]
   {\includegraphics[width=1.48in,height=1.10in]{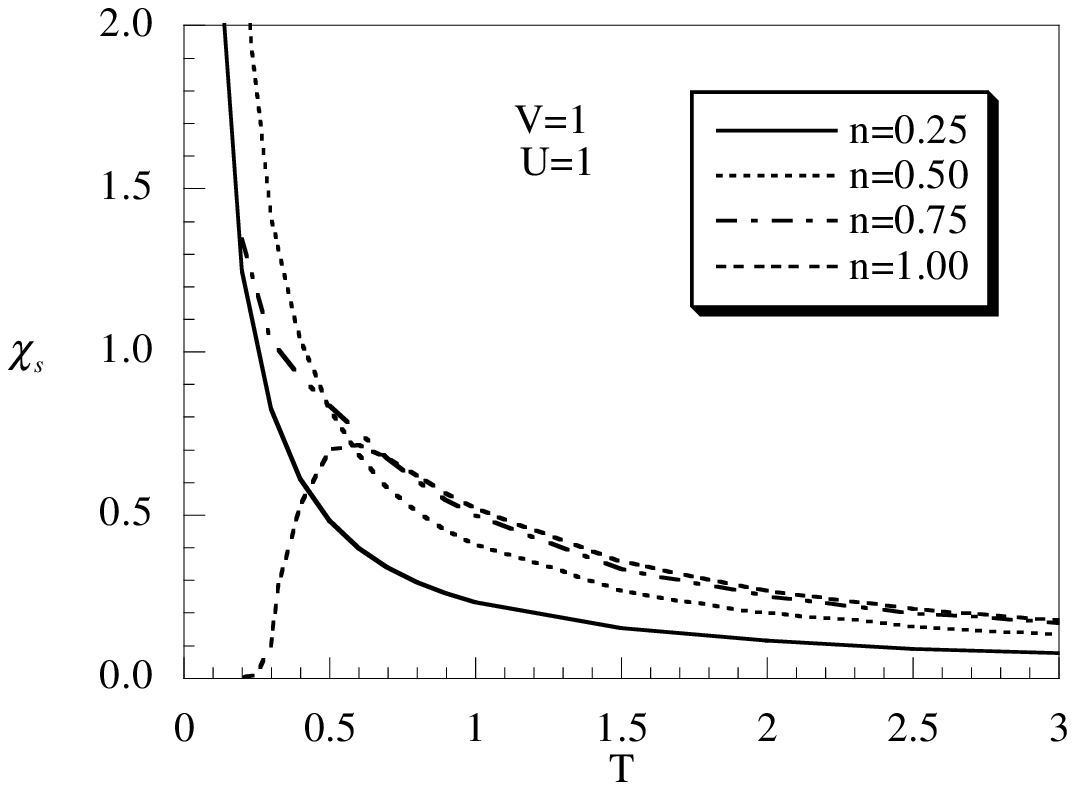}}
\hspace{1mm}
 \subfigure[]
   {\includegraphics[width=1.48in,height=1.10in]{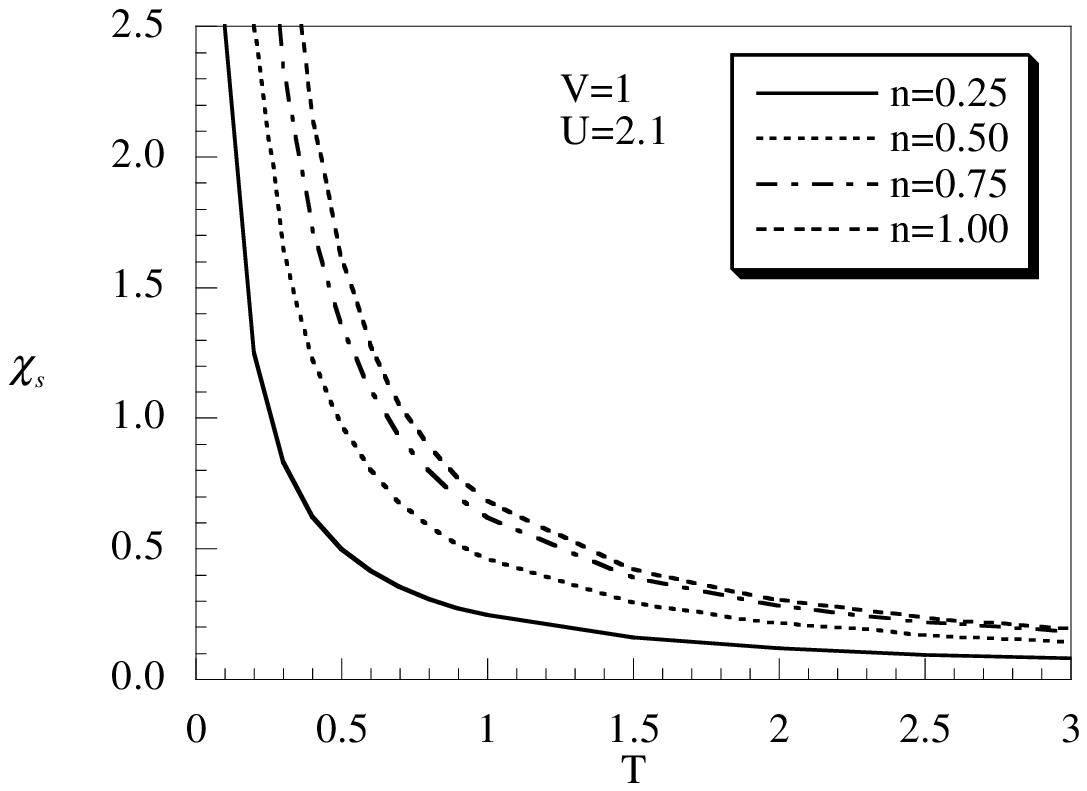}}
 \caption{\label{fig26} The spin susceptibility as a
function of the temperature for $V=1$, $n=0.25$, 0.5, 0.75, 1 and
 for (a) $U=-1$ ; (b) $U=1$; (c) $U=2.1$.}
 \end{figure}
In Figs. \ref{fig26}a-c we plot the spin susceptibility as a
function of the temperature for three representative values of $U$
($U=-V$, $U=V$ and $U=2.1V$, respectively) and for different
values of the filling ($n=0.25$, 0.5, 0.75, 1). In the first
region [$U<0$, phase ($a$)], the spin susceptibility vanishes at
zero temperature for all values of the filling: at $T=0$ all
electrons are paired and no alignment of the spin is possible. By
increasing $T$, the thermal excitations break some of the doublons
and a small magnetic field may induce a finite magnetization:
$\chi_s$ augments by increasing $T$ up to a maximum, then
decreases. Calculations show that, in the limit of vanishing
temperature, the behavior of $\chi_s $ in the phase ($a$) is
\begin{equation*}
\lim_{T\to 0} \chi_s =\frac{n}{T}e^{-\beta (\mu -U)}.
\end{equation*}
The exponential behavior exhibited by the spin susceptibility is
easily understood as a finite amount of energy is required to
break the doublons. In the second region [$0<U<2V$, phases ($b$)
and ($c$)], $\chi_s$ diverges at $T=0$ for all values of the
filling, but $n=1$. At zero temperature, for $n<1$ all the
electrons singly occupy a macroscopic number of sites and their
spins will align when the magnetic field is turned on. In the
limit of vanishing temperature, $\chi_s$ diverges, according to
the value of the filling, as
\begin{equation*}
\lim_{T\to 0} \chi _s =\left\{
\begin{array}{cc}
 \frac{n}{T}\quad  & \mbox{for $0\le n\le 0.5$} \quad \quad \mbox{phase
 ($b$)} \\
 \frac{1-n}{T} \quad & \mbox{for $0.5\le n<1$} \quad \quad \mbox{phase ($c$)}.
\end{array}
\right.
\end{equation*}
By increasing $T$, the alignment is disturbed by the thermal
excitations and, as a consequence, $\chi_s$ decreases. On the
other hand, for $n=1$ at $T=0$, all electrons are paired and
$\chi_s= 0$. At finite temperatures, $\chi_s$ increases linearly
up to a maximum, then decreases. In the third region [$2V<U$,
phases ($b$) and ($d$)] at $T=0$ and for all values of the
filling, all the sites are singly occupied: the spin
susceptibility decreases with $T$ and diverges at $T=0$. It can be
shown that in this region the spin susceptibility diverges as $T
\to 0$ according to the law $ \lim_{T \to 0} \chi_s = n/T. $
\begin{figure}[t]
 \centering
 \subfigure[]
   {\includegraphics[width=1.48in,height=1.10in]{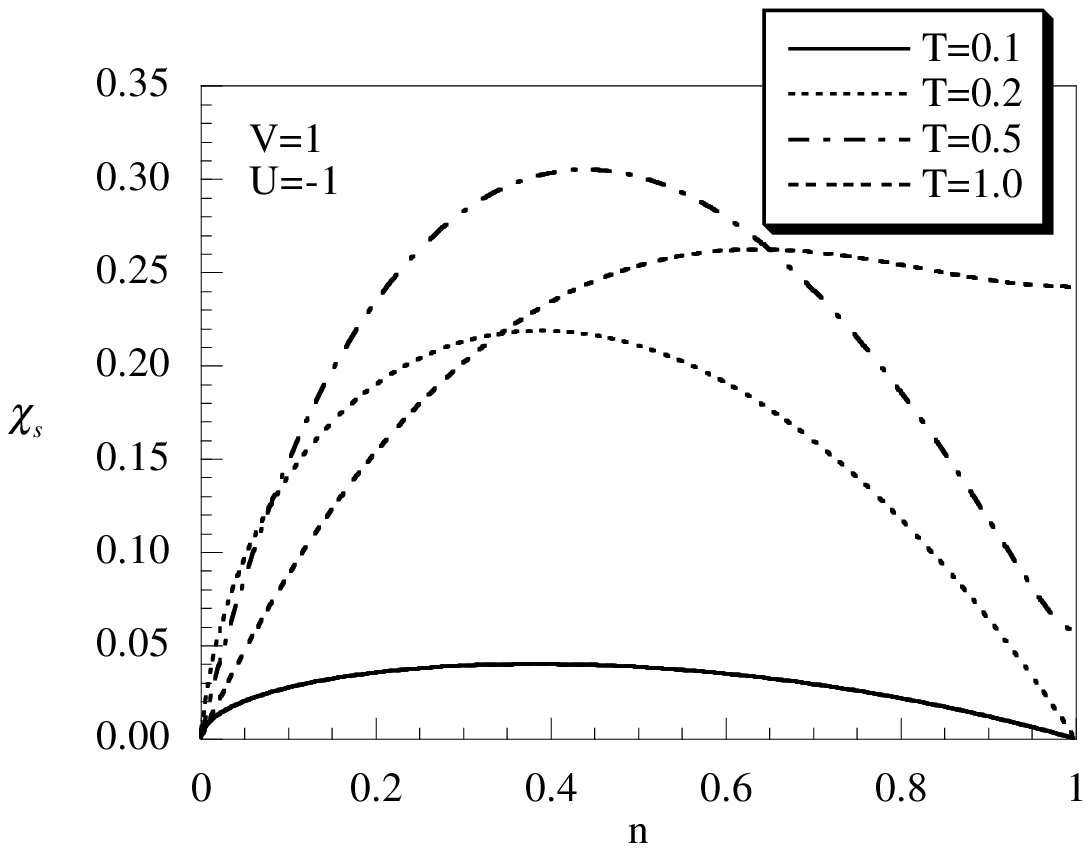}}
 \hspace{1mm}
 \subfigure[]
   {\includegraphics[width=1.48in,height=1.10in]{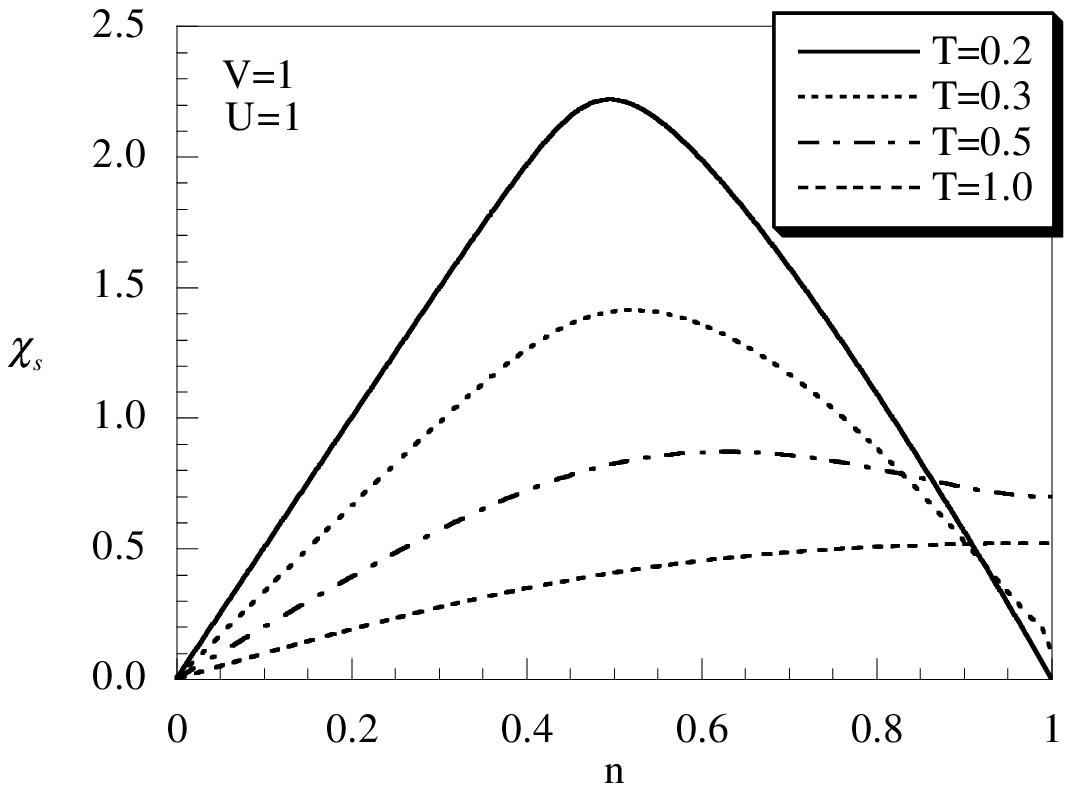}}
\hspace{1mm}
 \subfigure[]
   {\includegraphics[width=1.48in,height=1.10in]{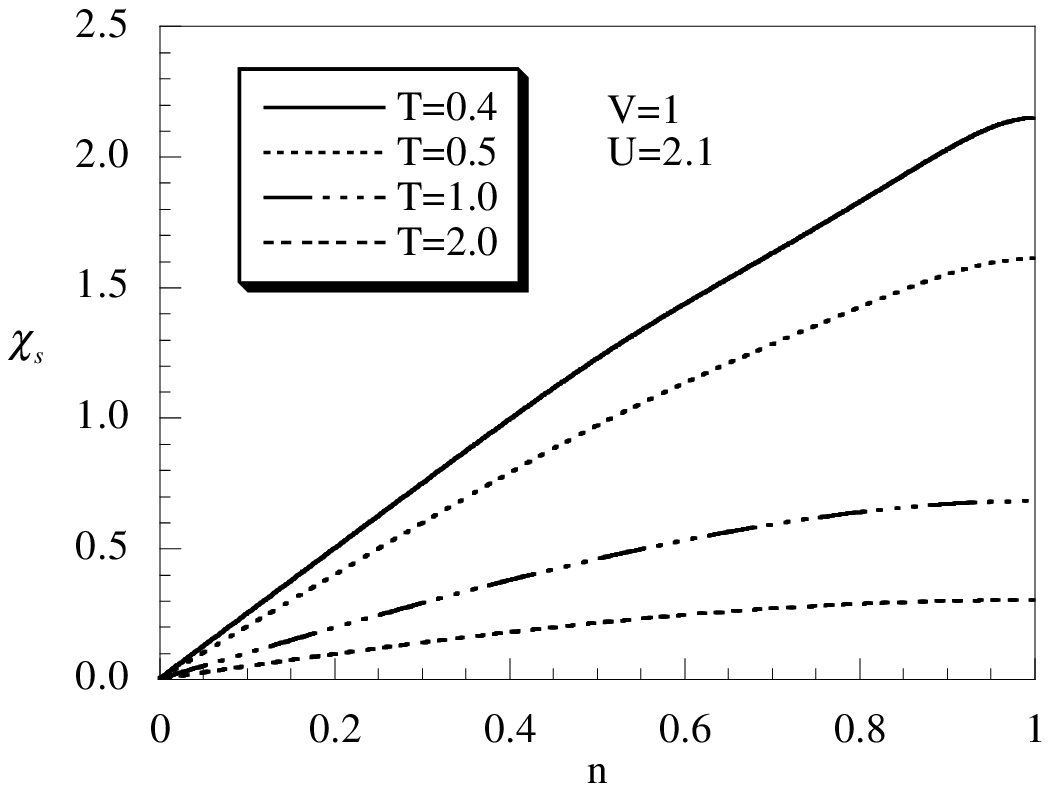}}
 \caption{\label{fig27} The spin susceptibility as a
function of the particle density for $V=1$ and different values of
the temperature, and  for (a) $U=-1$ ; (b) $U=1$; (c) $U=2.1$.}
 \end{figure}
It is easy to check that, for high temperatures, the spin
susceptibility decreases with the Curie law $\chi _s \approx T^{ -
1}$ in the entire ($U,n$) plane:
\begin{equation*}
\lim_{T \to \infty } \chi_s = \frac{a(n)}{T},
\end{equation*}
where $\alpha(n)$ is the same $U$-independent function appearing
in Eq. \eqref{EHM_36}. As a consequence, in the limit of high
temperatures, the ratio $\chi_c/ \chi_s$ is a universal function
of $T$:
\begin{equation*}
\lim_{T \to \infty } \left( {\frac{\chi _c }{\chi _s }} \right) =
T .
\end{equation*}
In Figs. \ref{fig27}a-c we plot the spin susceptibility as a
function of the particle density, again for the three
representative values of $U$ ($U=-V$, $U=V$ and $U=2.1V$,
respectively) and for different temperatures. In the first two
regions $\chi_s$ has a similar behavior: it increases with $n$ up
to a maximum around $n=0.5$, then decreases, exhibiting a minimum
at $n=1$, where at $T=0$ all electrons are paired. In the third
region $\chi_s$ is an increasing function of $n$ and presents a
maximum at $n=1$ where, at $T=0$, the maximum number of singly
occupied sites is present.

\subsection{Density of states}

\begin{figure}[t]
 \centering
 \subfigure[]
   {\includegraphics[width=1.48in,height=1.10in]{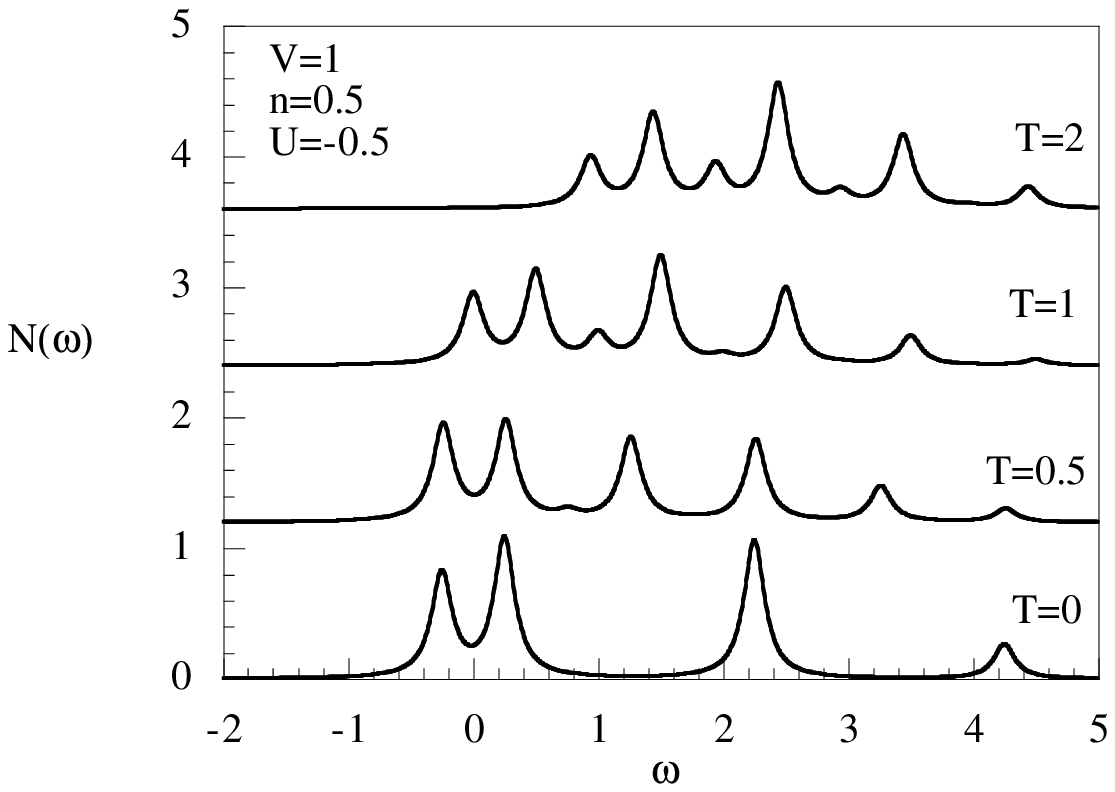}}
 \hspace{1mm}
 \subfigure[]
   {\includegraphics[width=1.48in,height=1.10in]{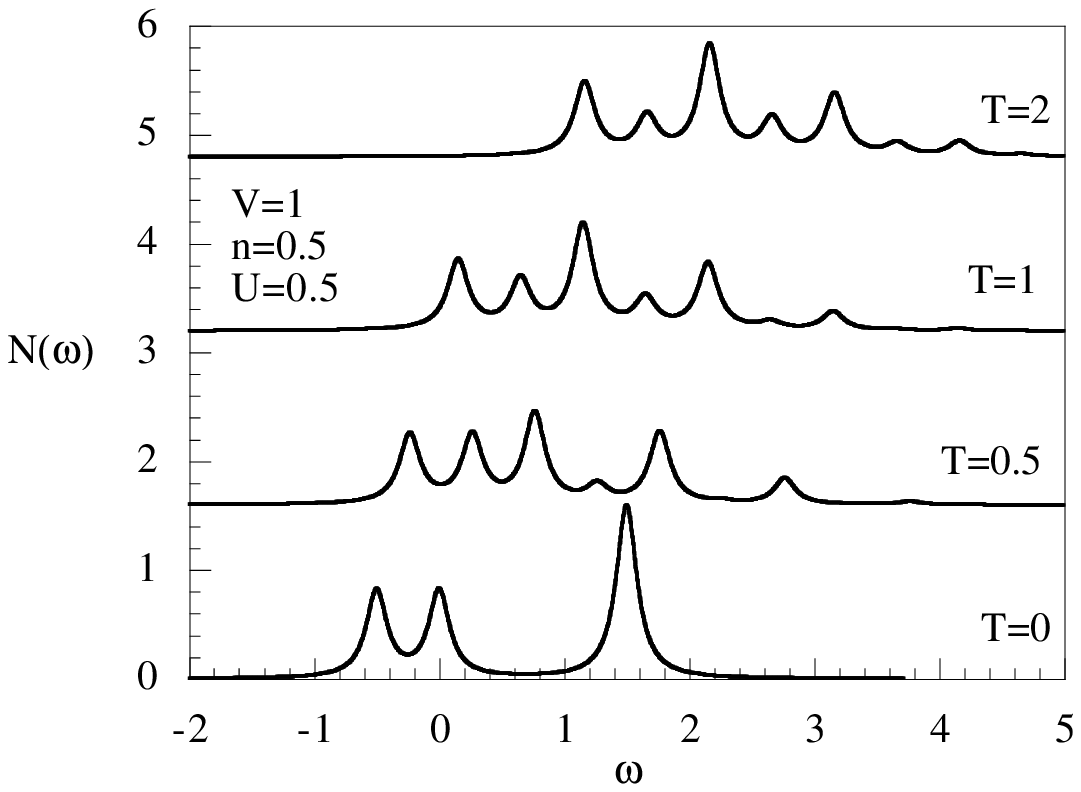}}
 \caption{\label{fig28} The density of states $N(\omega)$ as a
function of the frequency $\omega$ at quarter filling, $V=1$,
$T=0$, 0.5, 1 and 2 for (a) $U=-1/2$ ; (b) $U=1/2$.}
 \end{figure}

The behavior of the density of states is, obviously, richer at
finite temperatures since the thermal excitations favor the
emergence of new peaks. As one can clearly see from Fig.
\ref{fig28}, by increasing $T$ the number of peaks increases
leading to a larger number of states which the electrons can
occupy. In fact, for $T \neq 0$, the weights $\Sigma$ appearing in
Eq. \eqref{EHM_29} are all finite. Here we report, as an example,
only the DOS at quarter filling in the neighborhood of $U=0$. One
obtains similar diagrams also for other values of the filling.

\subsection{Specific heat}
\label{spec_heat}

The specific heat is given by
\begin{equation}
C=\frac{dE}{dT}
\label{EHM_37},
\end{equation}
where the internal energy $E$ can be computed as the thermal
average of the Hamiltonian \eqref{EHM_1} and it is given by
\begin{equation}
E=UD+2V\lambda ^{(1)}.
\label{EHM_38}
\end{equation}
The specific heat exhibits a very rich structure in correspondence
of the zero-temperature phase diagram shown in Fig. \ref{fig1}.
The possible excitations of the ground state are creation and
annihilation of particles in either a singly or a doubly occupied
site, induced by the Hubbard operators $\psi^{(\xi)}$ and
$\psi^{(\eta)}$, respectively. The corresponding transition
energies are given in Eq. \eqref{EHM_7} and the peaks exhibited by
the specific heat correspond to these transitions. One may
distinguish between the transitions induced by $\psi^{(\xi)}$ or
by $\psi^{(\eta)}$ by looking at the position of the peaks, i.e.,
if the position changes or remains constant by varying $U$.
 \begin{figure}[t]
 \centering
 \subfigure[]
   {\includegraphics[width=1.48in,height=1.10in]{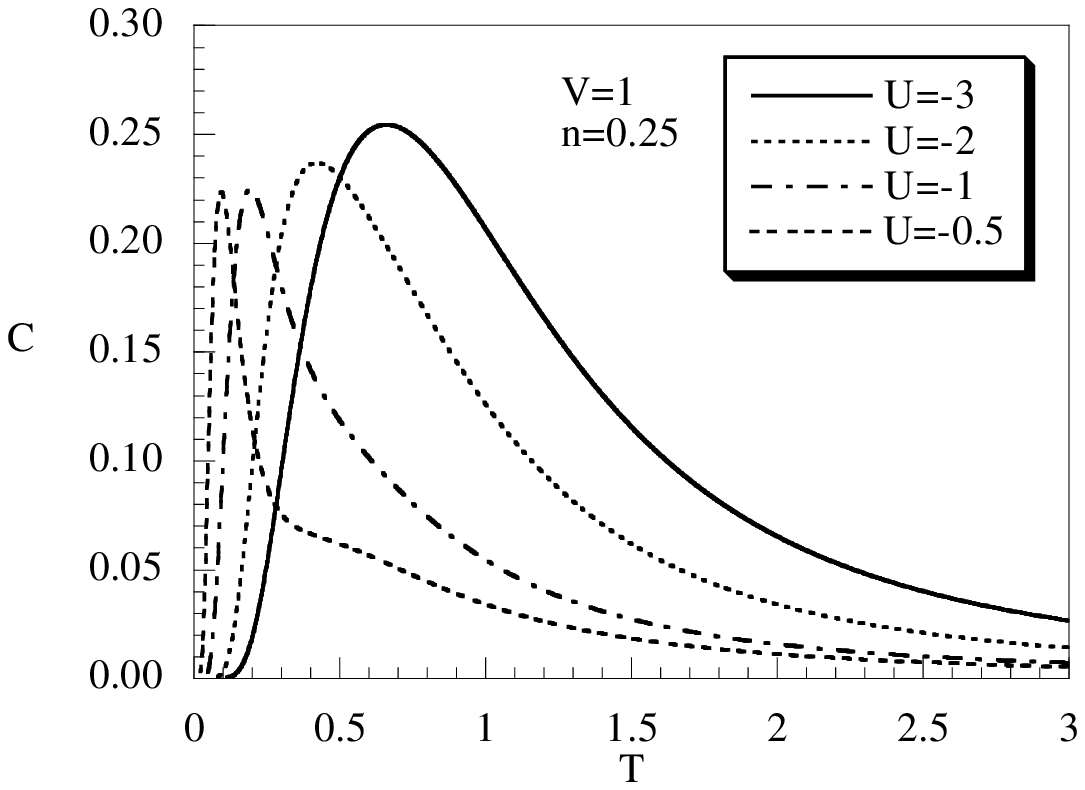}}
 \hspace{1mm}
 \subfigure[]
   {\includegraphics[width=1.48in,height=1.10in]{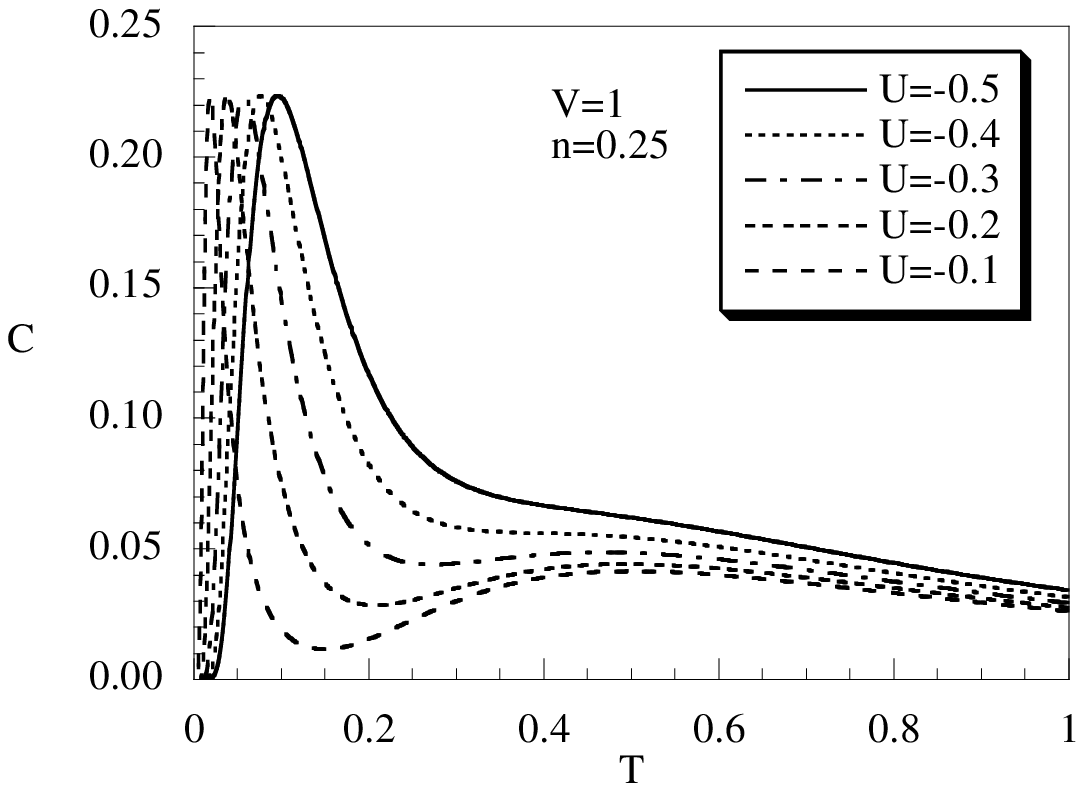}}\\
   \vspace{-4mm}
   \subfigure[]
   {\includegraphics[width=1.48in,height=1.10in]{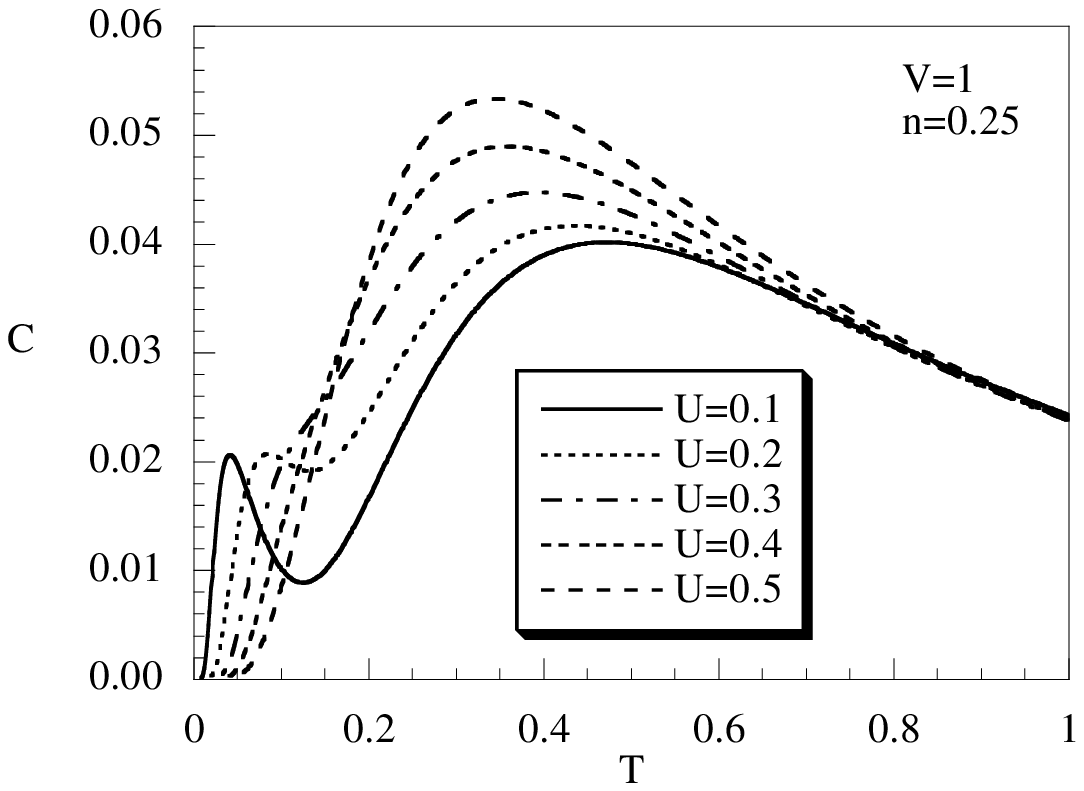}}
 \hspace{1mm}
 \subfigure[]
   {\includegraphics[width=1.48in,height=1.10in]{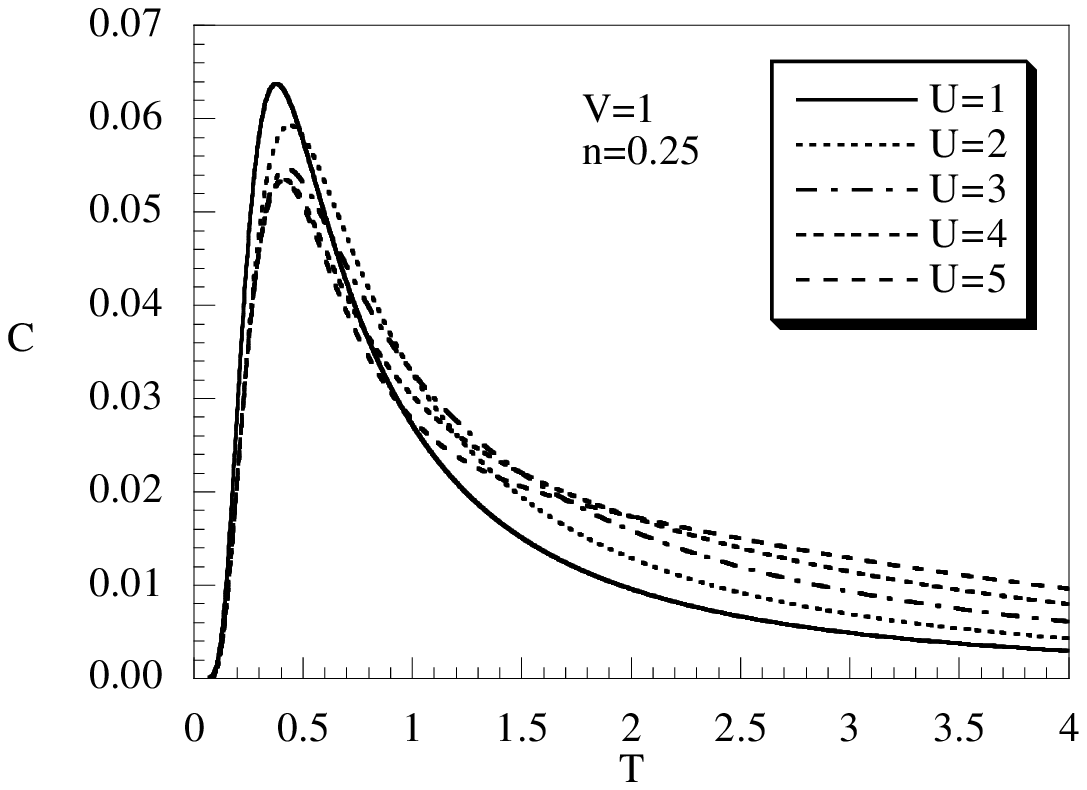}}
     \vspace{-4mm}
 \caption{\label{fig29} The specific heat $C$ as a function of the
temperature $T$ for $V=1$, $n=0.25$ and (a): $U=-3$,..., $-0.5$,
(b): $U=-0.5$,..., $-0.1$; (c): $U=0.1$,..., $0.5$; (d): $U=1$,
..., $5$.}
\end{figure}
The behavior of the specific heat at $n=0.25$ as a function of the
temperature is shown in Figs. \ref{fig29}a-d. The specific heat
presents one pronounced maximum for large negative values of $U$
at a temperature $T_1$ (see Fig. \ref{fig29}a). The position of
the maximum decreases linearly with $U$ until $U=0$. Around
$U\approx -0.3V$ a second peak appears at a higher temperature
(see Fig. \ref{fig29}b). In this region, both peaks represent
transitions mainly induced by $\psi^{(\eta)}$. The height of the
first peak decreases and suddenly goes to zero at $U=0$, where it
vanishes. At $U=0$ there is only one peak located at $T_2 \approx
0,46 V$. For $U>0$ there is a small region where the double peak
structure is again present (see Fig. \ref{fig29}c). When $U>0.2V$
the first peak disappears while the position of the second peak
slightly increases with $U$ up to $U\approx 2V$ where it assumes a
constant value (see Fig. \ref{fig29}d). The behavior of $T_1$ and
$T_2$ as a function of $U$ is shown in Fig. \ref{fig30}. The
behavior of the specific heat for this value of the filling is
easily understood if one recalls that at $T=0$, by varying $U$, a
PT is observed at $U=0$ from a state with doubly occupied sites to
a state with singly occupied ones (see Figs. \ref{fig3} and
\ref{fig8}). Accordingly, at finite temperatures, for $U<0$ the
excitations are mainly induced by the Hubbard operator
$\psi^{(\eta)}$ ($T_1$ varies linearly), while for $U>0$ the
excitations are mainly induced by $\psi^{(\xi)}$ ($T_2$ is
constant for $U>2V$).
\begin{figure}[t]
\centerline{\includegraphics[scale=0.45]{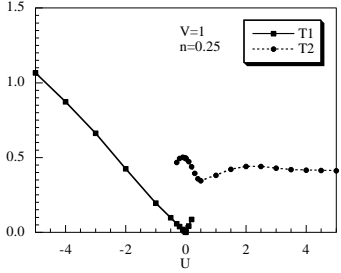}}
\caption{\label{fig30} The temperatures $T_1$ and $T_2$ at which
the specific heat exhibits a peak are plotted as functions of $U$
for $n=0.25$.}
\end{figure}

\begin{figure}[t]
 \centering
 \subfigure[]
   {\includegraphics[width=1.48in,height=1.10in]{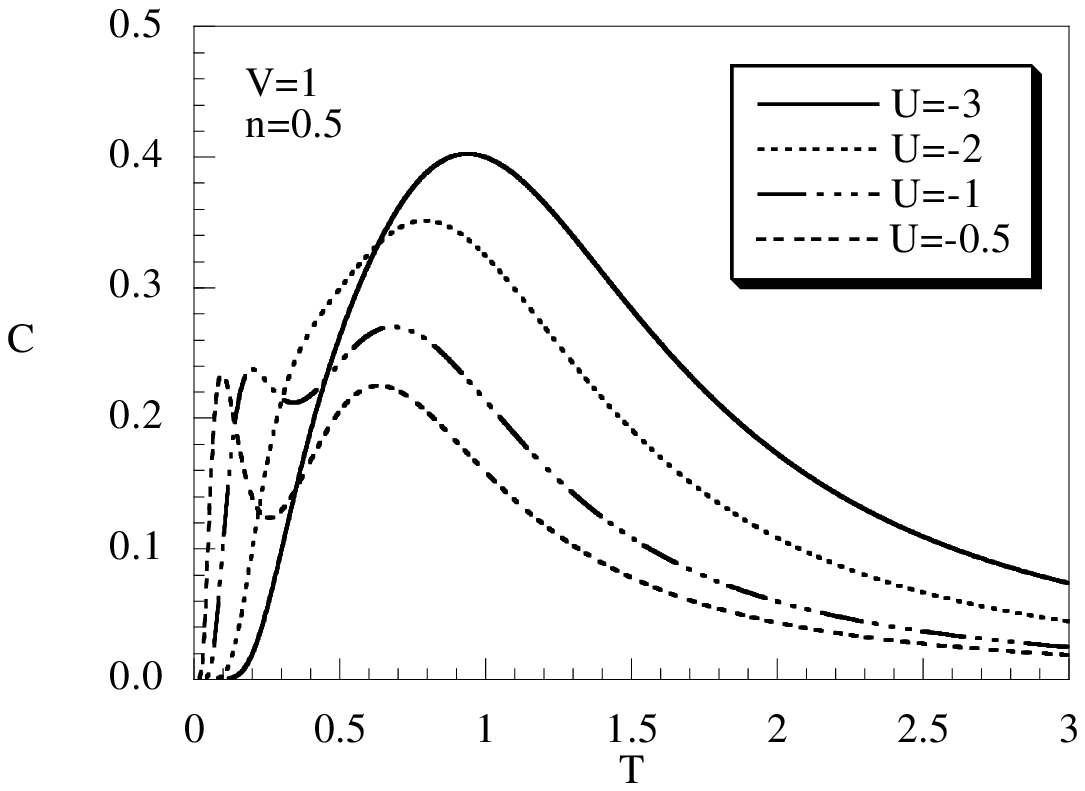}}
 \subfigure[]
   {\includegraphics[width=1.48in,height=1.10in]{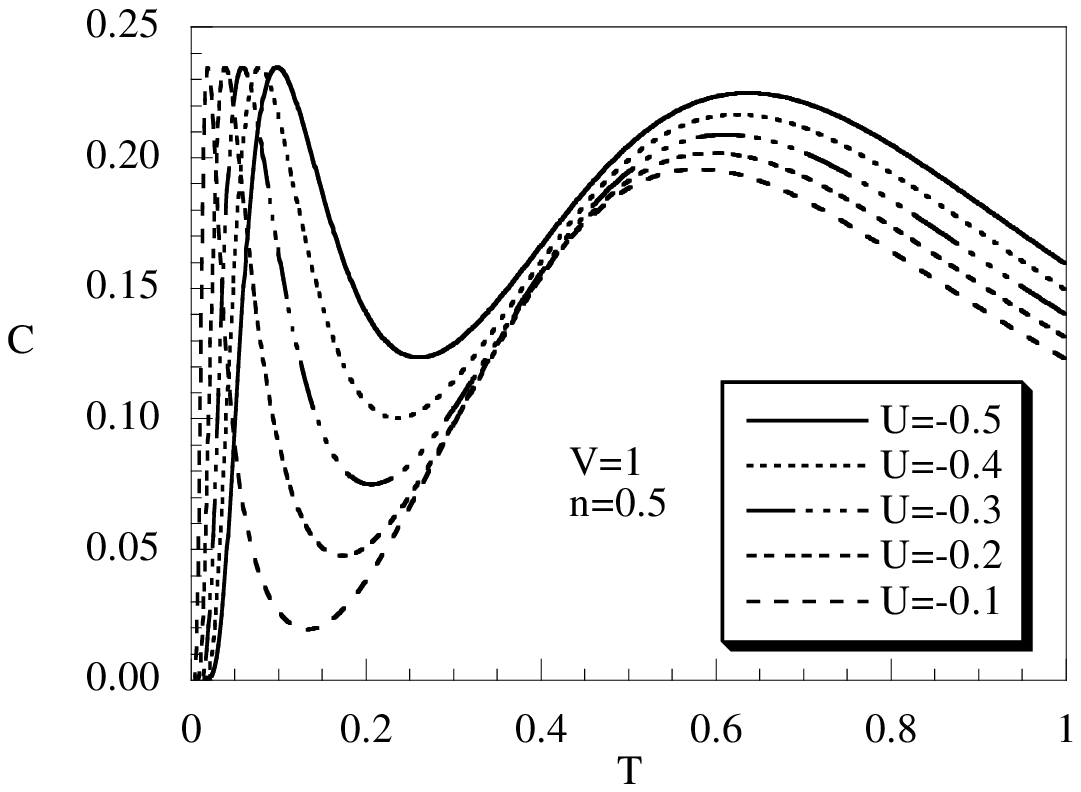}}\\
   \subfigure[]
   {\includegraphics[width=1.48in,height=1.10in]{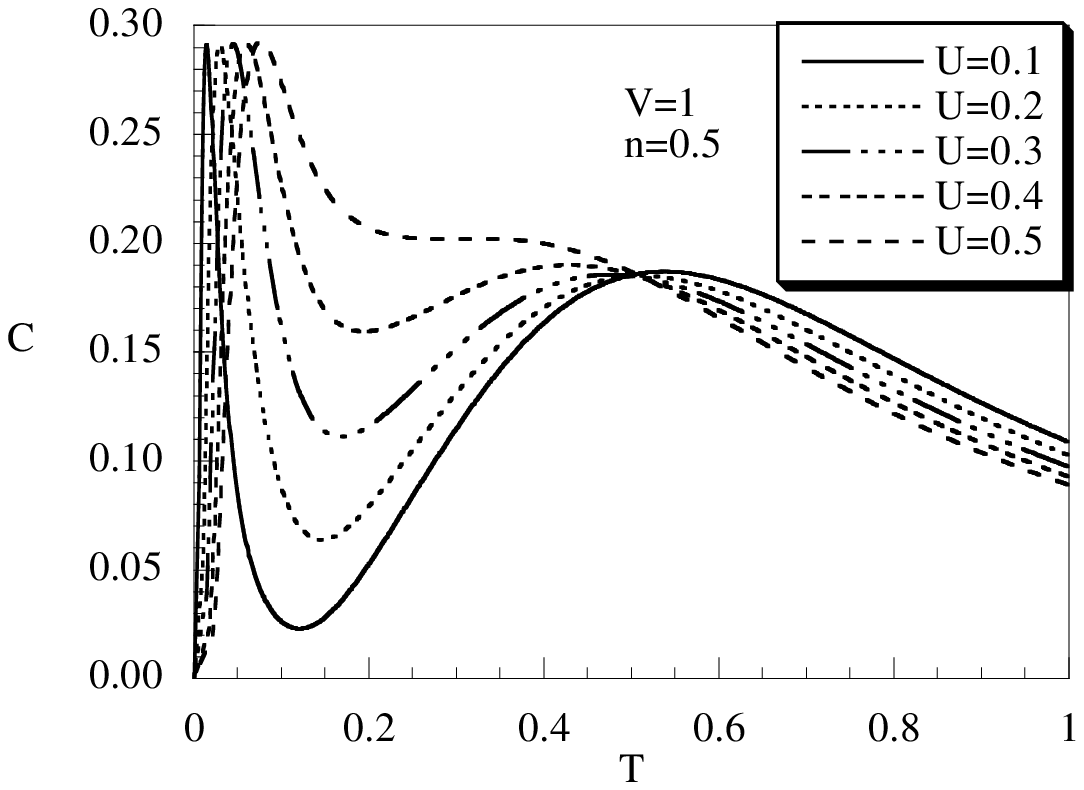}}
 \subfigure[]
   {\includegraphics[width=1.48in,height=1.10in]{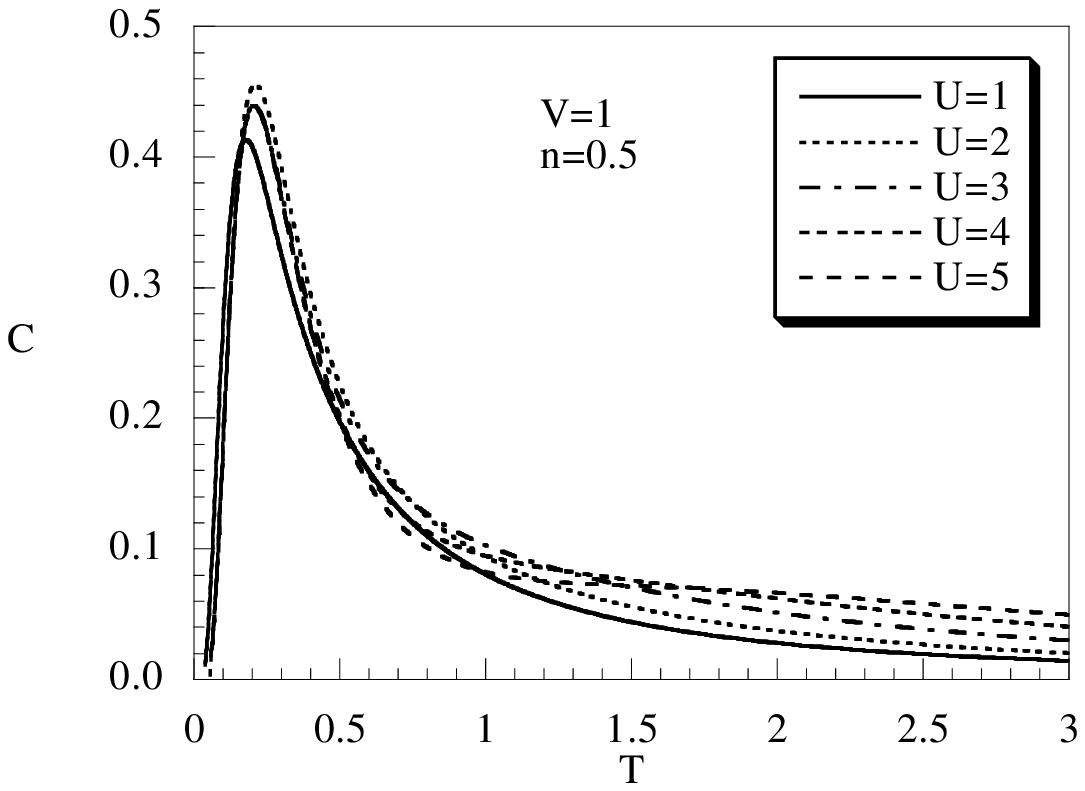}}
     \vspace{-4mm}
 \caption{\label{fig31} The specific heat $C$ as a function of the
temperature $T$ for $V=1$, $n=0.5$ and (a): $U=-3$, ..., $-0.5$;
(b): $U=-0.5$,...,$-0.1$; (c): $U=0.1$,...,$0.5$; (d):
$U=1$,..,$5$.}
 \end{figure}

\begin{figure}[b]
\centerline{\includegraphics[scale=0.45]{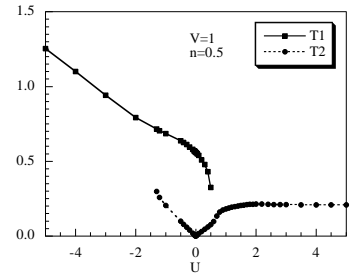}}
\caption{\label{fig32} The temperatures $T_1$ and $T_2$ at which
the specific heat exhibits a peak are plotted as functions of $U$
for $n=0.5$.}
\end{figure}

In Figs. \ref{fig31}a-d the temperature dependence of the specific
heat is shown for $n=0.5$. The first peak $T_1$, appearing for
large negative values of $U$ (see Fig. \ref{fig31}a), decreases
following a linear law up to $U\approx -2V$. At $U\approx -1.2V$,
a second peak appears at a temperature $T_2$ lower than $T_1$. By
decreasing $\vert $U$ \vert$, $T_2$ decreases linearly (see Fig.
\ref{fig31}b). At $U=0$ there is only one peak situated at $T_1
\approx 0.56V$. For small positive values of $U$, the specific
heat has again two peaks (see Fig. \ref{fig31}c). By increasing
$U$, $T_2$ increases following a linear law up to $U\approx 0.5V$;
further increasing $U$, $T_2$ deviates from the linear behavior
and tends to the constant value $T_2 =0.21$ for $U>2V$ (see Fig.
\ref{fig31}d). The position $T_1$ of the first peak decreases by
increasing $U$ and, for $U>0.5V$, the peak is no longer
observable. The behavior of $T_1$ and $T_2$ as a function of $U$
is shown in Fig. \ref{fig32}. It is worth noticing that, for
positive $U$, the double peak structure is associated with the
presence of a crossing point in the specific heat curves.
A double peak structure, as well as the appearance of
quasiuniversal crossing points, was already noticed in Refs.
\cite{moreo_97,juttner_98} and in Sec. 3.4.2 of Ref.
\cite{manciniavella} in the Hubbard model, in Ref.
\cite{dolcini_02} in an extended Hubbard-like model, and also in
some experimental data \cite{exp_data}. When $U>V/2$ the double
peak structure and the crossing point disappear. As for the case
$n=0.25$, the elementary excitations which mainly contribute to
$C$ for $n=0.5$ are those induced by $\psi^{(\eta)}$ for $U<0$ and
by $\psi^{(\xi)}$ for $U>0$.

\begin{figure}[t]
 \centering
 \subfigure[]
   {\includegraphics[width=1.48in,height=1.10in]{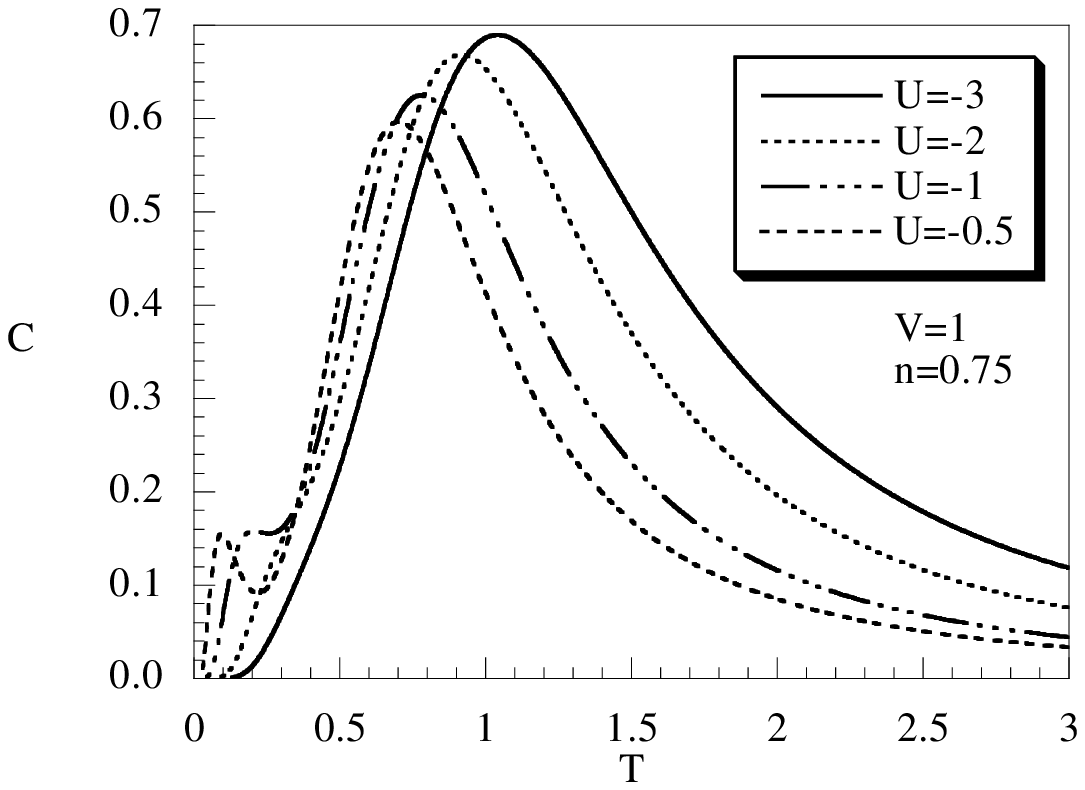}}
 \hspace{1mm}
 \subfigure[]
   {\includegraphics[width=1.48in,height=1.10in]{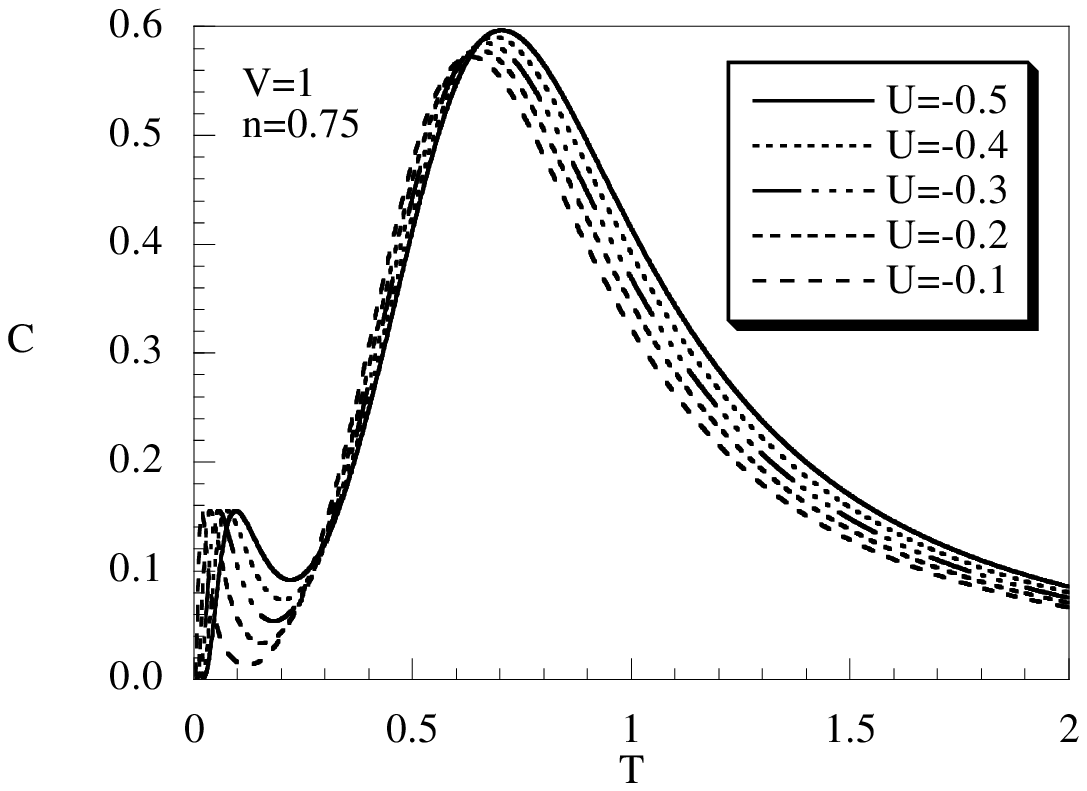}}\\
  \vspace{-4mm}
   \subfigure[]
   {\includegraphics[width=1.48in,height=1.10in]{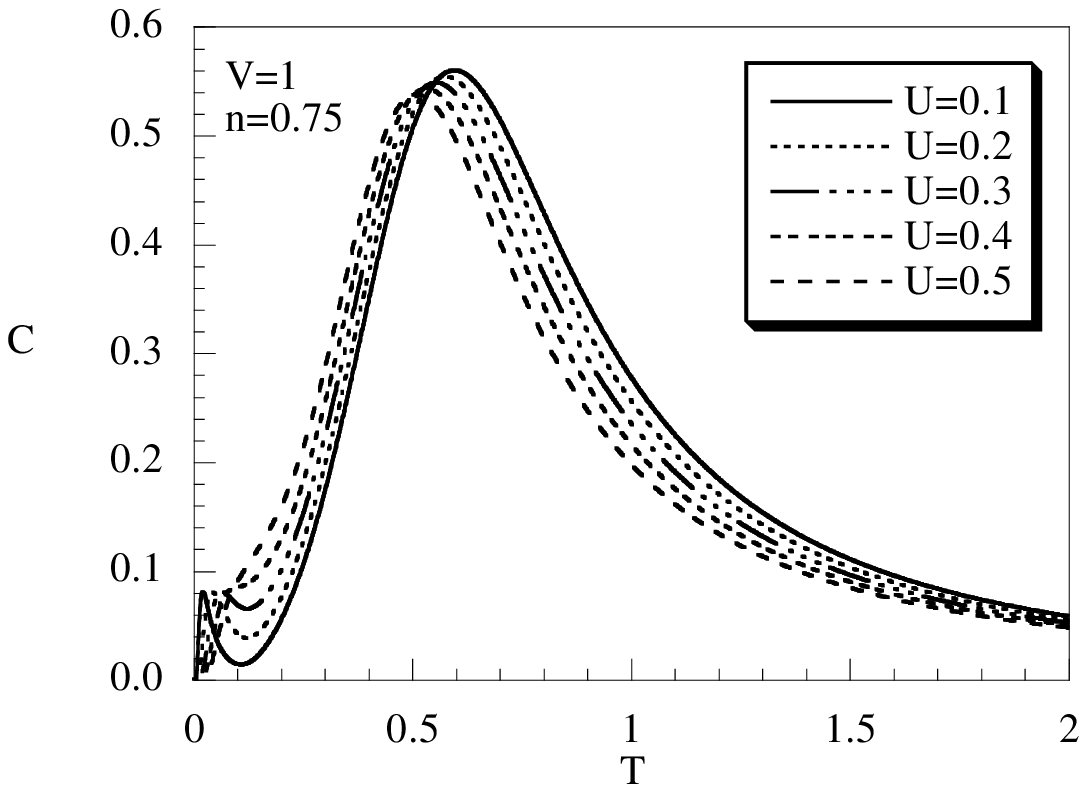}}
 \subfigure[]
   {\includegraphics[width=1.48in,height=1.10in]{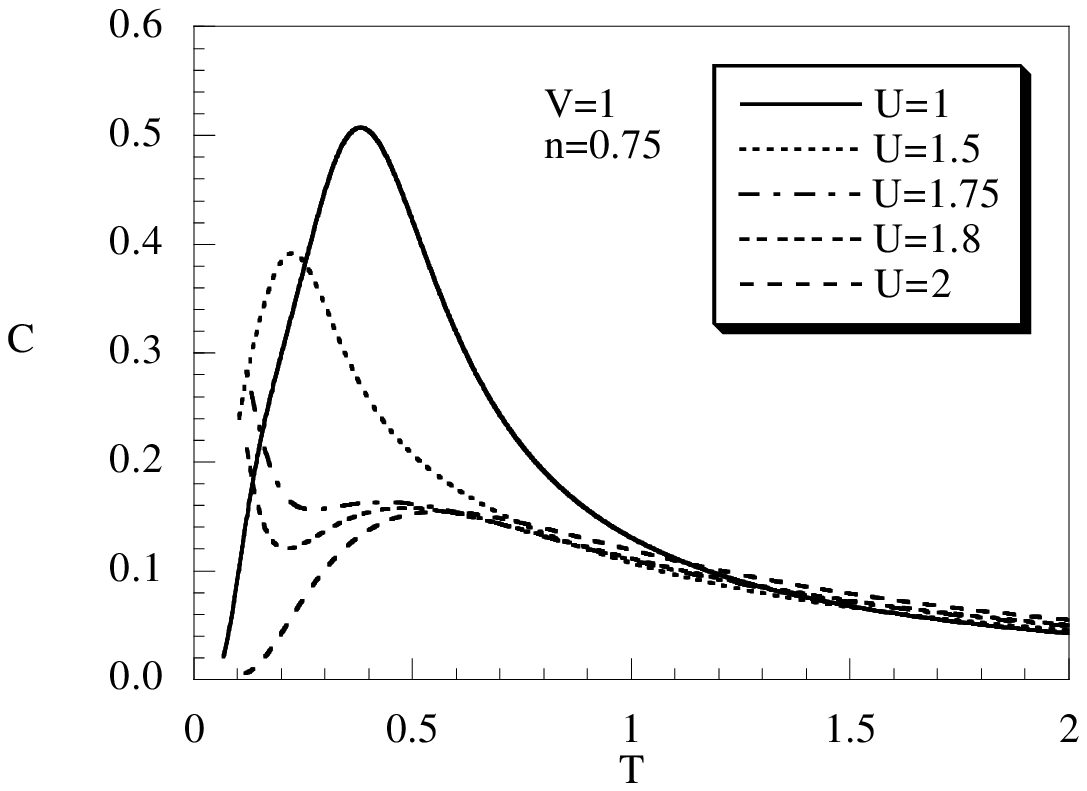}}
 \subfigure[]
   {\includegraphics[width=1.48in,height=1.10in]{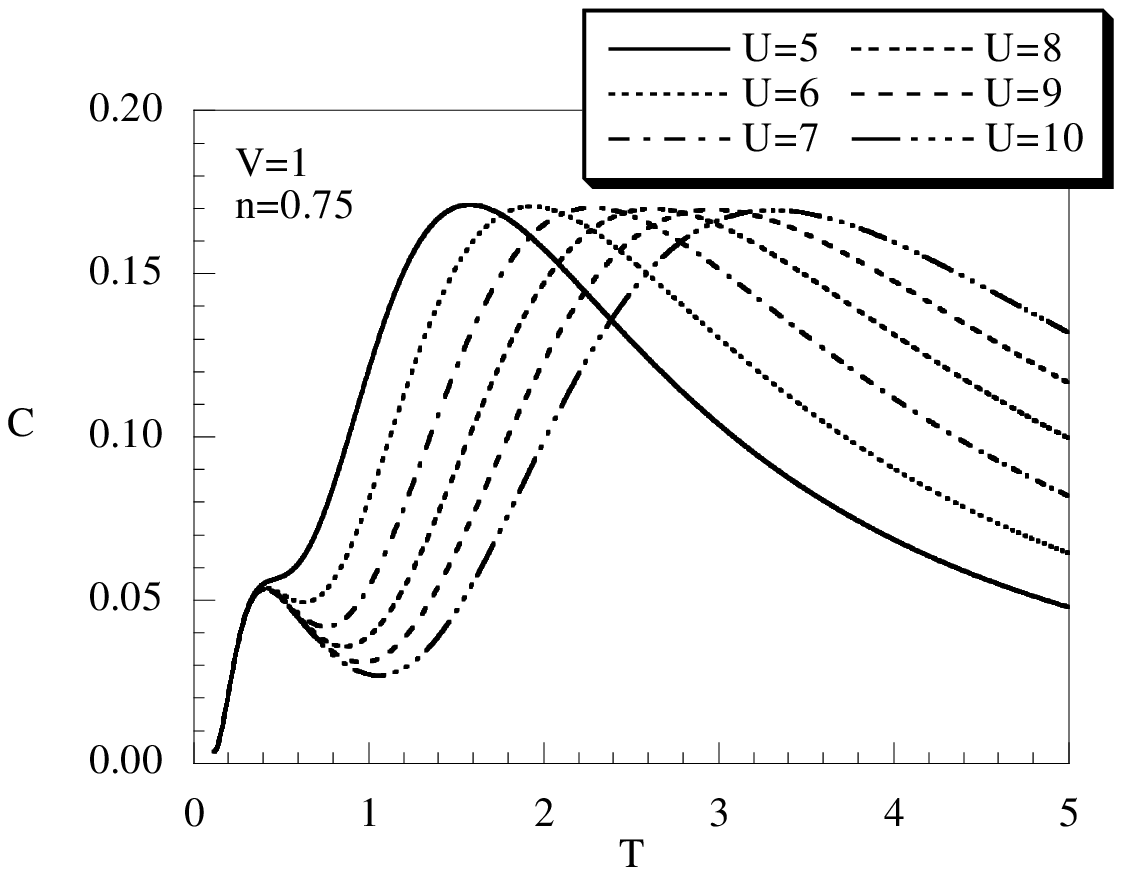}}
     \vspace{-4mm}
 \caption{\label{fig33} The specific heat $C$ as a function of the
temperature $T$ for $V=1$, $n=0.75$ and (a): $U=-3$,...,-0.5; (b):
$U=-0.5$,...,$-0.1$; (c): $U=0.1$,...,$0.5$; (d): $U=1$, $1.5$,
$1.75$, $1.8$ and $2$; (e): $U=5$,...,10.}
 \end{figure}

\begin{figure}[t]
\centerline{\includegraphics[scale=0.45]{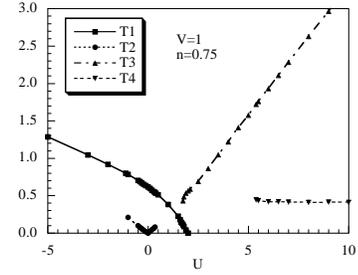}}
\caption{\label{fig34}  The temperatures $T_1$, $T_2$, $T_3$ and
$T_4$ at which the specific heat exhibits a peak are plotted as
functions of $U$ for $n=0.75$.}
\end{figure}

\begin{figure}[ht]
 \centering
 \subfigure[]
   {\includegraphics[width=1.48in,height=1.10in]{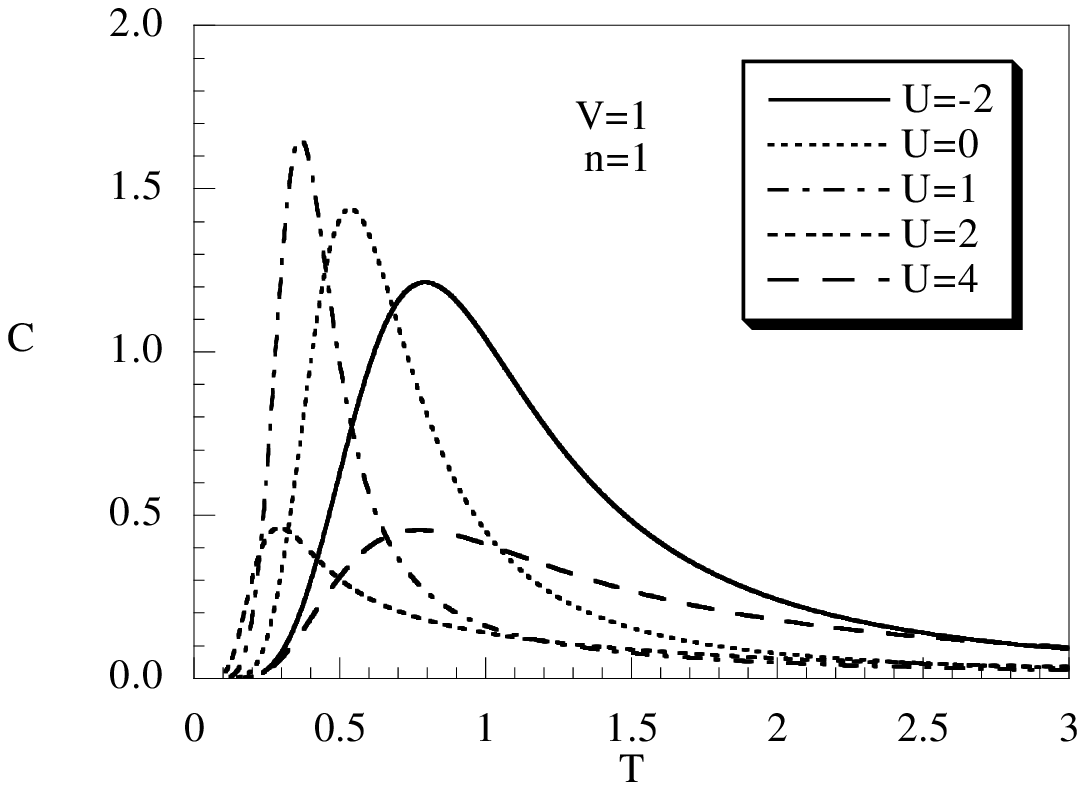}}
 \hspace{1mm}
 \subfigure[]
   {\includegraphics[width=1.48in,height=1.10in]{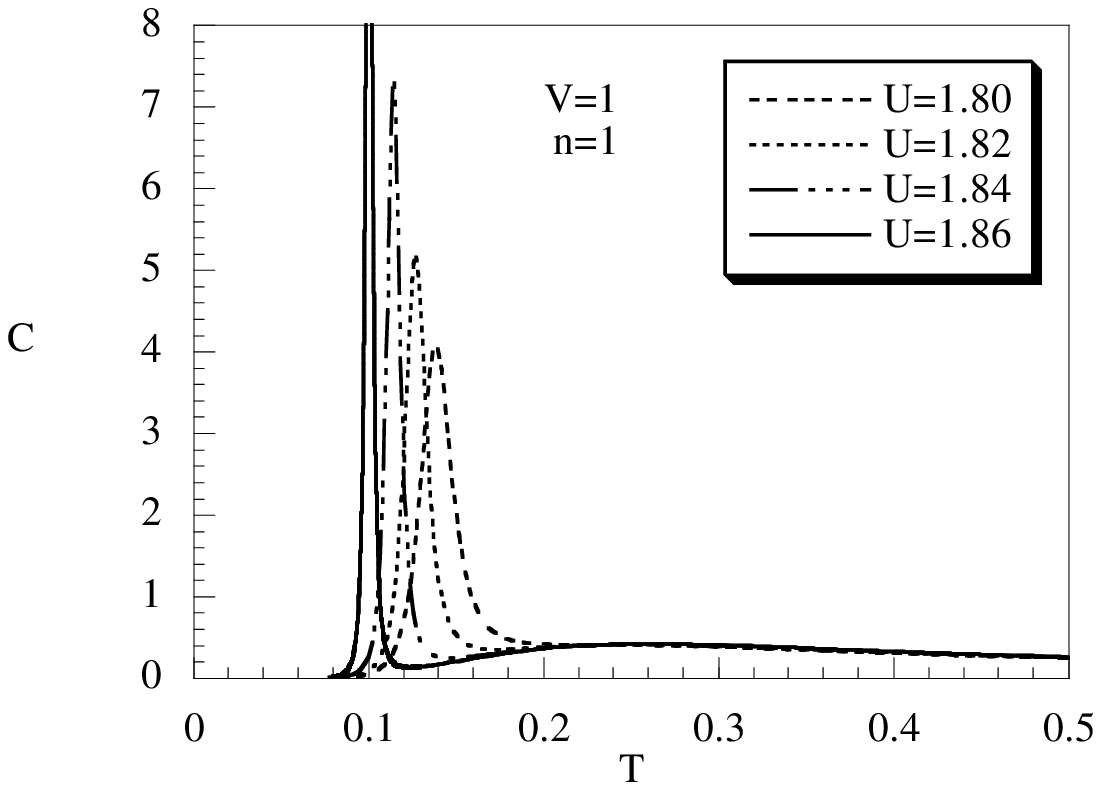}}
     \vspace{-4mm}
 \caption{\label{fig35}(a) The specific heat $C$ as a
function of the temperature $T$ for $V=1$, $n=1$ and (a) $U=-2$,
$0$, $1$, $2$, and $4$; (b)  $U=1.8$, $1.82$, $1.84$ and $1.86$.}
 \end{figure}

\begin{figure}[t]
 \centering
 \subfigure[]
   {\includegraphics[width=1.48in,height=1.10in]{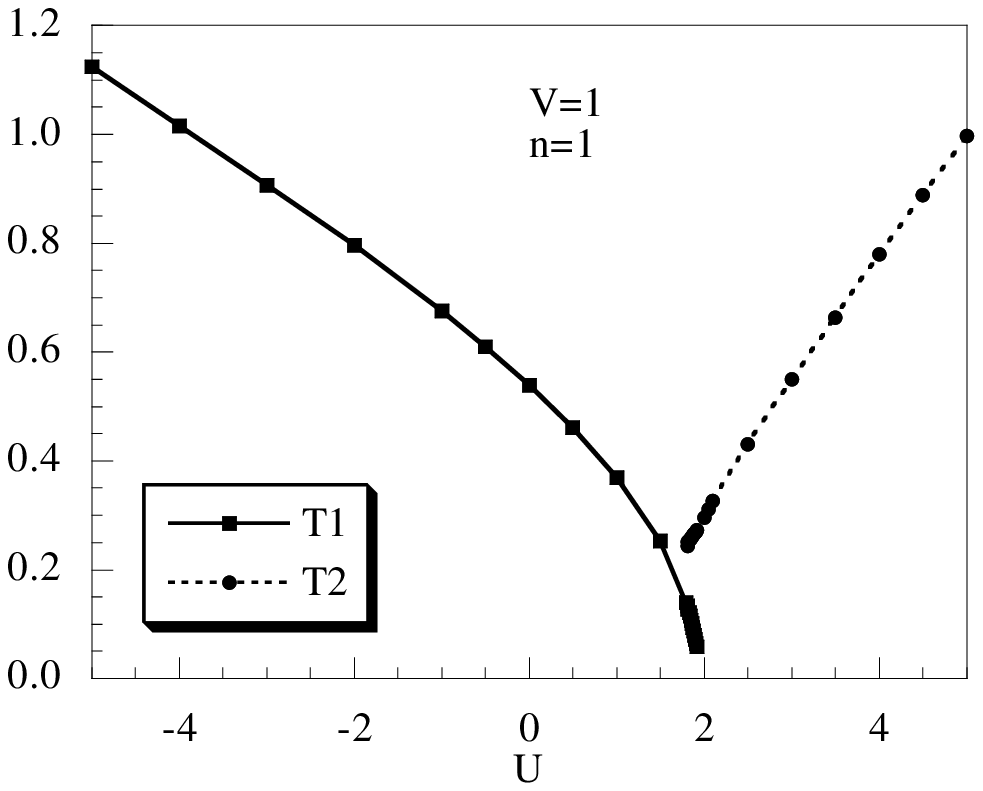}}
 \hspace{1mm}
 \subfigure[]
   {\includegraphics[width=1.48in,height=1.10in]{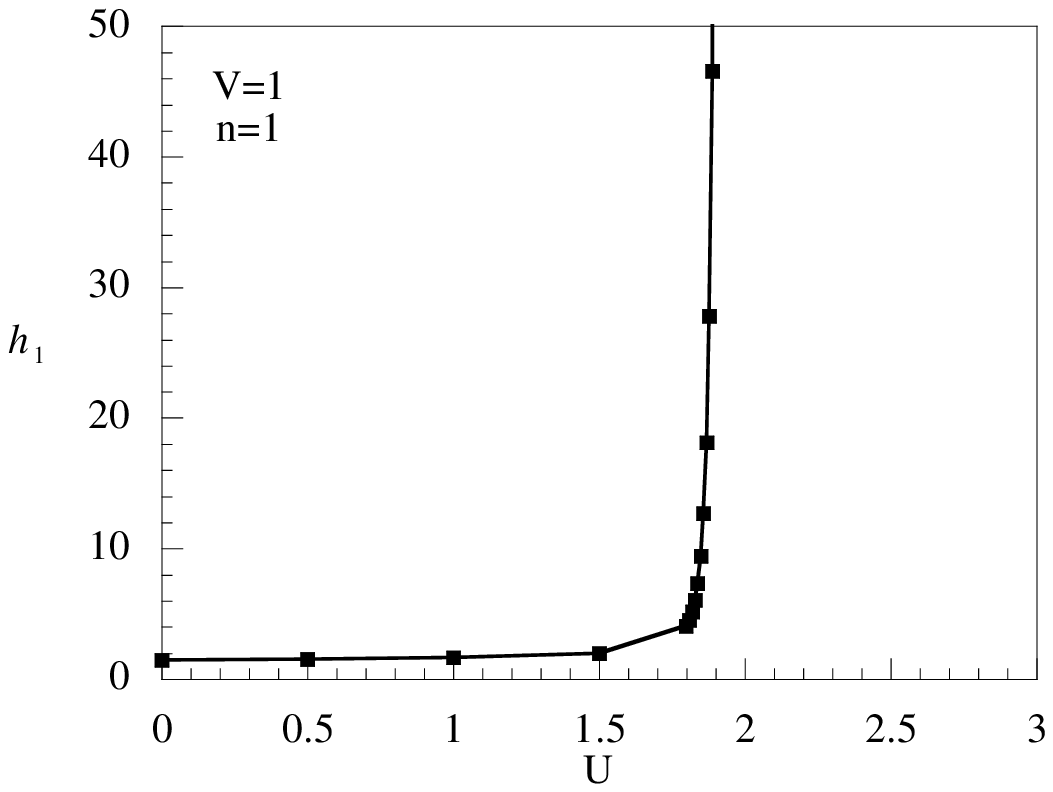}}
     \vspace{-4mm}
 \caption{\label{fig36}(a) The temperature $T_1$ and $T_2$
as functions of the on-site potential for $V=1$, $n=1$. (b) The
height $h_1$ as a function of the on-site potential for $V=1$,
$n=1$.}
\end{figure}

For $n=0.75$, as it is shown in Figs. \ref{fig33}a-e, one observes
a double peak scenario both around $U=0$ and $U=2V$, in
correspondence of the two different transitions observed at $T=0$.
For negative values of $U$, the specific heat has a peak at a
temperature $T_1$ (see Fig. \ref{fig33}a) which decreases
following a linear law until $U\approx 0$. Around $U\approx -V$,
at a lower temperature $T_2$ one finds  a second peak which
decreases following a linear law (see Fig. \ref{fig33}b).
At $U=0$ there is only one peak situated at $T_1 \approx 0.62V$.
For small positive values of $U$ the specific heat has again two
peaks (see Fig. \ref{fig33}c). By increasing $U$, $T_2$ increases
following a linear law up to $U\approx 0.35V$, where this peak is
no longer observable. On the other hand, the position of the first
peak $T_1$ decreases by increasing $U$ up to $U=2V$, where this
peak disappears. When $U$ approaches the value $U=2V$ a third peak
appears at some temperature $T_3$ (see Fig. \ref{fig33}d). We
recall that at $U=2V$ and $T=0$ (see Figs. \ref{fig8} and
\ref{fig12}) there is a transition from an ordered state
characterized by alternating empty and occupied sites to a
disordered state with only singly occupied sites. $T_3$ starts at
$U\approx 1.75V$ and increases linearly with $U$. Further
increasing the on-site potential one finds that at $U\approx 5.4V$
a fourth peak, situated at a lower temperature $T_4 $, appears
(see Fig. \ref{fig33}e). As a result, in the region $U>5.4V$ the
specific heat exhibits a double peak structure, with $T_3$
increasing linearly while $T_4$ remains constant at the value $T_4
\approx 0.41V$. In Fig. \ref{fig34} we show the behavior of $T_1$,
$T_2$, $T_3$, $T_4$ as functions of $U$. In contrast to what
happens in the range $0 \le n \le 0.5$, for $n=0.75$ the
excitations induced by $\psi^{(\eta)}$ and by $\psi^{(\xi)}$ are
both present for $U>2V$: indeed, at $T=0$, one observes both
singly and doubly occupied sites.
The behavior of the specific heat at $n=1$ as a function of
temperature is shown in Figs. \ref{fig35}a and \ref{fig35}b. When
$T=0$ and $U=2V$, there is a PT from an ordered state with doubly
occupied sites organized on a checkerboard distribution to a
homogeneous state, where all the sites are singly occupied. The
possible excitations are creation and annihilation of doubly
occupied states. One observes a peak for negative values of $U$ at
a temperature $T_1$, as shown in  Fig. \ref{fig36}a. $T_1$
decreases linearly until $U\approx 0$.
As $U$ gets close to $2V$, $T_1$ further decreases, while its
height $h_1$ exponentially increases and vanishes at $U=2V$. $T_2$
appears at $U \approx 1.81V$ and increases linearly with $U$. When
$U>2V$, further increasing the on-site potential leads to an
increase of the thermal energy required to excite an $\eta$
electron above the ground state. Thus, the peak in the specific
heat shifts towards higher temperatures as it is clear from Figs.
\ref{fig35}a and \ref{fig36}a. In Figs. \ref{fig36}a-b we report
the behavior of $T_1$, $T_2$ and $h_1$ as a function of $U$. Both
$T_1$ and $T_2$ vary linearly with $U$, according to the fact that
the only possible excitations are those induced by
$\psi^{(\eta)}$.

\begin{figure}[t]
\includegraphics[scale=0.45]{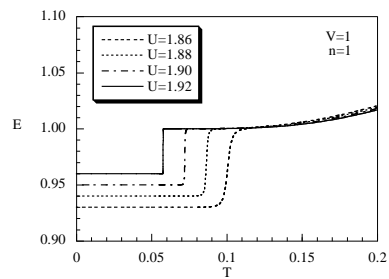}
\caption{\label{fig37}The internal energy $E$ as a function of $T$
for $V=1$, $n=1$ and various values of $U$.}
\end{figure}

The divergence of the specific heat in the limit $U \to 2V$, when
$T \to 0^+$, hints at a first order phase transition. Another way
of looking at this is to study the behavior of the internal
energy.
In Fig. \ref{fig37} we plot the internal energy $E$ as a function
of $T$ for several values of $U$ just below $U=2V$. One sees that
$E$ exhibits a jump around the related $T_1$ temperatures. By
increasing $U$, the jump becomes steeper and, at the same time,
smaller. In the limit $U\to 2V$, $E$ exhibits a discontinuity with
$T_1 \to 0$ and the specific heat diverges. On the other hand, at
$U=2V$, the specific heat is finite with a single peak. Sharp
peaks in the specific heat at half filling, in the neighborhood of
the phase transition, have also been observed in Refs.
\cite{Beni74,Glocke07}, although with approximate methods.

To end this subsection, we want now to comment on the issue
related to the two-peak structure exhibited by the specific heat.
This issue has been largely discussed in the literature in the
context of the ordinary Hubbard model, where the common
interpretation explains the two peaks as due to spin and charge
excitations. This explanation does not apply to the AL-EHM where
only charge excitations are present. Our exact solution of the
model points out that the possible excitations [recall the energy
spectrum given in Eq. \eqref{EHM_7}] are due to a redistribution
of the charge density induced by the creation and annihilation of
electrons under the constraint imposed by the preexisting charge
distribution. For instance, the operator
$\xi^{\dag}(i)n^{\alpha}(i)$ creates an electron at the site $i$,
provided that this site is empty and the neighbor site(s) is (are)
occupied.

Our extensive study of the $U$ dependence of the temperature
maxima $T_n$ (at which the specific heat exhibits a peak) clearly
indicates that the observed peak structure is due to charge
excitations induced by the two operators $\psi^{(\xi)}$ and
$\psi^{(\eta)}$. An explanation of the two peak structure, as due
solely to charge excitations, has also been proposed in Ref.
\cite{dolcini_02}, where an exact solution of a Hubbard-like model
has been given.

\subsection{The entropy}

The standard way to compute the entropy is via the derivative of
the specific heat:
\begin{equation}
\label{EHM_39}
 S(T)=S(0)+\int_0^T
\frac{C(T')}{T'} \, dT'.
\end{equation}
This expression requires the knowledge of the zero temperature
entropy $S(0)$, which can be computed by means of the formula
$S(0)=k_B \ln g$, where $k_B$ is the Boltzmann's constant and $g$
is the number of accessible microscopic states, i.e., all the
states with energy equal to the ground state energy. Now, if the
lowest energy level is non degenerate, $g=1$ and $S(0)=0$. On the
other hand, if the lowest energy level is degenerate $g\ne 1$ and
$S(0)\ne 0$. This latter situation is the one occurring in the
AL-EHM: as it is shown in Figs. \ref{fig3}, \ref{fig8}, and
\ref{fig12}, at $T=0$, every phase (except for $n=1$ and $U<2V$)
exhibits a macroscopic degeneracy. As a result, the zero
temperature entropy is different from zero. One statement of the
third law of thermodynamics is that, as the temperature approaches
zero, the entropy becomes independent of the external parameters.
This still holds for the AL-EHM confined in one of the possible
phases. As we will show later, $S(0)$ depends on the particle
density and on the on-site potential since, by varying $n$ or $U$,
the system can undergo a PT.

In some special cases, it easy to compute $S(0)$. For example,
from Figs. \ref{fig3}, \ref{fig8} and \ref{fig12} it is easy to
see that:

(i) for $U>0$ and $n=0.5$, the ground state corresponds to a
checkerboard distribution of singly occupied sites with arbitrary
spin; the ground state energy is independent on the spin
orientation and the number of accessible states is $g=(2)^{N/2}$;
therefore $S(0)=\frac{N}{2}\ln (2)$.

(ii) for $U<2V$ and $n=1$, the ground state corresponds to a
checkerboard distribution of doubly occupied sites. There is no
degeneracy and, as a result, the number of accessible states is
$g=1$ and $S(0)=0$.

(iii) for $U>2V$ and $n=1$, the ground state corresponds to an
homogeneous distribution of singly occupied sites with arbitrary
spin; the ground state energy is independent on the spin
orientation and the number of accessible states is $g=(2)^N$;
therefore $S(0)=N \ln (2)$.

\begin{figure}[t]
 \centering
 \subfigure[]
   {\includegraphics[width=1.48in,height=1.10in]{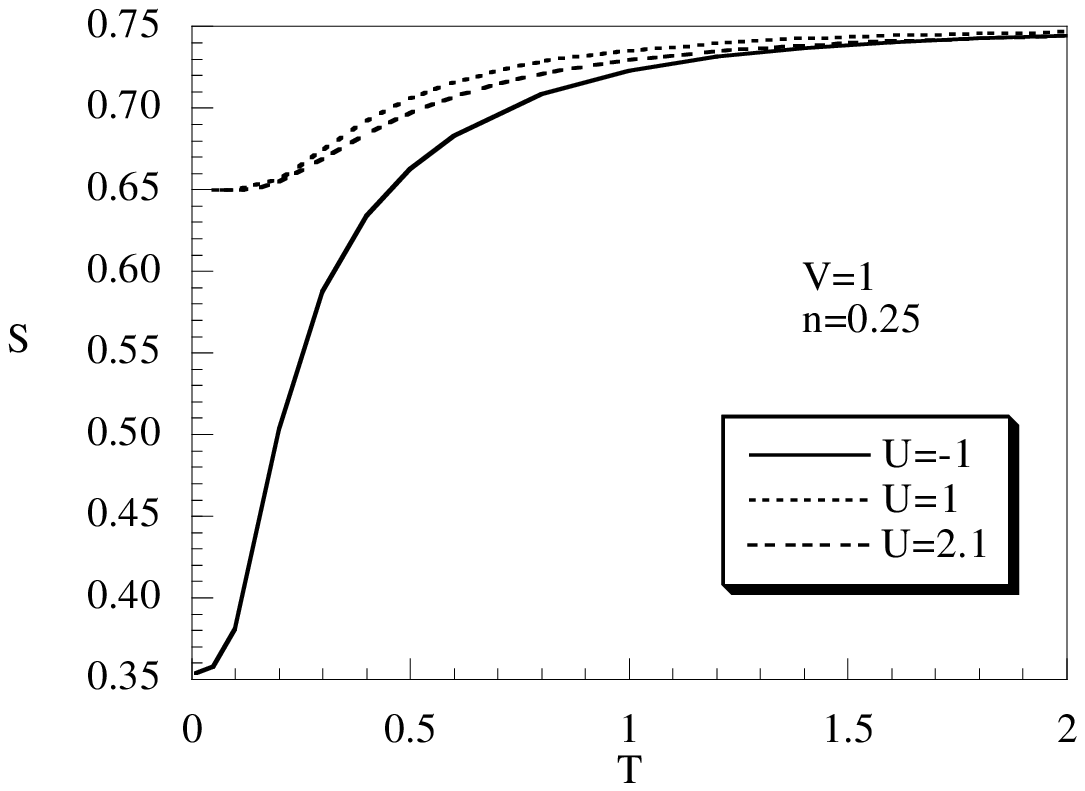}}
 \hspace{1mm}
 \subfigure[]
   {\includegraphics[width=1.48in,height=1.10in]{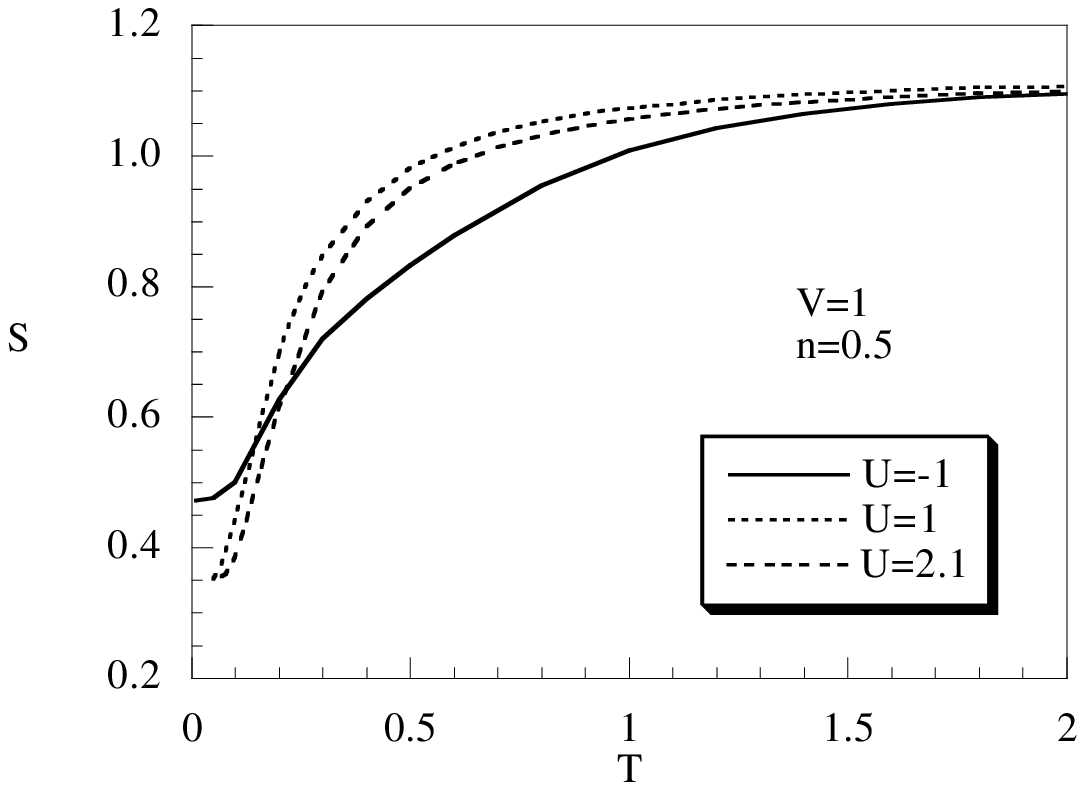}}
   \\
   \subfigure[]
   {\includegraphics[width=1.48in,height=1.10in]{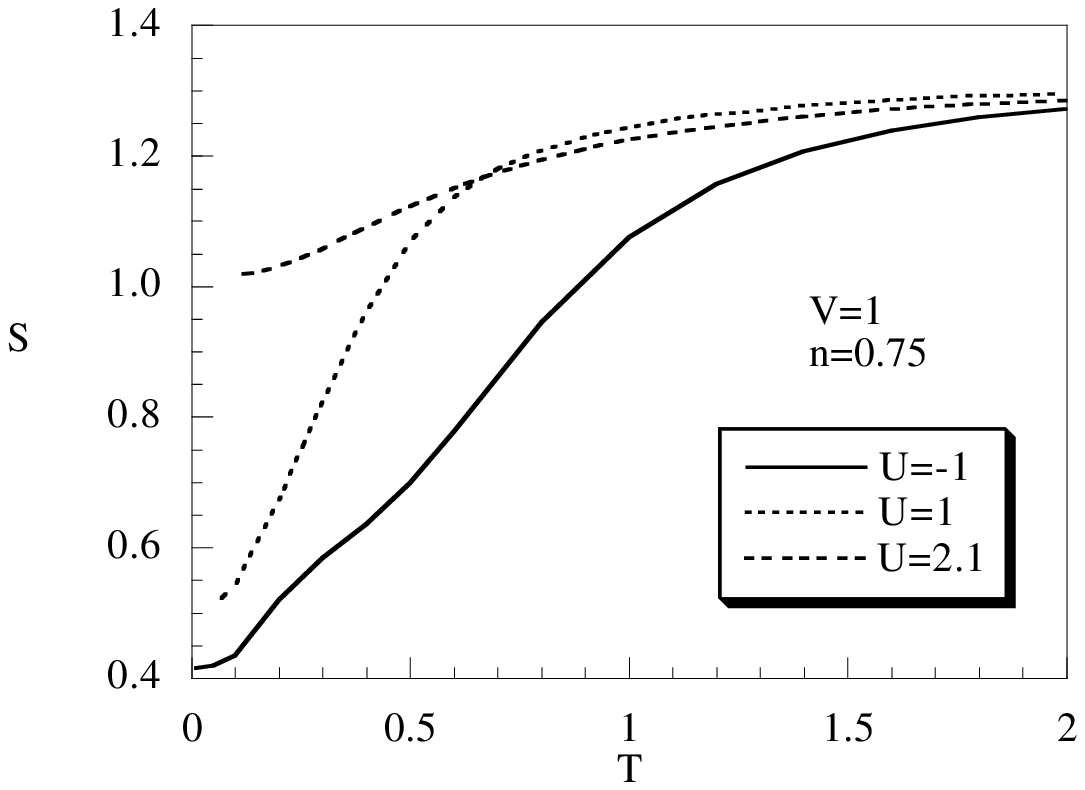}}
 \hspace{1mm}
 \subfigure[]
   {\includegraphics[width=1.48in,height=1.10in]{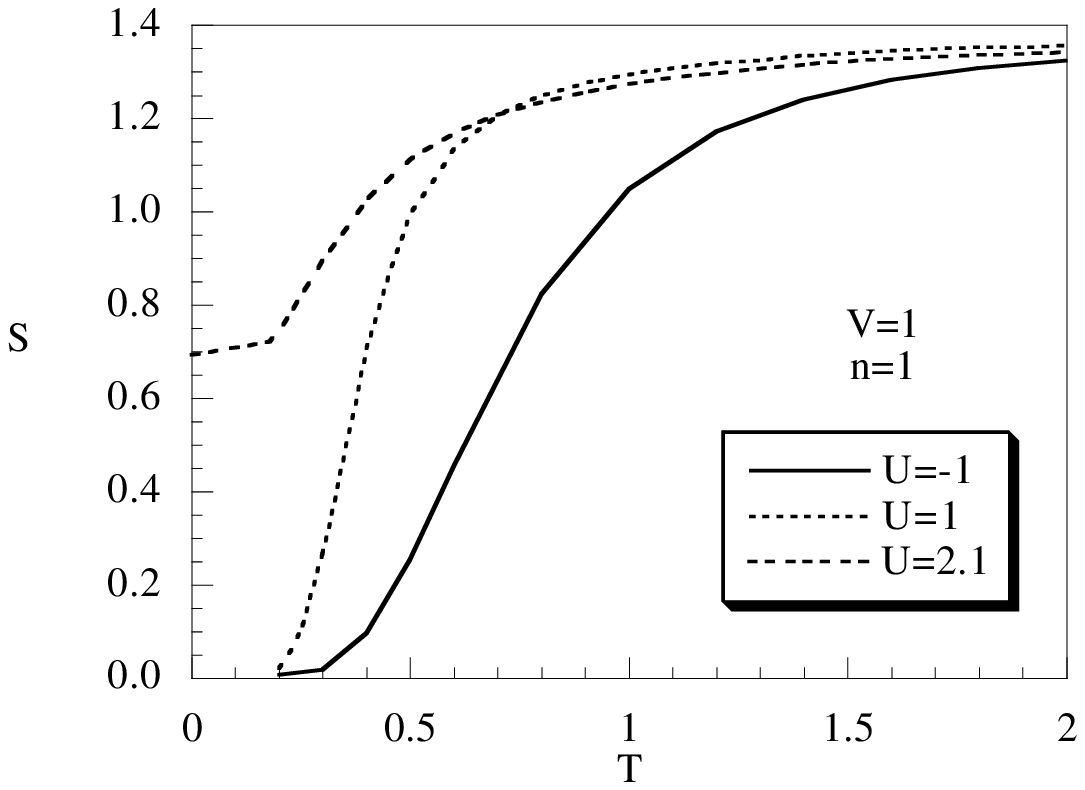}}
     \vspace{-4mm}
 \caption{\label{fig38} The entropy $S$ as a function of the temperature $T$
for $V=1$, $U=-1$, $1$, $2.1$ and (a): $n=0.25$, (b): $n=0.5$,
(c): $n=0.75$, (d): $n=1$.}
 \end{figure}

\begin{figure}[ht]
\centerline{\includegraphics[scale=0.45]{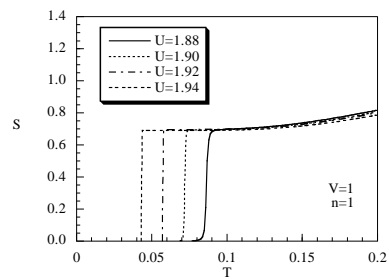}}
\caption{\label{fig39} The entropy as a function of the
temperature for $V=1$, $n=1$ and different values of $U$ close to
$2V$. }
\end{figure}

Discarding few simple cases, it is generally not easy to compute
$S(0)$. As a consequence, one has to use different formulas to
compute the entropy. In the limit of infinite temperature, the
entropy can be computed as $S(\infty)=k_B \ln g_{\infty}$, where
$g_{\infty}$ is the total number of states. For the model under
consideration it is easy to show that
\begin{equation}
\label{EHM_40} \lim_{T \to \infty} S(T)= 2 \log 2-
n\log(n)-(2-n)\log(2-n).
\end{equation}
Here and in the following we set the Boltzmann's constant $k_B =1$
and consider the entropy per site. For general values of the
temperature, recalling that $(n, -\mu)$ are thermodynamically
conjugated variables, one can express the entropy per site as
\cite{manciniavella}
\begin{equation}
\label{EHM_41}
\begin{split}
 S(n,T,U)
 &= -\int_0^n \frac{\partial \mu (n',T,U)}{\partial T}dn'
 \\
&=\frac{1}{T}\left[ E(n,T,U)-\int_0^n \mu (n',T,U)dn' \right].
\end{split}
\end{equation}
This expression requires only the knowledge of the internal energy
and of the chemical potential. By making use of Eq.
\eqref{EHM_41}, we shall present in the following several results
for the entropy as a function of $T$, $n$, and $U$. In Figs.
\ref{fig38} we plot the entropy as a function of the temperature
for relevant values of the particle density $n$. One clearly sees
that $S$ is always a decreasing function of the temperature and
that, for $n<1$, $S$ does not vanish for $T\to 0$. For instance,
for $n=0.75$ at $T=0$ three phases are present by varying $U$:
accordingly, the entropy for $T\to 0$ tends to three different
constants (see Fig. \ref{fig38}c) which depend only on the
degeneracy of the related ground state. Furthermore, it is easy to
check that, in the limit $T\to \infty$, the entropy tends to the
value given in Eq. \eqref{EHM_40}.
It is interesting to note, as it is shown in Fig. \ref{fig39},
that at half filling, tuning the parameter $U$ in the neighborhood
of the critical value $U=2V$, the entropy suddenly vanishes at the
same temperature $T_1$ at which the specific heat increases
exponentially (see Fig. \ref{fig35}b). This can be easily
understood by recalling that $C=T \cdot \partial S/\partial T$:
the divergence of the specific heat is a consequence of the
discontinuity of the entropy. One can think of this divergence as
induced by a change from a non-degenerate state to an infinitely
degenerate state.

\begin{figure}[ht]
 \centering
 \subfigure[]
   {\includegraphics[width=1.48in,height=1.10in]{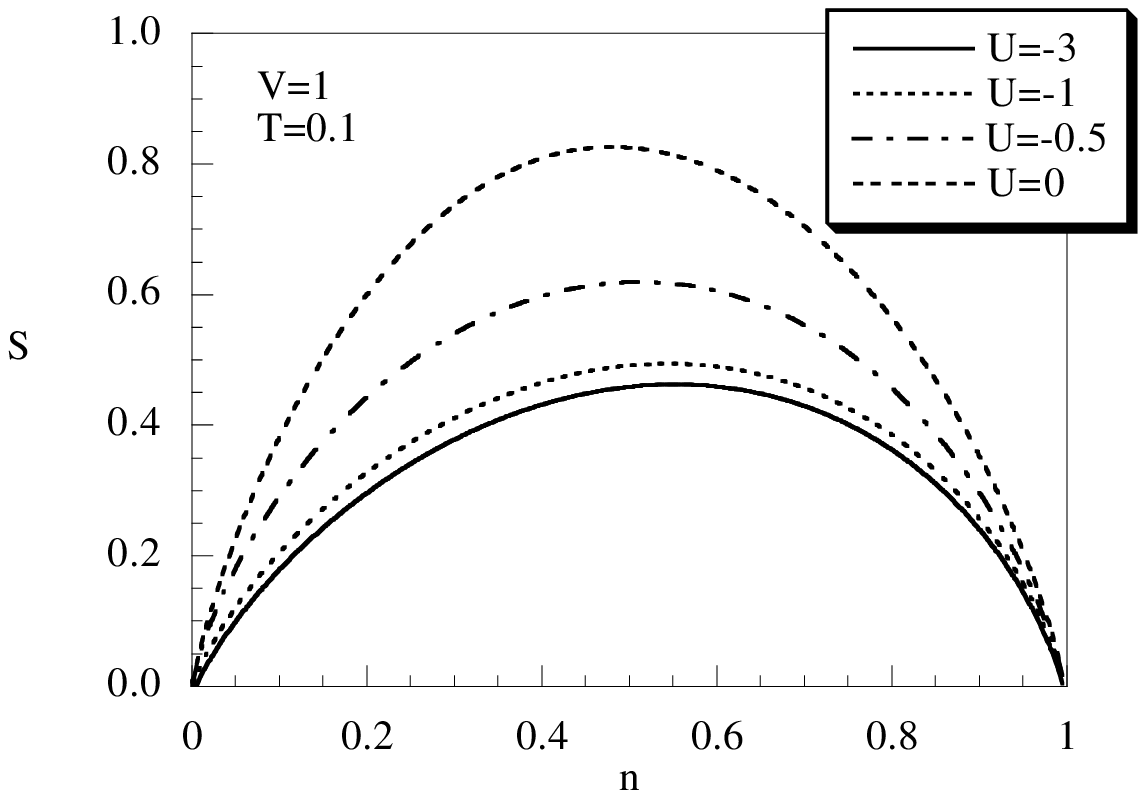}}
 \hspace{1mm}
 \subfigure[]
   {\includegraphics[width=1.48in,height=1.10in]{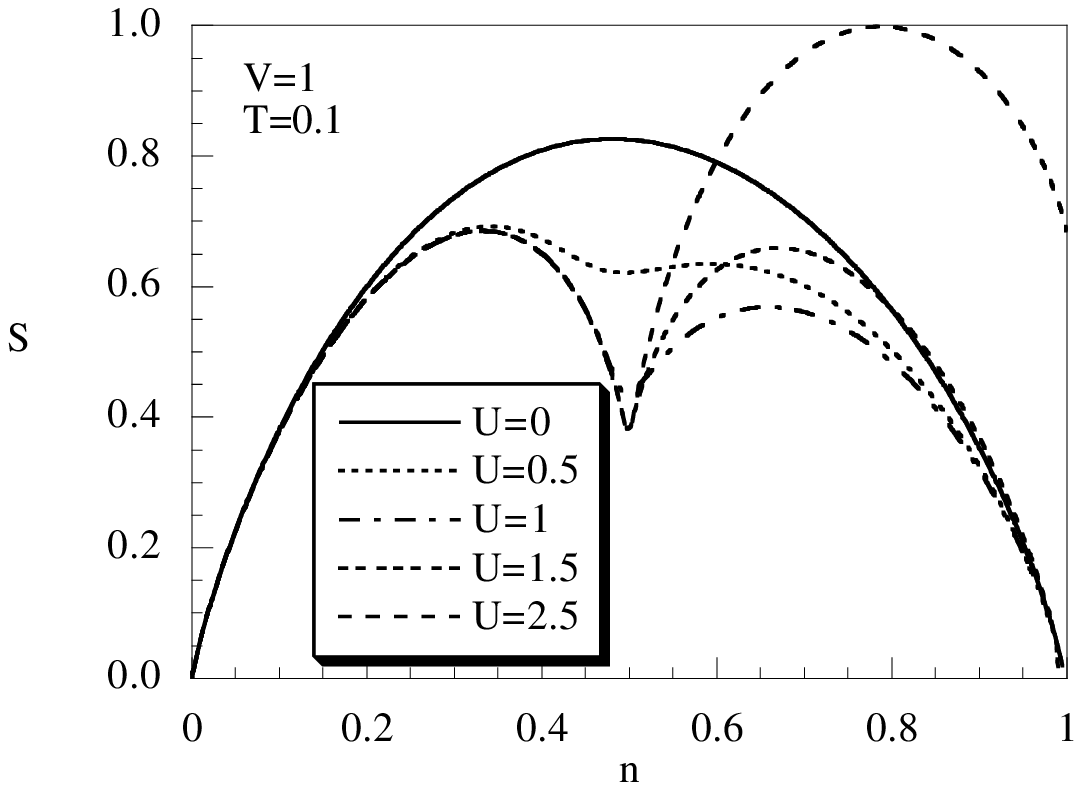}}
 \caption{\label{fig40}The entropy $S$ as a function of the particle
 density $n$ for $V=1$ and $T=0.1$ (a): $U=-3$, $-1$, $-1/2$, and $0$;
 (b): $U=0$, $1/2$, $1$, $3/2$ and $5/2$.}
 \end{figure}

When plotted as a function of the particle density (see Fig.
\ref{fig40}), the entropy shows a maximum around quarter filling
when one considers an attractive on-site interaction. At low
temperatures, the entropy increases by increasing the particle
density; it has a maximum around quarter filling and then
decreases vanishing, as expected, at half filling. On the other
hand, considering a repulsive on-site potential, the entropy
presents a minimum at quarter filling and two maxima around
$n=0.25$ and $n=0.75$. As evidenced in Fig. \ref{fig40}b, when
$U>2V$ the entropy does not vanish at half filling. The increasing
and decreasing of the entropy by varying the particle density can
be understood as due to the increase and decrease of the number of
accessible states, as it is easy to check by looking at Figs.
\ref{fig3}, \ref{fig8}, and \ref{fig12}. It is worthwhile to
observe that the minimum values of the entropy are observed in
correspondence of the ordered states, where the number of
accessible states is, of course, minimum.

\begin{figure}[t]
 \subfigure[]
   {\includegraphics[width=1.48in,height=1.10in]{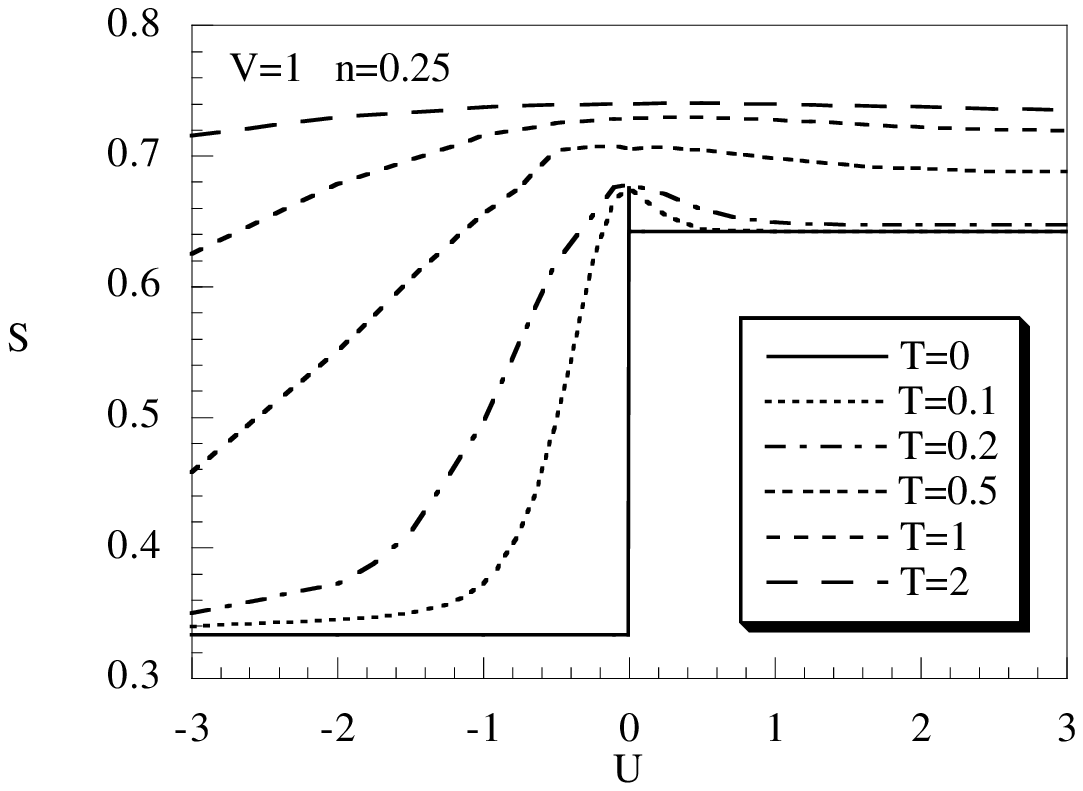}}
 \subfigure[]
   {\includegraphics[width=1.48in,height=1.10in]{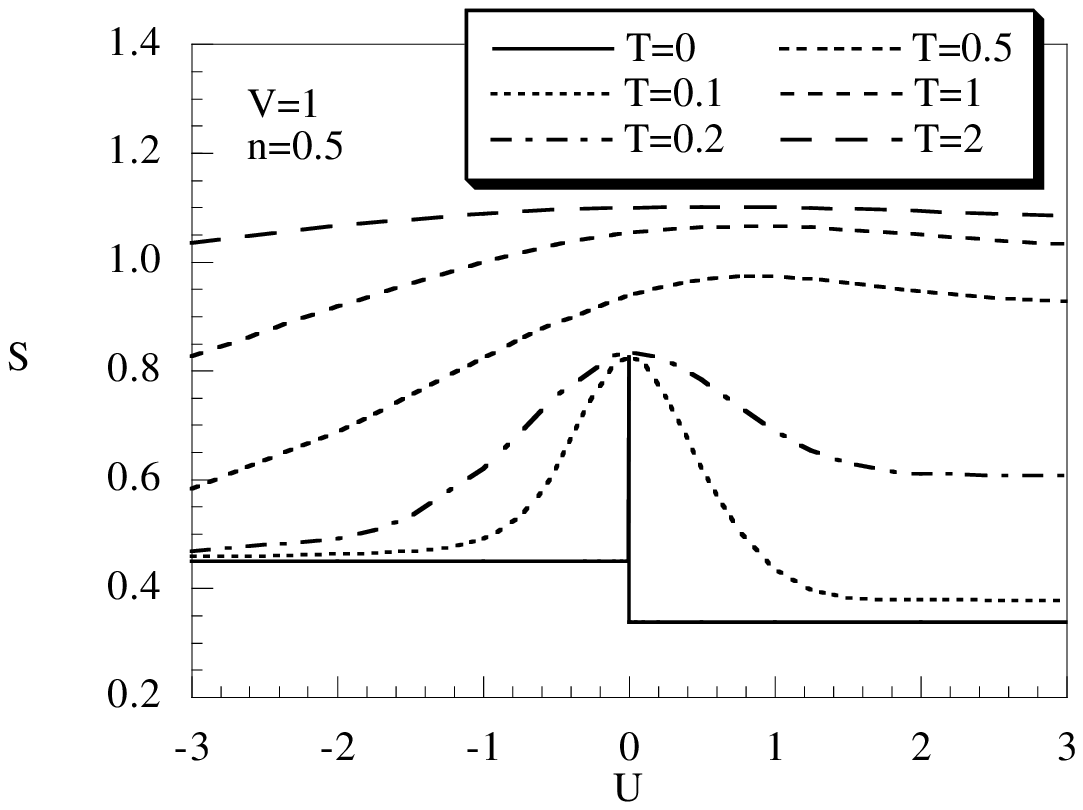}}
   \subfigure[]
   {\includegraphics[width=1.48in,height=1.10in]{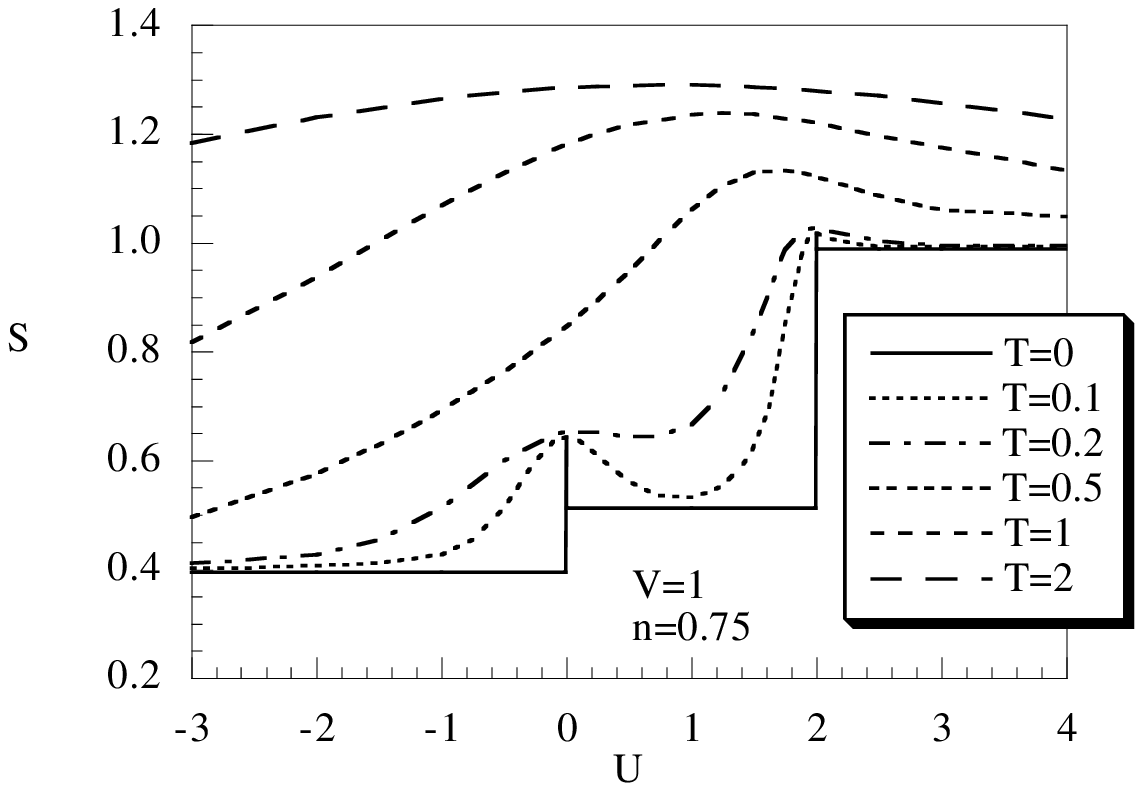}}
 \subfigure[]
   {\includegraphics[width=1.48in,height=1.10in]{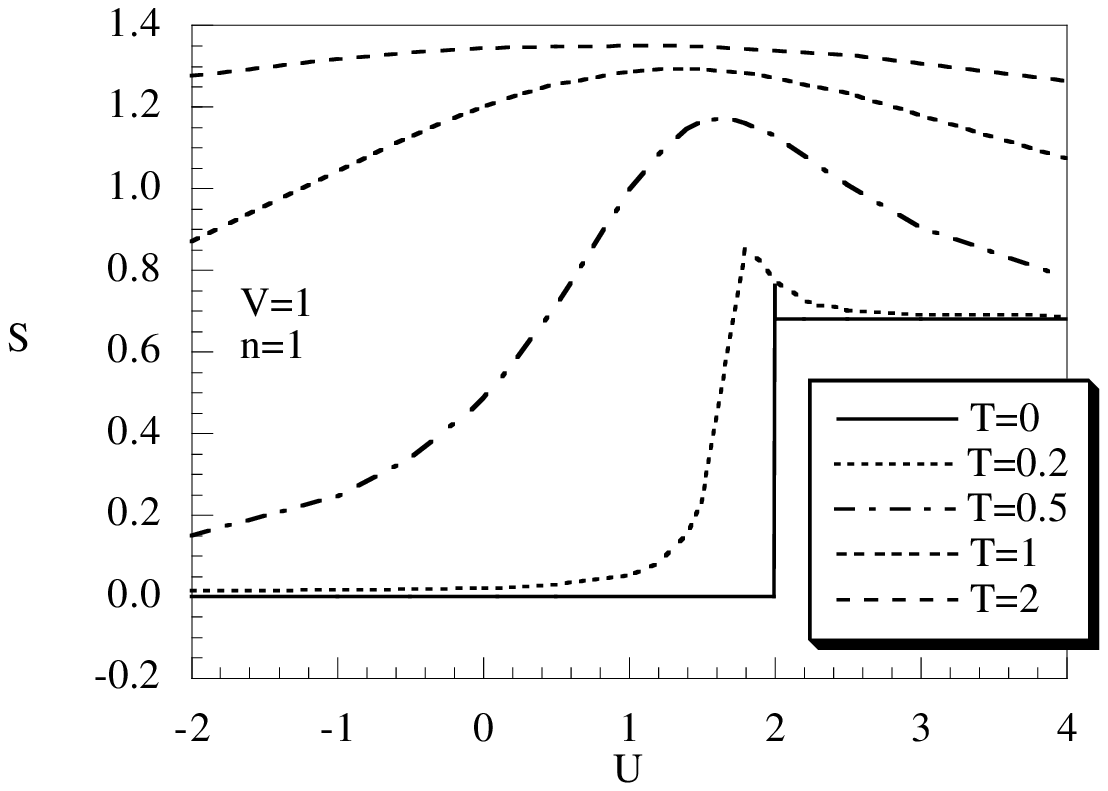}}
  \vspace{-4mm}
 \caption{\label{fig41} The entropy $S$ as a function of $U$ for $V=1$, for
temperatures varying in the interval (0,2) and (a): $n=0.25$, (b):
$n=0.5$, (c): $n=0.75$, (d): $n=1$.}
 \end{figure}

A useful representation of the finite-temperature phase diagram is
obtained by plotting the entropy as a function of the on-site
potential $U$, as it is shown in Figs. \ref{fig41}a-d. The $U$
dependence of the entropy is rather dramatic in the neighborhood
of the values at which a zero temperature transition occurs. At
low temperatures, the entropy presents a steplike behavior and
becomes rather insensitive to variations in $U$ for sufficiently
large on-site repulsive and attractive interactions.

When $n \le 0.5$, the entropy presents a peak around $U=0$ which
becomes more pronounced as the temperature decreases. At quarter
filling, one observes a decrease of the entropy by varying $U$
since, in the limit $T \to 0$, the electrons are organized in an
ordered checkerboard distribution of singly occupied sites. In the
region $0.5<n<1$, at low temperatures, one finds three regions of
constant entropy corresponding to the three phases present at
$T=0$ when one moves, at fixed particle density, from negative to
positive values of $U$. At half filling, as expected, one finds a
rather sharp increase of the entropy only near $U=2V$.
In Figs. \ref{fig41} we report the value $S(0)$ obtained by
subtracting Eqs. \eqref{EHM_39} and \eqref{EHM_41}. The study of
$S(0)$ constitutes another way to detect the zero temperature
transition from thermodynamic data. In fact, as it is clearly
shown in Fig. \ref{fig41}, for each value of the electronic
filling, the zero temperature entropy presents a discontinuity
right at the values of the on-site potential at which one observes
a phase transition.

\section{concluding remarks}
\label{sec_V}

We have evidenced how the use of the Green's function and equation
of motion formalism leads to the exact solution of the
one-dimensional extended Hubbard model in the atomic limit. We
provided a comprehensive and systematic analysis of the model by
considering all the relevant correlation functions and
thermodynamic quantities in the whole space of parameters $n$,
$T$, and $U$ (having chosen $V=1$ as the unity of energy),
including attractive on-site interactions. This study has shown
that, at zero temperature, the model exhibits phase transitions
for specific values of $n$ and $U$. In particular, we have
identified four phases in the ($U,n$) plane and PTs are observed
at the borders of these phases. Various types of long-range charge
ordered states have been observed: (i) at half filling and $U<2V$,
a checkerboard distribution of doubly occupied sites; (ii) at
quarter filling and $U>0$, a checkerboard distribution of
arbitrary-spin singly occupied sites; (iii) in the entire region
$0.5\le n\le 1$ and $0<U<2V$, an ordered state with alternating
empty and occupied sites. The onset of a CO is signalled by the
divergence of the correlation length in the charge correlation
function. It is worthwhile to mention that the PT observed at half
filling and $U \to 2V$ is a first order transition, as confirmed
by the divergence of the specific heat and by the discontinuity in
the entropy.

We derived the phase diagram by analyzing at $T=0$ various local
correlators which provide information on how the electrons are
distributed on the sites of the chain. As shown in details in
Secs. III and IV, the phase diagram and the observed CO states can
be equivalently studied by analyzing the behavior at $T=0$ of
other quantities. The chemical potential and the double occupancy
show discontinuities where PTs occur. The existence of a
long-range order is well evidenced by the periodicity of the
charge and double occupancy correlation functions. Another way to
detect the zero temperature phase diagram is provided by the study
of the behavior in the limit $T \to 0$ of the charge and spin
susceptibilities and of the entropy. In particular, the entropy at
zero temperature is non zero (except for $n=1$ and $U<2V$) and is
determined only by the degeneracy of the ground state. In the
limit $T \to 0$ the spin susceptibility diverges exponentially or
with a power law, depending on the phase.

In Sec. IV we have studied the model at non zero temperatures. At
finite but low temperatures the properties of the system are
strongly determined by the zero temperature solution: a finite
range order persists for a wide range of $T$, as evidenced by the
behavior of the correlation functions. As a function of $n$, the
charge susceptibility and the entropy exhibit a two-peak structure
with minima at $n=0.5$ and $n=1$, where PTs are observed.

The use of equation of motion and the introduction of a closed set
of composite operators $\psi^{(\xi)}$ and $\psi^{(\eta)}$,
eigenoperators of $H$, allowed for an exact determination of the
elementary excitations, $E_n^{(\xi)}$ and $E_n^{(\eta)}$, and of
the relative weights $\Sigma_n^{(\xi)} $ and $\Sigma_n^{(\eta)}$.
As a result, the density of states has been exactly computed for
any temperature, and we performed a detailed analysis of the
specific heat $C$. In particular, by studying the specific heat in
the parameter regions corresponding to the four $T=0$ phases, it
has been possible to identify the excitations contributing to $C$
as due to a redistribution of the charge density induced by the
operators $\psi^{(\xi)}$ and $\psi^{(\eta)}$ through the creation
and annihilation of electrons, provided that a charge distribution
preexisted. The presence of a two-peak structure, observed in some
regions of $U$, is explained as due to charge excitations induced
by different components of the multiplet field operator
$\psi^{(\eta)}$.

\section*{Acknowledgements}
We thank A. Avella for stimulating discussions and a careful
reading of the manuscript.

\appendix

\section{The parameters $\kappa ^{(k)}$ and
$\lambda ^{(k)}$} \label{App.A}

As already mentioned in Sec. \ref{sec_II}, the use of the
formalism of Green's functions and equations of motion requires
two ingredients: composite operators and algebraic properties. In
this Appendix we shall summarize the main results which can be
obtained by exploiting the algebra satisfied by the operators. By
means of the algebraic rules
\begin{equation}
\begin{split}
[n(i)]^k &= n(i)+a_k D(i) ,\\
[D(i)]^k &= D(i) ,\\
[n(i)]^k D(i)&= 2^k D(i) ,
\end{split}
\label{A.1}
\end{equation}
where $a_k =2^k-2$ for$k\ge 1$, it is possible to derive the
recursion rule
\begin{equation}
[n^\alpha (i)]^k=\sum_{m=1}^4 A_m^{(k)} [n^\alpha (i)]^m ,
\label{A.2}
\end{equation}
where the coefficients $A_m^{(k)}$ are rational numbers, given by
\begin{equation}
\begin{split}
 A_1^{(k)} &=-6+2^{3-k}-2^{k-1}+2^{3-k}\cdot 3^{k-1},
  \\
 A_2^{(k)} &=\frac{1}{3\cdot 2^{k+1}}(-104+57\cdot 2^{k+1}-56\cdot
3^k+11\cdot 4^k),
 \\
 A_3^{(k)} &=\frac{1}{3\cdot 2^{k-1}}(18-3\cdot 2^{k+3}+14\cdot
3^k-3\cdot 4^k),
 \\
A_4^{(k)} &=\frac{1}{3\cdot 2^{k-1}}(-4+3\cdot 2^{k+1}-4\cdot
3^k+4^k) .
 \end{split}
 \label{A.3}
 \end{equation}
It is easy to show that the relation $\sum_{m=1}^4 A_m^{(k)}=1$
holds. As it has been shown in Sec. \ref{II_C}, by using the
$H_0$-representation, the expectation value of a generic operator
$O$ can be expressed as
\begin{equation}
\langle O\rangle =\frac{\langle O e^{-\beta H_I }\rangle_0
}{\langle e^{-\beta H_I }\rangle_0 }.
 \label{A.4}
 \end{equation}
Recalling that $H_I =2Vn(i)n^\alpha (i)$ and by using the
recursion rule \eqref{A.2} one obtains
\begin{equation}
e^{-\beta H_I }=1+n(i)\sum\limits_{m=1}^4 f_m [n^\alpha
(i)]^m+D(i)\sum\limits_{m=1}^4 g_m [n^\alpha (i)]^m , \label{A.5}
 \end{equation}
where
\begin{equation}
\begin{split}
 f_m &=\sum\limits_{n=1}^\infty (-1)^n\frac{1}{n!}A_m^{(n)} (2\beta
 V)^n,
\\
 g_m &=\sum\limits_{n=2}^\infty (-1)^n\frac{1}{n!}a_n A_m^{(n)} (2\beta
 V)^n.
\end{split}
 \label{A.6}
 \end{equation}
By recalling Eq. \eqref{A.3}, one immediately sees that the
coefficients $f_m$ and $g_m$ are polynomials of $K=e^{-\beta V}$
and they are explicitly given by
\begin{equation}
\begin{split}
 f_1 &=-\frac{1}{2}K^4+\frac{8}{3}K^3-6K^2+8K-\frac{25}{6}, \\
 f_2
 &=\frac{11}{6}K^4-\frac{28}{3}K^3+19K^2-\frac{52}{3}K+\frac{35}{6},
\\
 f_3 &=-2K^4+\frac{28}{3}K^3-16K^2+12K-\frac{10}{3} ,\\
 f_4
 &=\frac{2}{3}K^4-\frac{8}{3}K^3+4K^2-\frac{8}{3}K+\frac{2}{3},
\end{split}
 \label{A.7}
 \end{equation}
 and
\begin{equation}
\begin{split}
 g_1
&=-\frac{1}{6} (K-1)^5 (25 + 29 K + 15 K^2 + 3 K^3),
\\
 g_2
&=\frac{1}{6} (-35 + 208 K - 332 K^2 + 112 K^3 \\
&+ 92 K^4 - 56 K^6 + 11 K^8),
\\
 g_3
&= - \frac{2}{3}(K-1)^3 (5 - 21 K - 12 K^2 + 4 K^3 \\
&+ 9 K^4 + 3 K^5),
\\
 g_4
&=\frac{2}{3}(K-1)^4 (-1 + 4 K + 6 K^2 + 4 K^3 + K^4).
\end{split}
 \label{A.8}
 \end{equation}
One can now easily compute the parameters $\kappa ^{(k)}$ and
$\lambda ^{(k)}$. By using Eqs. \eqref{A.4} and \eqref{A.5} and
recalling the properties of the $H_0$ representation
\cite{mancini05b} one has
\begin{equation}
\begin{split}
 \kappa ^{(k)}&=\frac{1}{\Upsilon _0}\{\langle
[n^\alpha (i)]^k\rangle_0 +\sum\limits_{p=1}^4 (B_1 f_p +B_2
g_p)\langle [n^\alpha(i)]^{p+k}\rangle_0 \},
 \\
 \lambda ^{(k)}&=\frac{1}{2 \Upsilon _0}\{B_1
\langle [n^\alpha (i)]^k\rangle _0
\\
&+\sum\limits_{p=1}^4 [(B_1 +2B_2 )f_p +2B_2 g_p ]\langle
[n^\alpha (i)]^{p+k}\rangle_0 \} ,
\end{split}
 \label{A.9}
 \end{equation}
where
\begin{equation}
\begin{split}
\Upsilon _0= \langle e^{-\beta H_I }\rangle _0
=1+\sum\limits_{p=1}^4 (B_1 f_p +B_2 g_p )[n^\alpha
(i)]^p ,\\
 B_1 =\langle n(i)\rangle _0 =\frac{2e^{\beta \mu }(1+e^{\beta (\mu
-U)})}{1+2e^{\beta \mu
}+e^{\beta (2\mu -U)}}, \\
 B_2 =\langle D(i)\rangle _0 =\frac{e^{\beta (2\mu -U)}}{1+2e^{\beta \mu
}+e^{\beta (2\mu -U)}} .
 \end{split}
 \label{A.10}
 \end{equation}
By using the properties of the $H_0$ representation, the
expectation values $\langle [n^\alpha (i)]^k\rangle _0$ can be
written as
\begin{equation}
\begin{split}
 \langle [n^\alpha (i)]^2\rangle _0 &=\frac{1}{2}X_1 +X_2
+\frac{1}{2}X_1 ^2 ,\\
 \langle [n^\alpha (i)]^3\rangle _0 &=\frac{1}{4}X_1 +\frac{3}{2}X_2
+\frac{3}{4}X_1
^2+\frac{3}{2}X_1 X_2 , \\
 \langle [n^\alpha (i)]^4\rangle _0 &=\frac{1}{8}X_1 (1+7X_1
)+\frac{7}{4}X_2 +\frac{9}{2}X_1 X_2 +\frac{3}{2}X_2 ^2 ,
\end{split}
 \label{A.11}
 \end{equation}
where the two parameters $X_1$ and $X_2$ are defined as
\begin{equation}
\begin{split}
 X_1 &=\langle n^\alpha (i)\rangle_0, \\
 X_2 &=\langle D^\alpha (i)\rangle_0 .
 \end{split}
 \label{A.12}
 \end{equation}
Then, one has
\begin{equation*}
\begin{split}
\Upsilon_0 &=1-B_1 +B_2 +(B_1 -2B_2 )(1+aX_1 +a^2X_2 )^2 \\
&+B_2 (1+dX_1 +d^2X_2 )^2 ,
 \end{split}
\end{equation*}
with $a=K-1 \quad d=K^2-1$. The correlators $\langle [n^\alpha
(i)]^k\rangle_0$ with $k>4$ can be computed by using the recursion
rule \eqref{A.2}. The previous formulas show that the local CFs
$\kappa ^{(k)}$ and $\lambda^{(k)}$ are exactly known in terms of
the two basic parameters $X_1$ and $X_2$. A similar procedure can
be used for computing the other parameters $\pi ^{(k)}=\langle
[D^\alpha (i)]^k\rangle $, $\delta ^{(k)}=\langle n(i)[D^\alpha
(i)]^k\rangle /2$, $\theta ^{(k)}=\langle D(i)[D^\alpha
(i)]^k\rangle /2$. In particular, one obtains
\begin{equation}
\begin{split}
 \theta ^{(1)}&=\frac{B_2 X_2 }{2\Upsilon_0
}(1+4a+4b+c)[1+(2a+b)X_1
+(c+2b)X_2 ], \\
 \pi ^{(2)}&=\frac{1}{3}\kappa ^{(4)}-\kappa ^{(3)}+\frac{11}{12}\kappa
^{(2)}-\frac{1}{4}n+\frac{1}{2}D,
\\
\delta ^{(1)}&=\frac{1}{2}\langle n(i)[D^\alpha
(i)]\rangle=\frac{X_2}{2 \langle e^{-\beta H_I}\rangle_0 }\{B_1 (1+2a+b)
\\ &\cdot(1+aX_1 +bX_2 ) + 2B_2 [6a^2X_1 +b^2(4X_1 +7X_2 )
\\ &+c(1+X_2 +cX_2 )+ a(2+X_1 +11bX_1 +2cX_1 \\
&+6bX_2 +4cX_2 )+b(3+X_1 +cX_1 +X_2 +6cX_2 )]\},
 \end{split}
 \label{A.15}
 \end{equation}
where $b=(K-1)^2$ and $c=4K-4K^2+K^4-1$.

\section{Nonlocal correlation functions}
\label{App.B}

One can define the following nonlocal correlation functions:
\begin{equation}
\label{B.1}
\begin{split}
 \Lambda ^{(k)}(i,j)&=\langle n(i)[n^\alpha (i)]^k n(j)\rangle , \\
 K^{(k)}(i,j)&=\langle [n^\alpha (i)]^kn(j)\rangle , \\
 \Theta ^{(k)}(i,j)&=\langle D(i)[D^\alpha (i)]^kn(j)\rangle  ,\\
 \Pi ^{(k)}(i,j)&=\langle [D^\alpha (i)]^kn(j)\rangle,
\\
 M^{(k)}(i,j)&=\langle D(i)[D^\alpha (i)]^kD(j)\rangle , \\
 P^{(k)}(i,j)&=\langle [D^\alpha (i)]^kD(j)\rangle ,\\
 Q^{(k)}(i,j)&=\langle n(i)[n^\alpha (i)]^kD(j)\rangle  ,\\
 R^{(k)}(i,j)&=\langle [n^\alpha (i)]^kD(j)\rangle ,
\end{split}
\end{equation}
where $j$ is a site $m$-steps $(m\ge 1)$ distant from the site
$i$. By using the $H_0 $-representation, the relevant, for our
purposes, correlation functions \eqref{B.1} can be written as:
\begin{equation}
\label{B.2}
\begin{split}
 \Lambda ^{(0)}(i,j)&=\langle n(i)n(j)\rangle =\sum_{\ell=1}^3 a_{\ell}
Y_{\ell} ,\\
 K^{(0)}(i,j)&=\langle n(j)\rangle =\sum_{\ell=1}^3 c_{\ell} Y_{\ell,} \\
 K^{(1)}(i,j)&=\langle n^\alpha (i)n(j)\rangle =\sum_{\ell=1}^3 d_{\ell}
Y_{\ell} , \\
 \Theta ^{(0)}(i,j)&=\langle D(i)n(j)\rangle =\sum_{\ell=1}^3 b_{\ell}
Y_{\ell} , \\
 \Pi ^{(1)}(i,j)&=\langle D^\alpha (i)n(j)\rangle =\sum_{\ell=1}^3
m_{\ell} Y_{\ell},
\\
 M^{(0)}(i,j)&=\langle D(i)D(j)\rangle =\sum_{\ell=1}^3 b_{\ell}
Z_{\ell}, \\
 P^{(0)}(i,j)&=\langle D(j)\rangle =\sum_{\ell=1}^3 c_{\ell}
 Z_{\ell},
\\
 P^{(1)}(i,j)&=\langle D^\alpha (i)D(j)\rangle =\sum_{\ell=1}^3 m_{\ell}
Z_{\ell} , \\
 Q^{(0)}(i,j)&=\langle n(i)D(j)\rangle =\sum_{\ell=1}^3 a_{\ell}
Z_{\ell},  \\
 R^{(1)}(i,j)&=\langle n^\alpha (i)D(j)\rangle =\sum_{\ell=1}^3 d_{\ell}
 Z_{\ell},
\end{split}
\end{equation}
where, by assuming unitary lattice spacing, the function
$Y_{\ell}$ and $Z_{\ell}$ are given by
\begin{equation}
\label{B.3}
\begin{split}
Y_1 &=\langle n(j)\rangle _0  ,\\
Y_2 &=\langle n(i+1)n(j)\rangle _0, \\
Y_3 &=\langle D(i+1)n(j)\rangle _0,
\end{split}
\quad \,
\begin{split}
Z_1 &=\langle D(j)\rangle _0 , \\
Z_2 &=\langle n(i+1)D(j)\rangle _0  ,\\
Z_3 &=\langle D(i+1)D(j)\rangle _0 .
\end{split}
\end{equation}
The coefficients $a_{\ell}$, $b_{\ell}$, $c_{\ell}$, $d_{\ell}$
and $m_{\ell}$ appearing in Eqs. \eqref{B.2} can be explicitly
computed by means of the formulas \eqref{A.4}, \eqref{A.10} and by
using the properties of the $H_0$-representation. As an example,
the coefficients $a_{\ell}$ are given by:
\begin{equation}
\label{B.4}
\begin{split}
 a_1 &=\frac{B_1 (1+aX_1 +bX_2 )+2B_2 [(a+b)X_1 +(b+c)X_2 ]}{
  \Upsilon_0  }, \\
 a_2 &=\frac{a^2(B_1 +6 B_2 )X_1 +2bB_2 [1+cX_2 +b(X_1 +2X_2 )]}{
 \Upsilon_0 } \\
 &+\frac{a[B_1 (1+bX_2 )+2B_2 (1+4bX_1 +3bX_2 +2cX_2 )]}{
 \Upsilon_0} ,\\
 a_3 &=\frac{b^2(4B_2 X_1 +B_1 X_2 +6B_2 X_2 )+2cB_2 (1+2aX_1 +cX_2
)}{\Upsilon_0 } \\
 &+\frac{b[B_1 (1+aX_1 )+2B_2 (1+3aX_1 +cX_1 +4cX_2 )]}{
 \Upsilon_0 },
 \end{split}
\end{equation}
where the coefficients $a$, $b$ and $c$ have been defined in
Appendix \ref{App.A}. To determine the functions $Y_1$, $Y_2$,
$Y_3$ and $Z_1$, $Z_2$, $Z_3$ one requires translational
invariance which leads to the following equations:
\begin{equation}
\label{B.5}
\begin{split}
 \langle n(i)\rangle =\langle n(j)\rangle \, &\Rightarrow \,
n=\sum_{\ell=1}^3 c_{\ell} Y_{\ell}, \\
 \langle D(i)\rangle =\langle D(j)\rangle \, &\Rightarrow \,
D=\sum_{\ell=1}^3 c_{\ell} Z_{\ell}, \\
 \langle D(i)n(j)\rangle =\langle n(i)D(j)\rangle \, &\Rightarrow
\, \sum_{\ell=1}^3 b_{\ell} Y_{\ell}
=\sum_{\ell=1}^3 a_{\ell} Z_{\ell}, \\
 \langle D^\alpha (i)n(j)\rangle =\langle n^\alpha (i)D(j)\rangle \,
&\Rightarrow \, \sum_{\ell=1}^3 m_{\ell} Y_{\ell} =\sum_{\ell=1}^3
d_{\ell} Z_{\ell} .
 \end{split}
\end{equation}
The charge and double occupancy correlation functions can more
conveniently written in terms of the relative distance $m=\vert
i-j \vert$ as
\begin{equation}
\begin{split}
 \Lambda ^{(0)}(m)&=\langle n(i)n(i+m)\rangle =\sum_{\ell=1}^3 a_{\ell}
Y_{\ell} ,\\
 M^{(0)}(m)&=\langle D(i)D(i+m)\rangle =\sum_{\ell=1}^3 b_{\ell}
 Z_{\ell} .
  \end{split}
 \label{B.6}
\end{equation}
By solving the system \eqref{B.5}-\eqref{B.6} one finds
\begin{equation}
\begin{split}
 Y_{\ell} &=u_{\ell} \, n+v_{\ell} \,  D+w_{\ell} \Lambda
^{(0)}(m)+z_{\ell} M^{(0)}(m), \\
 Z_{\ell} &=p_{\ell} \, n+q_{\ell} \,  D+r_{\ell} \Lambda
^{(0)}(m)+s_{\ell} M^{(0)}(m),
  \end{split}
 \label{B.7}
\end{equation}
where the coefficients $u_{\ell}$, $v_{\ell}$, $w_{\ell}$,
$z_{\ell}$ and $p_{\ell}$, $q_{\ell}$, $r_{\ell}$, $s_{\ell}$ are
just functions of the coefficients $a_{\ell}$, $b_{\ell}$,
$c_{\ell}$, $d_{\ell}$ and $m_{\ell}$. Next, we note that by
definition
\begin{equation}
\begin{split}
 K^{(1)}(i,j)&=\frac{1}{2} \left[\Lambda ^{(0)}(m+1)+\Lambda
 ^{(0)}(m-1)\right],
\\
 P^{(1)}(i,j)&=\frac{1}{2} \left[M^{(0)}(m+1)+M^{(0)}(m-1)\right].
  \end{split}
 \label{B.8}
\end{equation}
Upon substituting Eqs. \eqref{B.2} and \eqref{B.7} in Eq.
\eqref{B.8} one gets
\begin{equation}
\begin{split}
& \Lambda ^{(0)}(m+1)+\Lambda ^{(0)}(m-1) \\
&=2\sum_{\ell=1}^3
d_{\ell}
 \left[u_{\ell}\, n+v_{\ell} \,  D+w_{\ell} \Lambda
 ^{(0)}(m)+z_{\ell}
M^{(0)}(m)\right] , \\
 &M^{(0)}(m+1)+M^{(0)}(m-1)
 \\
 &=2\sum_{\ell=1}^3 m_{\ell} [p_{\ell} \,
n+q_{\ell} \,  D+r_{\ell} \Lambda ^{(0)}(m)+s_{\ell} M^{(0)}(m)] .
 \end{split}
 \label{B.9}
\end{equation}
These equations give the sought recurrence relations for the
correlation function $\Lambda ^{(0)}(m)$ and $M^{(0)}(m)$. As a
result, one can write these equations as
\begin{equation}
\begin{split}
 \Lambda ^{(0)}(m+1) &=W_0 \Lambda ^{(0)}(m)-\Lambda ^{(0)}(m-1)
 \\
 &+W_1
M^{(0)}(m)+W_2 , \\
 M^{(0)}(m+1) &=\Theta _0 M^{(0)}(m)-M^{(0)}(m-1)
 \\
 &+\Theta _1 \Lambda
^{(0)}(m)+\Theta _2 ,
 \end{split}
 \label{B.10}
\end{equation}
where one has defined
\begin{equation}
\label{B.11}
\begin{split}
 W_0 &=2\sum_{\ell=1}^3 d_{\ell} w_{\ell},  \\
 W_1 &=2\sum_{\ell=1}^3 d_{\ell} z_{\ell}, \\
 W_2 &=2\sum_{\ell=1}^3 d_{\ell} [u_{\ell} \, n+v_{\ell} \,  D],
\end{split}
\quad \;
\begin{split}
 \Theta_0 &=2\sum_{\ell=1}^3 m_{\ell} s_{\ell} , \\
 \Theta_1 &=2\sum_{\ell=1}^3 m_{\ell} r_{\ell} , \\
 \Theta_2 &=2\sum_{\ell=1}^3 m_{\ell} [p_{\ell} \, n+q_{\ell} \,  D] .
\end{split}
\end{equation}
Now, upon defining
\begin{equation}
\begin{split}
 G(m)&=\Lambda ^{(0)}(m)-n^2 ,\\
 F(m)&=M^{(0)}(m)-D^2 ,
  \end{split}
 \label{B.13}
\end{equation}
one can rewrite Eq. \eqref{B.10} as
\begin{equation}
\begin{split}
 G(m+1)&=W_0 G(m)-G(m-1)+W_1 F(m)
 \\
 &+W_2 +W_0 n^2-2n^2+W_1 D^2, \\
 F(m+1)&=\Theta _0 F(m)-F(m-1)+\Theta _1 G(m)
 \\
 &+\Theta _2 +\Theta _0
D^2-2D^2+\Theta _1 n^2 .
 \end{split}
 \label{B.12}
\end{equation}
It is not difficult to show that the following identities do hold
for any $n$:
\begin{equation}
\begin{split}
 W_2 +W_0 n^2-2n^2+W_1 D^2&=0 ,\\
 \Theta _2 +\Theta _0 D^2-2D^2+\Theta _1 n^2&=0 .
 \end{split}
 \label{B.14}
\end{equation}
Therefore, Eqs. \eqref{B.12} simplify as
\begin{equation}
\begin{split}
 G(m+1)&=W_0 G(m)-G(m-1)+W_1 F(m) ,\\
 F(m+1)&=\Theta _0 F(m)-F(m-1)+\Theta _1 G(m) .
 \end{split}
 \label{B.15}
\end{equation}
The results \eqref{B.14} are just a manifestation of the ergodic
conditions
\begin{equation}
\begin{split}
\lim_{m\to \infty} \Lambda ^{(0)}(m)&=\lim_{m\to \infty} \langle
n(i)n(i+m)\rangle =\langle n(i)\rangle ^2 ,
 \\
\lim_{m\to \infty } M^{(0)}(m)&=\lim_{m\to \infty } \langle
D(i)D(i+m)\rangle =\langle D(i)\rangle ^2 ,
 \end{split}
 \label{B.16}
\end{equation}
which break down when there is a long range order. Examples of
this scenario are shown in Sec. \ref{sec_III}. Equations
\eqref{B.15} together with Eq. \eqref{EHM_22} allow one to compute
the correlation functions $\Lambda ^{(0)}(m)$ and $M^{(0)}(m)$ for
any value of $m$. Then, by means of Eq. \eqref{B.7} one can
compute the functions $Y_1$, $Y_2$, $Y_3$, and $Z_1$, $Z_2$,
$Z_3$, whose knowledge allows one to compute any correlation
functions, by using Eq. \eqref{B.2} or relative formulas.
Equations \eqref{B.15} show that $G(m)$ and $F(m)$ have the
following form
\begin{equation}
\begin{split}
 G(m)=Ap^m+Bq^m, \\
 F(m)=Cp^m+Eq^m ,
  \end{split}
 \label{B.17}
\end{equation}
where the coefficients $A$, $B$, $C$, $E$, $p$, and $q$ are given
by
\begin{equation}
\begin{split}
 A&=\frac{G_0 (F_2 G_0 -F_0 G_2 )+2G_1 (F_0 G_1 -F_1 G_0 )+G_0 Q}{2Q}, \\
  B&=G_0 -A, \\
 C&=\frac{F_0 (F_2 G_0 -F_0 G_2 )+2F_1 (F_0 G_1 -F_1 G_0 )+F_0 Q}{2Q}, \\
  E&=F_0 -C, \\
 p&=\frac{F_2 G_0 -F_0 G_2 -Q}{2(F_1 G_0 -F_0 G_1 )}, \\
 q&=\frac{F_2 G_0 -F_0 G_2 +Q}{2(F_1 G_0 -F_0 G_1 )},
 \end{split}
 \label{B.18}
\end{equation}
with
\begin{equation}
\begin{split}
Q &=\big[(F_2 G_0 - F_0 G_2)^2 \\
&+ 4 (F_1 G_0 - F_0 G_1) (F_1 G_2-F_2 G_1) \big]^{1/2},
 \end{split}
 \label{B.19}
 \end{equation}
and
\begin{equation*}
\begin{split}
 G_0 &=G(0)=n+2D-n^2, \\
 G_1 &=G(1)=2\lambda ^{(1)}-n^2, \\
 G_2 &=G(2)=2\kappa ^{(2)}-n-2D-n^2,
 \end{split}
 \end{equation*}
and
\begin{equation*}
\begin{split}
 F_0 &=F(0)=D-D^2 ,\\
F_1 &=F(1)=2\theta ^{(1)}-D^2, \\
 F_2 &=F(2)=2\pi ^{(2)}-D-D^2 .
\end{split}
 \end{equation*}
One can rewrite Eq. \eqref{B.17} as
\begin{equation}
\begin{split}
 G(m)&=A \,[\mbox{sign}(p)]^m \, e^{-m/\xi_n}+B\, [\mbox{sign}(q)]^m \, e^{-m/\xi_D }, \\
 F(m)&=C \,[\mbox{sign}(p)]^m \, e^{-m/\xi_n}+E\, [\mbox{sign}(q)]^m \, e^{-m/\xi_D
 },
  \end{split}
 \label{B.21}
\end{equation}
where the two correlation lengths $\xi_n$ and $\xi_D$ are defined
as
\begin{equation}
\begin{split}
 \xi_n &=\left[ {\ln \left( {\frac{1}{\vert p \vert}} \right)}
\right]^{-1},
\\
\xi _D &=\left[ {\ln \left( {\frac{1}{\vert q \vert}} \right)}
\right]^{-1}
 \end{split}
 \label{B.22}
\end{equation}
As it has been shown in Sec. \ref{sec_III}, in the limit of zero
temperature the correlation length $\xi_n$ diverges in some
regions of the $(U,n)$ plane, signalling the presence of a phase
with long-range order.

\end{document}